\documentclass[nofootinbib,aps,prx,twocolumn]{revtex4-1}

\usepackage{amsfonts}
\usepackage{amsmath}
\usepackage{calc}
\usepackage{color}
\usepackage[colorlinks=true, linkcolor=blue,urlcolor=true,citecolor=blue,anchorcolor=green]{hyperref}
\usepackage{graphicx} 
\usepackage{geometry}
\usepackage{enumerate}
\usepackage{wasysym}
\usepackage{amssymb}
\usepackage{amsthm}
\usepackage{algorithm,algorithmicx}
\usepackage[noend]{algpseudocode}
\usepackage[caption=false]{subfig}

\usepackage{todonotes}
\usepackage{comment}
\usepackage{framed}

\newcommand{\be}{\begin{eqnarray} \begin{aligned}}
\newcommand{\ee}{\end{aligned} \end{eqnarray} }
\newcommand{\benn}{\begin{eqnarray*} \begin{aligned}}
\newcommand{\eenn}{\end{aligned} \end{eqnarray*}}
\newcommand{\cancel}[1]{} 
\newcommand*{\cB}{\mathcal{B}}
\newcommand*{\cC}{\mathcal{C}}

\newcommand*{\cL}{\mathcal{L}}

\newcommand*{\cN}{\mathcal{N}}
\newcommand*{\cD}{\mathcal{D}}

\newcommand*{\cR}{\mathcal{R}}

\newcommand*{\cP}{\mathcal{P}}
\newcommand*{\cS}{\mathcal{S}}

\newcommand*{\cV}{\mathcal{V}}

\newcommand*{\tr}{\mathop{\mathrm{tr}}\nolimits}

\newcommand{\bc}{\begin{center}}
\newcommand{\ec}{\end{center}}

\usepackage{mathtools}
\mathtoolsset{showonlyrefs}

\usepackage{amsfonts}

\def\01{\{0,1\}}

\newcommand{\ket}[1]{|#1\rangle}
\newcommand{\bra}[1]{\langle#1|}

\newcommand*{\GKP}{\mathsf{GKP}}

\newcommand{\tvector}[2]{\left(#1\quad #2\right)^T}
\newcommand{\innerprod}[2]{{#1}^T{#2}}

\newcommand*{\gkpsquarezero}{\ket{\overline{0}}_{\square}}
\newcommand*{\gkpsquareone}{\ket{\overline{1}}_{\square}}
\newcommand*{\gkpsquarelogicalX}{\overline{X}_{\square}}
\newcommand*{\gkpsquarelogicalZ}{\overline{Z}_{\square}}

\newcommand*{\gkpreclogicalX}{\overline{X}_{r}}
\newcommand*{\gkpreclogicalZ}{\overline{Z}_{r}}

\newcommand*{\symplecticform}{J}

\newcommand*{\ehex}{e_{\hexagon}}

\begin{document}
\title{Enhanced noise resilience of the surface-GKP code via designed bias}
 \author{Lisa H\"anggli}
 \author{Margret Heinze}
 \author{Robert K\"onig}
 \affiliation{Institute for Advanced Study and Zentrum Mathematik, Technical University of Munich, Germany}

\begin{abstract}
We study the  code  obtained by 
concatenating the standard
single-mode Gottesman-Kitaev-Preskill
(GKP) code with the surface code. We show that the noise tolerance of this surface-GKP code with respect to (Gaussian) displacement errors improves when a single-mode squeezing unitary is applied to each mode assuming that the identification of quadratures with logical Pauli operators is suitably modified. We observe noise-tolerance thresholds of up to~$\sigma\approx 0.58$ shift-error standard deviation when the surface code is decoded without using GKP syndrome information. In contrast, prior results by Fukui et al.~and Vuillot et al.~report a threshold between~$\sigma\approx 0.54$ and~$\sigma\approx 0.55$ for the standard (toric- respectively) surface-GKP code. The modified surface-GKP code effectively renders the mode-level physical noise asymmetric, biasing the logical-level noise on the GKP-qubits. The code can thus benefit from the resilience of the surface code against biased noise. 
We use  the approximate maximum likelihood decoding algorithm of Bravyi et al.~to obtain our threshold estimates. 
Throughout, we consider an idealized scenario where measurements are noiseless and GKP states are ideal. 
 Our work demonstrates that Gaussian encodings of individual modes can enhance concatenated codes.
\end{abstract}
\maketitle
\tableofcontents

\section{Introduction}\label{sec:introduction}
We study a modified  surface-Gottesman-Kitaev-Preskill (surface-GKP) code subject to classical isotropic Gaussian displacement noise. The modification includes an additional encoding of each underlying bosonic mode by means of a single-mode Gaussian unitary, as well as an adaptation of the concatenation procedure of the two codes. Specifically, we study the case where the Gaussian unitary is given by single-mode squeezing -- this is not to be confused with the term squeezing in the context of non-ideal, normalizable GKP states. The squeezing effectively transforms the error from isotropic to anisotropic displacement noise. At the level of the logical GKP-qubits, the anisotropic displacement noise manifests itself as biased Pauli noise. To exploit this asymmetry, we concatenate with a surface code in such a way that the primary direction of the asymmetric noise is aligned with the preferred direction of biased Pauli noise  for surface code decoding. We show numerically that the bias designed this way causes an improvement of the noise tolerance threshold of this concatenated asymmetric surface-GKP code. Our result shows that -- contrary to a common belief -- encoding bosonic modes into other bosonic modes by means of a Gaussian unitary can be beneficial -- namely when considering concatenated codes.

Without concatenation, Gaussian encodings are indeed not beneficial.  At the level of a single bosonic mode, the  futility  of using a Gaussian unitary as an encoding map can  be illustrated using the random displacement channel (also called classical displacement noise channel)
\begin{align}\label{eq:intronoisechannel}
  \cN_f(\rho)&=\int_{\mathbb{R}^2} f(\nu) D(\nu)\rho D(\nu)^\dagger d^{2}\nu\ ,
\end{align}
where $f$ is a probability density function on the phase space and $D(\nu)$ is the (Weyl) displacement operator. Assume that we apply a single-mode squeezing unitary~$U=U_{S_r^{-1}}$ (cf.\ Eq.~\eqref{eq:squeezingsymplectic} below) with squeezing parameter~$r>1$ before the action of the noise~$\cN_f$. The resulting  channel can be written as $\cN_f(U\rho U^\dagger)=U\cN_{\tilde{f}}(\rho)U^\dagger$ where $\cN_{\tilde{f}}$ is  another random displacement channel with $\tilde{f}(\nu)=f(\mathsf{diag}(r^{1/2},r^{-1/2})\nu)$ a squeezed version of the  distribution~$f$. For example, if $f\equiv f_{\sigma^2}$ is a centred Gaussian with covariance matrix~$\sigma^2 I_2$ associated with an isotropic Gaussian displacement channel
\begin{align}
\cN_{f_{\sigma^2}}(\rho)&=\frac{1}{2\pi\sigma^2}\int_{\mathbb{R}^2} e^{-\frac{\Vert \nu\Vert^2}{2\sigma^2}} D( \nu)\rho D( \nu)^\dagger d^{2}\nu\ ,\label{eq:Gaussnoisechannel}
\end{align}
$\Vert \nu\Vert^2=\nu\cdot \nu=\nu_1^2+\nu_2^2$,
then~$\tilde{f}$ is a centred Gaussian with covariance matrix $\mathsf{diag}(r^{-1}\sigma^2,r\sigma^2)$. In particular, the strength of the ``worst noise'' -- as expressed by the maximal singular value of the covariance matrix of the noise -- cannot be lowered by the introduction of the encoding unitary~$U$.  This is a fundamental limitation on the way noise can be reshaped.

This kind of obstacle to error correction by Gaussian operations/encoding maps was formalized more generally in~\cite{niseketalnogoQerror09}: it was argued that Gaussian one-mode states cannot be protected by Gaussian operations (even against Gaussian errors). Such results fall in line with a number of related no-go results concerning e.g.,  entanglement distillation and entanglement swapping~\cite{eisertetal02Gaussiandistillation,fiurasekgaussiantransf02,giedkecirac02} by means of Gaussian operations. 
A related recent no-go-result~\cite{vuillotetal}, specifically dealing with Gaussian  CV-into-CV encodings, concerns~$k$ bosonic modes encoded into~$n\geq k$ modes by means of a Gaussian unitary: consider the problem of recovering from classical isotropic Gaussian displacement noise by means of syndrome measurements of linear combinations of the quadratures followed by  maximum likelihood decoding. Here the noise is described by a  random variable with centred normal distribution~$Z\sim\mathsf{N}({\bf 0},\sigma^2 I_{2n})$  on the phase space~$\mathbb{R}^{2n}$. The authors of~\cite{vuillotetal} show that the resulting effective logical noise is displacement noise with a centred normal distribution, whose covariance matrix has
eigenvalues $\{\sigma^{(j)}_P,\sigma^{(j)}_Q\}_{j=1}^k$ satisfying~$\sigma^{(j)}_P\sigma^{(j)}_Q=\sigma^2$ for every $j=1,\ldots, k$. In other words, the worst-case noise variance  (per quadrature) does not improve even at the 
logical (encoded) level when using a Gaussian encoding map. Overall, these results appear to suggest that readily available  Gaussian unitaries 
may not be used  to boost noise tolerance levels of bosonic error-correcting codes.

We show that when bosonic degrees of freedom are used to encode  logical qubits by means of concatenated codes, this apparent no-go theorem no longer applies. Indeed, we find that the introduction of additional single-mode Gaussian squeezing unitaries in surface-GKP codes increases  fault-tolerance error thresholds, implying that more noise can be tolerated.  Here the error-correction threshold of an infinite code family of bosonic codes is the maximum physical (single-mode) displacement noise standard deviation~$\sigma$
(for a centered normal distribution) such that the logical failure rate can be made arbitrarily small by using  sufficiently large code sizes.

A scenario where additional Gaussian unitaries improve a (unconcatenated) GKP code has been discussed previously in the literature: Recall that GKP codes  are constructed algebraically from  certain lattices in~$\mathbb{R}^{2n}$~\cite{gkp01}. The chosen lattice determines the code space dimension~$K$ as well as the code's robustness against displacements. In particular, the size (norm) of the smallest uncorrectable shift depends on the lattice: shifts inside the Voronoi cell of the dual lattice are correctable. As observed early on~\cite{gkp01}, this implies for example that in two dimensions the GKP code  based on a hexagonal lattice beats the ``standard'' square lattice GKP code in this respect, being able to tolerate larger displacements. (This is related to the fact that the hexagonal lattice permits the densest sphere packing in~$\mathbb{R}^2$, see~\cite{Fejes1942}.) More detailed estimates of the difference
between hexagonal and square lattice GKP codes were obtained in~\cite{nohetal18}  in terms of estimates of the logical error probability for a single encoded qubit in the presence of pure-loss noise. Furthermore, the characterization of correctable displacements in terms of Voronoi cells has also been used to derive bounds on the (quantum and classical)  capacity of the channel~\eqref{eq:intronoisechannel}, see \cite{gkp01} and~\cite{harringtonpreskill}.  Importantly, the hexagonal lattice GKP code is related to the square lattice GKP code by
a Gaussian unitary (see Section~\ref{sec:asymGKP}). Thus the square lattice GKP code can be enhanced by a simple application of a Gaussian unitary. This improvement may appear minor for a single mode (e.g., it changes a constant prefactor from $\pi/4$ to $\pi/(2\sqrt{3})$ in the logical error probability considered in~\cite{nohetal18}). However, as we argue here,  similar modifications  lead to dramatic improvements in the setting of concatenated codes.

The underlying mechanism which improves fault-tolerance properties in the setting of concatenated codes is not simply a matter of increasing the volume of a Voronoi cell. Instead,  the introduction of an additional encoding map has the effect of  artificially shaping the logical-level noise, yielding biased qubit noise. If the GKP code is appropriately concatenated with the surface code, this leads to improved fault-tolerance properties due to the fact that known decoders for the surface code can benefit from a noise bias. The same mechanism therefore applies to other codes resilient to biased noise. In this sense, our work on surface-GKP codes is merely a case study illustrating this principle.

In more detail, we seek to design optimized codes protecting
against single-mode i.i.d.~(independently and identically distributed) displacement noise of the form~\eqref{eq:Gaussnoisechannel} on each mode.
Starting from a standard  square lattice GKP code we apply a single-mode squeezing unitary to every mode yielding a rectangular, i.e., asymmetric lattice GKP code. As already mentioned, applying the one-mode squeezing operation with parameter $r>1$ (see Eq.~\eqref{eq:squeezingsymplectic} below), we are effectively dealing with anisotropic noise~$\cN_{f_{Z_r}}$, where~$Z_r\sim \mathsf{N}({\bf 0},\Sigma_r)$ is an anisotropically distributed Gaussian random variable with covariance matrix
 \begin{align}
 \Sigma_r
 &=
\begin{pmatrix}
\sigma^2/r & 0\\
0 & r\sigma^2
\end{pmatrix}=:\mathsf{diag}(\tilde{\sigma}^{2}_Q,\tilde{\sigma}^{2}_P)\ .\label{eq:tildeSigmaimprov}
 \end{align} In other words, this is equivalent to error correction with square lattice GKP codes subject to asymmetric noise (for $r>1$):
 the noise variance $\sigma^2_Q$ in the $Q$-quadrature is reduced, while the variance~$\sigma^2_P$ in the $P$-quadrature is increased. After decoding the GKP-qubit, this results in biased Pauli noise: a fact that we can exploit when considering surface-GKP codes. For this to work, the concatenation of codes needs to be done in such a way as to make the primary direction of the noise asymmetry align 
 suitably with the surface code, i.e., those (Pauli) axes where  the noise bias is favorable for the surface code decoder.

The fact that surface codes perform well under certain biased noise has been established recently~\cite{tuckettetal18,tuckettetal19}; we briefly review these results in Section~\ref{sec:priorbiased}. Our main contribution
is  a detailed discussion of how the surface-GKP code needs to be modified to
exploit this property of the surface code. This is detailed in Section~\ref{sec:engineered}. We also provide numerical evidence showing that this strategy indeed provides higher error thresholds (see Section~\ref{sec:numerical}).

\subsection*{Prior work}
 Among strategies to go beyond the no-go results on Gaussian encoding maps is the recent paper~\cite{nohgirvinjiang19}. Here Noh, Girvin and Jiang show that Gaussian CV-into-CV encodings are meaningful if combined with non-Gaussian resources. By using GKP states and modular quadrature measurements, they define new families of non-Gaussian CV-into-CV encoding maps derived from (qubit)  stabilizer codes. Here we take a different approach and consider DV-into-CV encoding maps obtained by concatenating surface codes with modified (asymmetric) square lattice GKP codes.

We note that the codes we study here are closely related to the surface-GKP and toric-GKP codes previously investigated in pioneering work by Fukui et al.~\cite{fukuietal18} and Vuillot et al.~\cite{vuillotetal} respectively. These authors study the concatenation of the GKP code with the surface/toric code,  providing error threshold estimates.
They consider a noise model which is i.i.d.~Gaussian noise with isotropic distribution $\mathsf{N}({\bf 0},\sigma^2 I_{2})$ on every mode, and, in~\cite{vuillotetal}, additional noise in the syndrome measurements.  Numerically, threshold estimates are obtained in~\cite{fukuietal18,vuillotetal} by 
level-by-level concatenated decoding: modular position- and momentum measurements are performed to extract GKP syndrome information for every GKP-qubit. By Bayesian
update, an effective Pauli-error distribution is computed, either conditioned on the particular GKP syndrome information (meaning that this prior GKP-information is taken into account when decoding) or averaged over the GKP measurement outcomes. Subsequently, Edmond's maximum matching algorithm~\cite{edmonds_1965} (in the following referred to as  the minimum weight matching decoder)  is applied to decode the surface code given this effective Pauli noise: for the case where GKP information is exploited, the latter is heuristically incorporated into a weighting of the edges. Additional (heuristic) decoders are studied in~\cite{vuillotetal} in the case where the GKP syndrome information (bosonic measurements) is noisy. Furthermore, the threshold error probability is related to a phase transition in a classical statistical model in this case. 
 
 Closely related to our work are approaches to exploiting asymmetry in (Gaussian) displacement noise by means of code concatenation, e.g., of cat-codes with the repetition code. An additional key ingredient we use is the behavior of the surface code with respect to biased Pauli noise. 
 In order to highlight the relationship to our work, we briefly discuss these prior works in Section~\ref{sec:priorbiased} after introducing some more terminology.

\subsection*{Outline}
We begin by presenting some background material in order to fix notation and set the stage. In Section~\ref{sec:surfacecodes} we review the surface code, maximum likelihood decoding, and the Bravyi-Suchara-Vargo decoder. 
In Section~\ref{sec:gkp}, we discuss different versions of the Gottesman-Kitaev-Preskill code as well as error recovery procedures with and without side information. 

We then present our main findings: In Section~\ref{sec:surfacegkp}, we introduce the (modified) surface-GKP code and show how the introduction of single-mode Gaussian unitaries effectively leads to biased Pauli noise. In Section~\ref{sec:numerical}, we present our numerical results for the noise tolerance threshold using modified surface-GKP codes. We conclude in Section~\ref{sec:conclusions}.

\section{The surface code\label{sec:surfacecodes}}
In this section, we review surface codes~\cite{bravyikitaev98,KITAEV20032}. These are CSS (stabilizer) codes with geometrically local generators in 2~dimensions, encoding one logical qubit into $n$~physical qubits. Subsequently, we discuss the decoding problem for these codes. In particular, we review the definition of  maximum likelihood decoding for general stabilizer qubit codes. We then  discuss the BSV-decoder (by Bravyi, Suchara, and Vargo~\cite{bsv14}) which approximates the maximum likelihood decoder for the surface code.

\subsection{Definition of the surface code}
Consider a $(L-1)\times (L-1)$ square lattice, i.e., a rectangular grid with $L-1$~edges on each side. If we first identify its vertical (left/right) boundaries, yielding a cylinder, and subsequently perform a vertical cut located in between any two vertical lines of the lattice, the result (mapped back to the plane) is the grid on which a $L\times L$ surface code with ``smooth'' top/bottom boundaries and ``rough'' left/right boundaries is defined, cf.\ Fig.~\ref{surfacecodedfour} below. These are the only surface codes we consider here for simplicitly. Note, however, that a~$r\times s$ surface code for arbitrary~$r,s\geq 1$ can be defined similarly, and also the boundary conditions may be chosen differently, see~\cite{bravyikitaev98} as well as~\cite{rotatedsurfacecodes} for the so-called rotated surface codes.

The surface code is now defined as follows: to every edge of the grid a qubit is attached, yielding a total number of $n=L^2+(L-1)^2$ physical qubits. Denote by $v$, $e$, and $p$ a vertex, edge and plaquette of the grid respectively. The (co-)boundaries $\delta v$, $\partial p$ consist of the edges incident on~$v$ and the edges bounding~$p$, respectively. By the usual convention, the stabilizer generators of the surface code are on the one hand given by tensor products $A_v\coloneqq\prod_{e \in \delta v}X_e$ of Pauli-$X$ operators acting on the edges incident to a vertex~$v$, and on the other hand by tensor products~$B_p\coloneqq\prod_{e\in \partial p}Z_e$ of Pauli-$Z$ operators acting on the edges surrounding a plaquette $p$. Denote by $\mathsf{V}$ the set of vertices, and by $\mathsf{P}$ the set of plaquettes of the grid. Since $|\mathsf{V}|=|\mathsf{P}|=L(L-1)$, we have $2L(L-1)$ independent stabilizer generators, yielding $k=n-2L(L-1)=1$ encoded qubits. The centralizer of the stabilizer group $\mathcal{S}\coloneqq\langle A_v,B_p\rangle$ with respect to the $n$-qubit Pauli group $\mathcal{P}_n$, where $v,p$ run over the sets $\mathsf{V},\mathsf{P}$, is given by
\begin{align}
\cC(\cS):=\{P\in \cP_n\ |\ PS=SP\textrm{ for all }S\in \cS\}\ . \label{eq:centralizerdefinition}
\end{align}
We can define logical Pauli-$X$ (Pauli-$Z$) operators, denoted by $\overline{X}$ ($\overline{Z}$), as the product of Pauli-$X$ (Pauli-$Z$) operators along the left (top) boundary of the lattice, cf.~Fig.~\ref{surfacecodedfour}. Then $\overline{X},\overline{Z}\in \mathcal{C}(\mathcal{S})\setminus \mathcal{S}$, and $\overline{X}\,\overline{Z}=-\overline{Z}\,\overline{X}$, as required. Moreover, the weight of both~$\overline{X}$ and $\overline{Z}$ is~$L$. These are minimal-weight logical operators, i.e., the code distance is~$d=L$. 

\subsection{Error recovery for the surface code\label{sec:errorrecoverysurfacecode}}
 Recall that recovery from errors (i.e., after the action of a noise channel) in a stabilizer code 
 proceeds by measurement of stabilizer operators
  (cf.~Section~\ref{sec:gkpdecoding}) and subsequent application of
  a recovery map (mapping back to the code space).

A noise channel in this context is a CPTP map $\cN\colon\cD((\mathbb{C}^2)^{\otimes n})\to\cD((\mathbb{C}^2)^{\otimes n})$ on the density operators $\cD((\mathbb{C}^2)^{\otimes n})$ on the Hilbert space $(\mathbb{C}^2)^{\otimes n}$. Here we consider probabilistic Pauli noise, i.e.,  noise channels of the form 
\begin{align}
\cN_{\pi}(\rho) &=\sum_{E\in\cP_n} \pi(E) E\rho E^\dagger\ ,\label{eq:errorchannelpi}
\end{align}
with $\pi$ a probability distribution on the Pauli group~$\cP_n$. Note that in our applications later on, we (predominantly) consider i.i.d.~channels of the form~$\cN_{p^n}(\rho)=\cN_{p=(p_I,p_X,p_Y,p_Z)}^{\otimes n}(\rho)$, where the single-qubit noise channel $\cN_{p}$ 
 represents random Pauli noise (see Eq.~\eqref{eq:qubitnoisechanneln} below). We  also consider  channels of the form $\cN_{\prod_{j=1}^n p^{(j)}}=\otimes_{j=1}^n \cN_{p^{(j)}}$ where each qubit $j\in \{1,\ldots,n\}$ experiences random Pauli  noise given by a distribution~$p^{(j)}=(p_I^{(j)},p_X^{(j)},p_Y^{(j)},p_Z^{(j)})$.  However, such a restriction does not need to be imposed at this point.

The stabilizers we measure in the surface code are the (commuting) vertex- and plaquette-type stabilizers~$\{A_v\}_{v\in\mathsf{V}}$,  $\{B_p\}_{p\in\mathsf{P}}$.
 This provides a {\em syndrome} (measurement outcome)~$s=\left(\{s_v\}_{v\in\mathsf{V}},\{s_p\}_{p\in\mathsf{P}}\right)\in \{0,1\}^{|\mathsf{V}|+|\mathsf{P}|}$ and simultaneously projects the state into the (common) eigenspaces of each $A_v$ associated with eigenvalue~$(-1)^{s_v}$, and the eigenspaces of each $B_p$ associated with eigenvalue~$(-1)^{s_p}$. A syndrome~$s$ is observed with probability
\begin{align}
p(s):&=\tr(\Pi(s)\cN_{q^n}(\rho))\qquad\textrm{ where }\quad\\
\Pi(s)&=\prod_{p\in\mathsf{P}}\frac{1}{2}(I+(-1)^{s_p} B_p)\prod_{v\in\mathsf{V}}\frac{1}{2}(I+(-1)^{s_v} A_v)  ,
\end{align}
and the post-measurement state is given by $p(s)^{-1}\Pi(s)\cN_{q^n}(\rho)\Pi(s)$. A recovery procedure mapping the post-measurement state back to the code space is described by a unitary correction operation~$C(s)$ depending on the syndrome~$s$. Choosing a recovery procedure (sometimes referred to as a ``decoder'') thus amounts to the choice of a  function $s\mapsto C(s)$ associating a correction operation to a given syndrome. 

The success probability of a given recovery strategy can be analysed by considering a single $n$-qubit Pauli error $E\in\cP_n$.
The above recovery procedure (using Pauli corrections~$C(s)$) successfully recovers from an error~$E$ if the following conditions are satisfied:
\begin{enumerate}[(i)]
\item \label{it:firstmldecodingprop}
First, the coset $C(s)\cC(\cS)$  of the chosen correction operation~$C(s)$ 
coincides with the coset~$E\cC(\cS)$ of the actual error~$E$. Here
$\cC(\cS)$ is the centralizer of~$\cS$ in the Pauli group, see~\eqref{eq:centralizerdefinition}. This means that $C(s)$ causes the same syndrome as~$E$, that is, $C(s)$ maps the corrupted state back to the code space.
\item\label{it:secondconditiondecodingprop}
Second, the chosen correction operation $C(s)$ belongs to the same coset of $\cS$ inside $E\cC(\cS)$. More precisely, for every syndrome~$s$, fix a ``representative error'' $E_s$ which causes syndrome~$s$. Then the coset~$E_s\cC(\cS)$ can be partitioned into
\begin{align}
E_s\cC(\cS)=E_s\cS\cup E_s\overline{X}\cS\cup E_s\overline{Y}\cS\cup E_s\overline{Z}\cS\ ,\label{eq:cosetpartition}
\end{align}
where all elements in each of the four subsets have the same logical action. 
Decoding is successful if for the syndrome~$s$ caused by~$E$ (that is, $E\cC(\cS)=E_s\cC(\cS)$), $E$ and the correction $C(s)$ belong to the same subset on the right hand side of~\eqref{eq:cosetpartition}.
\end{enumerate}

\subsubsection{Maximum likelihood decoding}
Given the available syndrome information~$s$, the optimal choice of recovery map~$C(s)$ is determined by the maximum likelihood decoding strategy: one sets
\begin{align}
\cC_{\textrm{ML}}(s):=\arg\max_{\cC\in \{E_s\cS, E_s\overline{X}\cS, E_s\overline{Y}\cS, E_s\overline{Z}\cS\}}\pi(\cC)\label{eq:cmldecoder}
\end{align}
i.e., finds the coset with maximal weight with respect to the error distribution~$\pi$ (cf.~\eqref{eq:errorchannelpi}), and subsequently selects a correction $C_{\textrm{ML}}(s)\in \cC_{\textrm{ML}}(s)$ belonging to this coset (arbitrarily). This guarantees that the correction satisfies~\eqref{it:firstmldecodingprop} and simultaneously maximizes the probability that it obeys~\eqref{it:secondconditiondecodingprop}. The resulting (averaged) success probability is then given by 
\begin{align}
P_{success}&=\sum_{s}\pi(\cC_{\textrm{ML}}(s))\ .
\end{align}

\subsubsection{The Bravyi-Suchara-Vargo (BSV) decoder\label{sec:bsvdecoder}}
While the maximum likelihood decoder is optimal from the point of view of decoding error probability, 
the determination of the coset~\eqref{eq:cmldecoder} is non-trivial (even for simple i.i.d.~error models) as computing the coset probabilities involves sums over sets of exponential size. In other words, the maximum likelihood decoder cannot be realized efficiently.

To overcome this obstacle, Bravyi, Suchara and Vargo~\cite{bsv14} have proposed a decoding strategy~$\tilde{C}^\chi_{ML}(s)$ which approximates the maximum likelihood decoder~$C_{ML}(s)$.  The decoding strategy depends on a parameter $\chi\in\mathbb{N}$ (the {\em bond dimension}) and becomes exact, i.e., maximum likelihood decoding, in the limit~$\chi\rightarrow\infty$. In contrast to the maximum likelihood decoder, it is efficiently computable (for small, i.e., typically constant~$\chi$): the computation of $\tilde{C}^\chi_{ML}(s)$ from $s$~involves $O(n\chi^3)$ basic arithmetic operations. The proposed decoder  generally applies to ``local'' stochastic error models for 2D~stabilizer codes with local generators, see e.g., the appendix of~\cite{tuckettetal19} for an application to 2D~color codes.

\begin{figure}
   \subfloat[Surface code with distance $d=4$.\label{surfacecodedfour}]{\includegraphics[width=.38\textwidth]{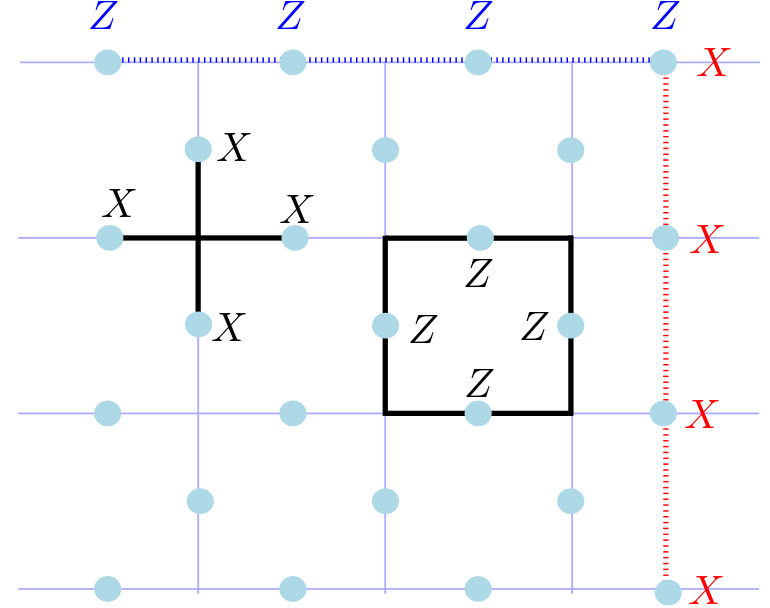}}\\
   
   \subfloat[Tensor network for coset probability $\pi(E\cS)$.\label{tnsprob}]{\includegraphics[width=.38\textwidth]{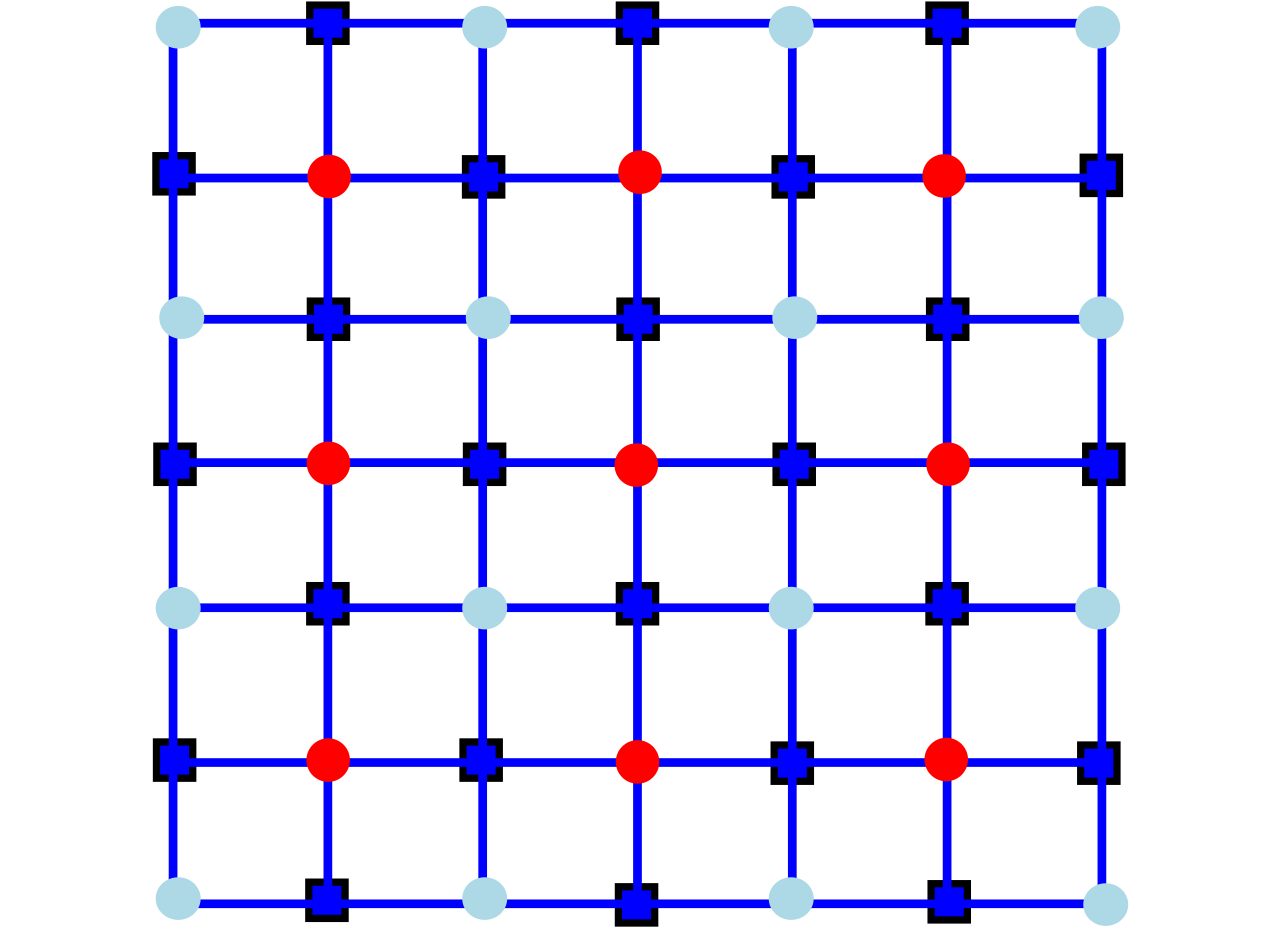}}\\
   
   \subfloat[Bulk tensors; the first two associated with a qubit at~$(j,k)$. \label{bulktensors} ]{\includegraphics[width=.32\textwidth]{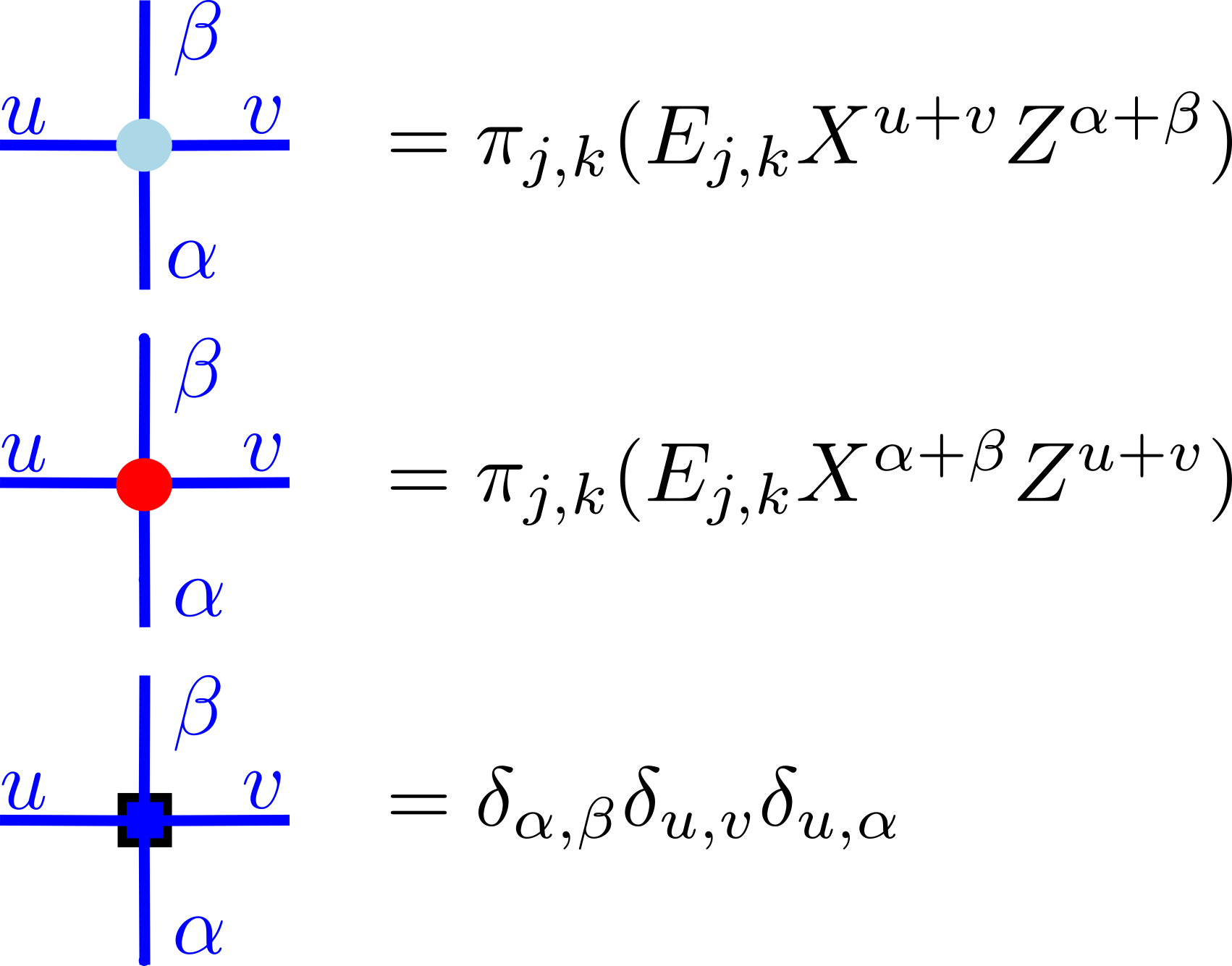}}
   
   \caption{Tensor network associated with the BSV decoder.
    Fig.~\ref{surfacecodedfour} shows the surface code with $d=4$, logical operators~$\overline{X}$ and $\overline{Z}$ and two stabilizer generators.
    Fig.~\ref{tnsprob} gives the tensor network whose contraction gives the value
    $\pi(E\mathcal{S})$ of a coset determined by a Pauli operator~$E=\otimes_{(j,k)}E_{j,k}$ acting with the single-qubit Pauli operator~$E_{j,k}$ on the qubit at location~$(j,k)$. Evaluating this scalar for $E\in\{E_s,E_s\overline{X},E_s\overline{Y},E_s\overline{Z}\}$, where $E_s$ is a representative error giving syndrome~$s$, allows to perform maximum likelihood decoding according to Eq.~\eqref{eq:cmldecoder}. Local tensors are defined as shown in Fig.~\ref{bulktensors}.
    (Tensors near the boundaries  are defined similarly.)
    A priori Pauli-error probabilities for a qubit located at $(j,k)$ are denoted $\pi_{j,k}(Q)=p_Q$, where $p_Q$ is the a priori probability of a Pauli error $Q\in \{I,X,Y,Z\}$.
}\label{fig:BSVdecoder}
\end{figure}
To briefly summarize the construction of the BSV-decoder, consider the case of Pauli noise where the distribution~$\pi$ over Pauli errors factorizes into a product of (not necessarily identical but independent) distributions over single-qubit errors. The key insight of~\cite{bsv14} is that coset probabilities $\pi(E\cS)$ can then be expressed as contractions of  planar 2D~tensor networks, see Fig.~\ref{fig:BSVdecoder}. Such tensor networks are known to be efficiently (approximately) contractible by contracting along a single axis, effectively rendering the problem $1$-dimensional. In more detail, a coset probability can be written as 
\begin{align}
\pi(E\cS)&=\bra{\Psi_1}M_1\cdots M_{2d-2} \ket{\Psi_2}\ ,
\end{align} 
where $\ket{\Psi_1},\ket{\Psi_2}\in (\mathbb{C}^2)^{\otimes 2d-1}$ are matrix product states (MPS) with bond dimension~$2$, and each~$M_j\in \cB((\mathbb{C}^2)^{\otimes 2d-1})$ is a certain matrix product operator (MPO)  of bond dimension~$2$. The idea then is to evaluate the right hand side stepwise by applying the MPOs in succession, and eventually computing the inner product of two MPS. However, successive multiplication of an MPS with constant-bond dimension MPOs generically leads to an MPS with exponential bond dimension (in the number of applications of MPOs), rendering an exact evaluation inefficient.  To avoid this overhead, the BSV decoder replaces intermediate MPS obtained during the computations by MPS approximations having bond dimension bounded by~$\chi$. The latter step involves a standard truncation procedure for MPS by Murg, Verstraete and Cirac~\cite{murgverstraetecirac}, which singles out the largest singular values across each cut.

In~\cite{bsv14}, this decoder is compared to an exact (efficiently realized) maximum likelihood decoder in the case of pure~$X$-noise, showing that moderate values of~$\chi$ suffice to achieve good accuracy. It was then used to show that the minimum weight matching decoder is typically suboptimal.  The BSV decoder was subsequently applied in~\cite{tuckettetal18,tuckettetal19} to study
the performance of surface codes under biased noise.

\section{Gottesman-Kitaev-Preskill (GKP) codes\label{sec:gkp}}
In this section, we  review  the pertinent facts about 
Gottesman-Kitaev-Preskill (GKP) codes~\cite{gkp01}. We begin with a general discussion of bosonic quantum systems in Section~\ref{sec:cvqsystems}. In Section~\ref{sec:displacementnoise}, we discuss displacement noise, a central noise model of interest in bosonic systems.
 We then review the general definition of
 a GKP code associated with a symplectically integral lattice (Section~\ref{sec:gkpcodes}) before specializing to square lattice (Section~\ref{sec:squarelatticeGKP}) and asymmetric (Section~\ref{sec:asymGKP})  GKP codes. Finally, in Section~\ref{sec:errorrecoveryGKP}, we discuss the decoding problem for GKP codes.

\subsection{Continuous-variable (CV) quantum systems\label{sec:cvqsystems}}
For concreteness, we focus on a single bosonic mode $(n=1)$; the generalization to $n>1$ modes is straightforward. A pure state of a single  mode, i.e., of a particle on a line, is given by a (equivalence class with respect to a global phase of a) tempered distribution~$\psi\in\mathcal{S}'(\mathbb{R})$.  We call the state normalizable if $\psi\in L^2(\mathbb{R})$.

 Let $Q,P$ denote the position- and momentum operators, also called quadratures, acting on $\cS'(\mathbb{R})$ and satisfying the canonical commutation relation\footnote{We work in units $\hbar=1$.} $[Q,P]=iI$. We collect them in a vector $R\coloneqq \tvector{Q}{P}$ such that the commutation relation takes the form $[R_j,R_k]=iJ_{j,k}I$. Here  the matrix 
\begin{equation}
	J=\begin{pmatrix}0 & 1\\ -1 & 0\end{pmatrix}\ 
\end{equation} 
is associated with the symplectic form $(\xi,\eta)\mapsto \innerprod{\xi}{J\eta}$ on $\mathbb{R}^2\times\mathbb{R}^2$.

For $\xi=\tvector{\xi_1}{\xi_2}\in\mathbb{R}^2$, the {\em displacement operator}~$D(\xi)$ is defined as 
\begin{equation}\label{eq:def_phase_space_translations}
	D(\xi)\coloneqq  e^{-i\innerprod{\xi}{\symplecticform R}}=e^{i(\xi_2Q-\xi_1P)}\ .
\end{equation}
These operators yield a representation $(\xi,\alpha)\mapsto e^{-i\alpha }D(\xi)$ 
 of the Weyl-Heisenberg group on~$\mathcal{S}'(\mathbb{R})$ which restricts to a unitary irreducible representation on $L^2(\mathbb{R})$. That is, we have 
\begin{equation}
	e^{-i\alpha}D(\xi)e^{-i\beta}D(\zeta)= e^{-i(\alpha+\beta+\frac{1}{2}\innerprod{\xi}{\symplecticform\zeta})}D(\xi+\zeta)\, ,
	\label{eq:weylrelations}
\end{equation}
for all $\alpha,\beta\in\mathbb{R}$, and $\xi,\zeta\in\mathbb{R}^2$. We note that the operator~$D(\xi)$ translates -- by conjugation -- the 
vector $\tvector{Q}{P}$ of mode operators by the amount $\xi\in \mathbb{R}^{2}$ in phase space.

The group~$\mathsf{Sp}(2,\mathbb{R})$
of symplectic linear maps on $\mathbb{R}^2$, i.e., linear maps preserving the symplectic form, is 
\begin{equation}
\mathsf{Sp}(2,\mathbb{R}) \coloneqq \{S\in \mathsf{Mat}_{2\times 2}(\mathbb{R})\ |\ S^T\symplecticform S=\symplecticform\}\ .
\end{equation}
The metaplectic representation $S\mapsto U_S$
associates a Gaussian unitary~$U_S$ on $\cS'(\mathbb{R})$ to every $S\in\mathsf{Sp}(2,\mathbb{R})$. The action of~$U_S$ (by conjugation) on the quadrature operators is linear and given by $S$, i.e., 
\begin{align}\label{eq:metaplecticrep}
    U_SR_jU_S^\dagger&=\sum_{k=1}^2S_{j,k}R_k\ ,\quad j=1,2\ .
    \end{align}
    In particular, $U_S$ preserves canonical commutation relations. One consequence is that
     for arbitrary $S\in \mathsf{Sp}(2,\mathbb{R})$  and $\xi\in\mathbb{R}^2$, we have 
\begin{align}\label{eq:sympl_trafo_displacements}
D(S\xi)&
=U_{S^{-1}}D(\xi)U_{S^{-1}}^\dagger\ .
\end{align}
In other words, $D(S\xi)$ is unitarily equivalent to $D(\xi)$ with the corresponding unitary conjugation only depending on $S$ but not on $\xi$. 

\subsection{Displacement noise in bosonic systems\label{sec:displacementnoise}}
GKP codes (named after the paper~\cite{gkp01} by Gottesman, Preskill
and Kitaev) isometrically embed a finite-dimensional quantum system~$\mathbb{C}^K$ into $L^2(\mathbb{R})$ (and more generally $L^2(\mathbb{R})^{\otimes n}$). Although their performance under various forms of noise has 
been studied in the original paper~\cite{gkp01} and a number of subsequent papers~\cite{GlancyKnill06,TerhalWeigand16}, the codes were originally designed primarily to protect against displacement noise, i.e., noise channels of the form
\begin{align}
  \cN_{f_Z}(\rho)&=\int_{\mathbb{R}^2} f_Z(\nu) D(\nu)\rho D(\nu)^\dagger d^{2}\nu\ .\label{eq:randomdisplacementnoisechannel}
\end{align}
Here 
$f_Z:\mathbb{R}^{2}\rightarrow [0,1]$ is a probability density function associated with a random variable~$Z$ on the phase space~$\mathbb{R}^{2}$ and $D(\nu)$ is the displacement operator.

Of specific interest is the case where~$Z\sim \mathsf{N}({\bf 0},\Sigma)$, i.e., is described by a centred normal distribution with (positive semidefinite) covariance matrix~$\Sigma$, hence
\begin{align}
f_Z(\nu)=(2\pi)^{-1} \det(\Sigma)^{-1/2}e^{-\frac{1}{2}\nu^T \Sigma^{-1}\nu}\ .\label{eq:densitygaussianprob}
\end{align}
In the isotropic case, $\Sigma=\sigma^2 I_{2}$, the channel~\eqref{eq:randomdisplacementnoisechannel} becomes the isotropic Gaussian displacement channel \eqref{eq:Gaussnoisechannel}. The single-parameter family of noise channels 
which results by varying~$\sigma$  provides a natural testbed for assessing the noise resilience of GKP and other codes (see e.g.,~\cite{harringtonpreskill}), as well as related capacity questions. In this context  a recent breakthrough~\cite{giovann15} has confirmed a long-standing conjecture introduced in the seminal work~\cite{HolevoWerner01}.

\subsection{The GKP code $\mathsf{GKP}(\cL)$}\label{sec:gkpcodes}
 GKP codes are constructed algebraically from certain lattices~$\cL\subset \mathbb{R}^{2n}$. We first restrict our attention to~$n=1$ which is sufficient for our purposes, i.e., we  consider GKP codes for a single mode encoding a $K$-dimensional system.  We  then  further restrict to $K=2$, i.e., codes encoding a single qubit into a single-mode bosonic system\footnote{Although we only treat the case $n=1$ here, the discussion in this section generalizes to arbitrary~$n\in\mathbb{N}$. Note that for $n>1$, $\mathsf{Sp}(2n,\mathbb{R})$ is a proper subgroup of $\mathsf{SL}(2n,\mathbb{R})$, however for $n=1$ we have $\mathsf{Sp}(2,\mathbb{R})=\mathsf{SL}(2,\mathbb{R})$.}.
We note that  the construction discussed here was previously known~\cite{BALAZS19891,HANNAY1980267} and rigorously discussed by Bouzouina and de Bi\`evre~\cite{bouzouina1996} in~1996. Its potential in terms of quantum error correcting codes was, however, only recognized by Gottesman, Kitaev and Preskill in 2001~\cite{gkp01}.

\subsubsection{Definition of the code space}
 Let us review the construction of a code space $\mathsf{GKP}(\cL)\subset \cS'$ based on a   lattice~$\cL\subset\mathbb{R}^2$ with certain properties. Specifically, $\cL$ needs to be {\em symplectically integral}:  there are vectors $\xi_1,\xi_2\in\mathbb{R}^2$ 
 and $K\in \mathbb{N}$ such that
\begin{equation}\label{eq:sympl_cond_lattice}
\innerprod{\xi_1}{\symplecticform\xi_2}=\pm 2\pi K\ ,
\end{equation} 
and $\cL:=\{n_1 \xi_1+n_2\xi_2\ |\ n_1,n_2\in\mathbb{Z}\}$ is the (integer) span of~$\{\xi_1,\xi_2\}$. Given such a lattice~$\cL$, the space~$\mathsf{GKP}(\cL)$ is defined as the (simultaneous)~$+1$ eigenspace of
all (pairwise commuting) operators $D(\xi)$, $\xi\in\cL$. (Condition~\eqref{eq:sympl_cond_lattice} implies that $\innerprod{\xi}{\symplecticform\xi'}\in 2\pi\mathbb{Z}$ for all $\xi,\xi'\in\cL$, which, because of~\eqref{eq:weylrelations}, ensures that that these operators commute.)   That is, the stabilizer group~$\cS$ of $\GKP(\cL)$ is given by 
\begin{align}
\cS=\{D(\xi)\ |\ \xi\in\cL\}=\langle D(\xi_1),D(\xi_2)\rangle\ .
\end{align}
The space~$\mathsf{GKP}(\cL)$ can be interpreted as the Hilbert space of a quantum system with phase space $\mathbb{R}^2/\mathcal{L}$: states belonging to~$\mathsf{GKP}(\cL)$ are invariant under lattice transformations  realized by displacements $D(\xi)$, $\xi\in\mathcal{L}$. 

We next discuss the effect of various (error) operators on the code space~$\GKP(\cL)$.
Because the displacements $D(\zeta)$, $\zeta\in\mathbb{R}^2$ form an operator basis, we restrict our attention to displacements (similar to the way qubit stabilizer codes are analyzed in terms of Pauli operators).
In other words, we are interested in the effect of an arbitrary displacement $D(\zeta)$,~$\zeta\in\mathbb{R}^2$. A first question is which of these  operators preserve the subspace~$\GKP(\cL)$: Ignoring irrelevant global phases, we are interested in the centralizer  of the stabilizer group in the Heisenberg-Weyl group, i.e., the  set 
\begin{align}
\cC(\cS):=\{ D(\zeta )\ |\ &\zeta\in\mathbb{R}^2\textrm{ such that }\\
&D(\zeta )S=S D(\zeta )\textrm{ for all }S\in\cS\}\ .
\end{align}
The set of these operators is characterized by  the set of vectors~$\zeta\in\mathbb{R}^2$ satisfying 
\begin{equation}
\innerprod{\zeta}{\symplecticform\xi}\in 2\pi\mathbb{Z}\qquad\textrm{ for all }\xi\in\cL\ ,
\end{equation}
 cf.~\eqref{eq:weylrelations}. These vectors form a lattice on their own: the (symplectically) dual lattice $\cL^{\perp}$. Because of~\eqref{eq:sympl_cond_lattice}, we can choose its generating vectors $\xi^\perp_1,\xi^\perp_2\in\mathbb{R}^2$ such that 
\begin{equation}\label{eq:dual_lattice}
\innerprod{\xi^\perp_i}{\symplecticform\xi_j}= 2\pi \delta_{ij}\ ,\quad i,j\in \{1,2\}\ .
\end{equation}
Thus the centralizer of the stabilizer group is given by
\begin{align}
\cC(\cS)=\{D(\zeta)\ |\ \zeta\in\cL^\perp\}=\langle D(\xi^\perp_1),D(\xi^\perp_2)\rangle\ .
\end{align}
Operators belonging to~$\cC(\cS)$ are logical, i.e., preserve the code space~$\GKP(\cL)$, but may or may not act non-trivially on it. Elements $D(\zeta),D(\zeta')\in\cC(\cS)$ have identical logical action if and only if they differ by multiplication by an element in~$\cS$, that is, if and only if $\zeta-\zeta'\in\cL$. Thus 
\begin{align}
\left\{D(\zeta)\ \big|\ [\zeta]\in \cL^\perp/\cL\right\}\ \label{eq:logicalerrorsall}
\end{align}
is a complete family of inequivalent logical errors indexed by the set $\cL^\perp/\cL=\{[\zeta]\ |\ \zeta\in\cL^\perp\}$ of cosets of~$\cL$, where $[\zeta]\coloneqq \zeta+\cL$ is a coset with representative~$\zeta\in\cL^\perp$.

\subsubsection{Voronoi cells and lattice modulo operation}
As discussed below, logical error probabilities in the GKP error recovery process heavily depend on properties of the underlying lattices~$\cL$ and $\cL^\bot$ -- especially the Voronoi cell of~$\cL^\bot$.  Let us briefly define the latter set for an arbitrary lattice~$\cL\subset\mathbb{R}^2$. For  $x\in\mathbb{R}^2$, the closest lattice point in $\cL$ to~$x$ is 
\begin{equation}
Q_{\cL}(x)\coloneqq\arg\min_{\xi\in\cL}\Vert x-\xi\Vert\ .
\end{equation}
Here the Euclidean distance is used and we assume  that ties are broken in a systematic manner (e.g.,~such that $Q_{\cL}(x)_i\geq x_i$, $i=1,2$). The Voronoi cell~$\cV$ of $\cL$ is the set of points closest to the origin ${\bf 0}\coloneqq \tvector{0}{0}$, i.e.,
\begin{align}
\cV:=\{x\in\mathbb{R}^2\ |\ Q_{\cL}(x)={\bf 0}\}\ .
\end{align}
For later use, we also define a lattice modulo operation by
\begin{align}
\begin{matrix}
(\ \cdot\pmod \cL): & \mathbb{R}^2&\rightarrow & \cV&\\
&x & \mapsto & x\pmod{ \cL}&\coloneqq x-Q_{\cL}(x)\ .
\end{matrix}
\end{align}

\subsection{Square lattice GKP codes\label{sec:squarelatticeGKP}}
In the following two sections we list several examples of lattices giving rise to GKP codes. We begin with a discussion of the ``standard'' square lattice GKP code. We  also refer to this as the ``symmetric'' GKP code code below.

\subsubsection{Square lattice GKP codes of dimension~$K$}
Let $K\in\mathbb{N}$ be arbitrary. Then the vectors
\begin{align}\label{eq:lattice_basis_square}
\begin{matrix*}[l]
\xi_1&=\tvector{\sqrt{2\pi K}}{{\bf 0}}\ ,\\
\xi_2&= \tvector{{\bf 0}}{\sqrt{2\pi K}}\ 
\end{matrix*}
\end{align}
satisfy the condition~\eqref{eq:sympl_cond_lattice} and span a symplectically integral lattice~$\cL$. 
Its dual lattice~$\cL^\bot$ is spanned by the vectors
\begin{align}\label{eq:dual_lattice_basis_square}
\begin{matrix*}[l]
\xi^\bot_1&=\tvector{{\bf 0}}{-\sqrt{2\pi/ K}}\ ,\\ \xi^\bot_2&=\tvector{\sqrt{2\pi/ K}}{{\bf 0}} .
\end{matrix*}
\end{align}
In particular, the set of cosets $\cL^\bot/\cL$ associated with logical errors (cf.~\eqref{eq:logicalerrorsall})
is given by
\begin{align}
    \begin{matrix*}[l]
    \cL^\bot/\cL=\big\{[n_1\xi_1^\bot+n_2\xi_2^\bot]\ \big| & n_1\in \{0,\ldots,K-1\},\\
    &n_2\in \{0,\ldots,K-1\}\big\}\ . \phantom{abc}
    \end{matrix*}\label{eq:explicitcosetssquarelattice}
\end{align}

A detailed description of the action of the associated logical errors can be given by fixing a basis of~$\mathsf{GKP}(\cL)$. Let $\psi\in\mathsf{GKP}(\cL)\subset \cS'$. Then 
the eigenvalue equation $D(\xi_2)\psi=e^{i\sqrt{2\pi K}Q}\psi=\psi$ 
implies that $\psi\in\mathcal{S}'$ is of the form
\begin{equation}
	\psi(x)= \sum_{n\in\mathbb{N}}c_n\delta\left(x-n\sqrt{2\pi/K}\right)\ , \label{eq:lattice_state}
\end{equation}
with $c_n\in\mathbb{C}$, and with $\delta$ denoting the Dirac-$\delta$-distribution. The eigenvalue equation $D(\xi_1)\psi=e^{-i\sqrt{2\pi K}P}\psi=\psi$ further implies that
\begin{equation}\label{eq:coeff_lattice_states}
c_n=c_{n+K}\qquad\textrm{ for all }n\in\mathbb{N}\ .
\end{equation}
From \eqref{eq:lattice_state} and~\eqref{eq:coeff_lattice_states} we conclude that the dimension of $\mathsf{GKP}(\cL)$ is $K$, consistent with the fact that~\eqref{eq:explicitcosetssquarelattice} shows the existence of $K^2$~linearly independent logical  operators. Furthermore, the set $\{e_j\}_{j=0}^{K-1}$ defined by
\begin{equation}\label{eq:basis_gkp}
e_j(x)=\sum_{n\in\mathbb{N}}\delta\left(x-n\sqrt{2\pi K}-j\sqrt{2\pi/ K}\right) 
\end{equation}  for $j=0,\ldots,K-1$,
defines a basis of~$\mathsf{GKP}(\cL)$. The action of logical error operators $D(n_1\xi_1^\bot+n_2\xi_2^\bot)\propto D(\xi_1^\bot)^{n_1}D(\xi_2^\bot)^{n_2}$ (cf.~\eqref{eq:explicitcosetssquarelattice})  on these basis vectors can be computed to be
\begin{align}
\begin{matrix}
	D(\xi^\bot_1)e_j&=&e^{\frac{-2\pi ij}{K}}e_j\\
	D(\xi^\bot_2)e_j&=&e_{(j+1)\mod K}\ 
	\end{matrix}\label{eq:logical_paulis_square}
\end{align}
for all $j=0,\ldots,K-1$. That is, $(D(-\xi^\bot_1),	D(\xi^\bot_2))$ are the (generalized) logical Pauli-$\overline{Z}$ and $\overline{X}$ operators of a $K$-dimensional system. They  generate what is sometimes referred to as a finite-dimensional Weyl system.

\subsubsection{Square lattice GKP codes encoding a qubit $(K=2$)}
In the following, we specialize to the square lattice GKP code encoding a qubit, i.e., the case where $K=2$. Explicitly, the square lattice and its dual generated by the vectors \eqref{eq:lattice_basis_square} respectively  \eqref{eq:dual_lattice_basis_square} are
\begin{align}\label{eq:squarelattice}
 \begin{matrix*}[l]
\cL_{\square}\coloneqq \left\{\left.
\tvector{2\sqrt{\pi}\ n_1}{2\sqrt{\pi}n_2}\ \right|\ n_1,n_2\in\mathbb{Z}\right\}\ ,\\
\cL_{\square}^\perp\coloneqq \left\{\left.
\tvector{\sqrt{\pi}\ n_1}{\sqrt{\pi}\ n_2}\ \right|\ n_1,n_2\in\mathbb{Z}\right\}\ .
\end{matrix*}
\end{align}
We refer to the corresponding code $\GKP(\cL_{\square})$ using the expression \emph{square lattice GKP code}. It is the most commonly used version of the GKP code, and often simply referred to as {\em the GKP code}. 
Labeling the basis elements~\eqref{eq:basis_gkp}
 as  
 \begin{align}
      \gkpsquarezero:=e_0\qquad\textrm{ and }\qquad 
     \gkpsquareone:=e_1\  ,\label{eq:standardlabelinggkpstates}
  \end{align} 
 one finds that the action~\eqref{eq:logical_paulis_square}
 is that of the standard Pauli operators (with respect to the computational basis). Thus one writes
\begin{align}
\gkpsquarelogicalX & \coloneqq D\left(\tvector{\sqrt{\pi}}{0}\right),\\
\gkpsquarelogicalZ &\coloneqq D\left(\tvector{0}{\sqrt{\pi}}\right)\  .
\end{align}
We emphasize that~\eqref{eq:standardlabelinggkpstates} is simply a (commonly used) convention. Indeed, as argued below, it is essential for our purposes to choose a slightly different mapping from GKP basis states to (logical) qubit states.

It is clear from the construction above that the basis elements $\gkpsquarezero,\gkpsquareone$ are not proper quantum mechanical states in the usual sense: they constitute an infinite uniform superposition of position (equivalently: momentum) eigenstates and are therefore not normalizable. To obtain a physically meaningful  description, one has to view the GKP states as a limit of approximate GKP states, in which the sum is replaced by one with weights according to a Gaussian envelope, and the delta distributions forming the position eigenstates are replaced by squeezed Gaussian states.

\subsection{Asymmetric GKP codes\label{sec:asymGKP}}
Applying a  Gaussian unitary~$U_{S^{-1}}$ to a GKP code~$\mathsf{GKP}(\cL)$ results in a GKP code~$\mathsf{GKP}(\cL')=U_{S^{-1}}\mathsf{GKP}(\cL)$,
where the lattice $\cL'=S\cL$ is obtained by applying the associated symplectic transformation~$S\in\mathsf{Sp}(2,\mathbb{R})$ to~$\cL$ (cf.\ \eqref{eq:sympl_trafo_displacements}).  Note that~\eqref{eq:sympl_cond_lattice} is invariant under symplectic transformations of the lattice, hence if $\xi_1,\xi_2$ span~$\cL$,  then $S\xi_1,S\xi_2$ 
are lattice basis vectors of~$\cL'$ satisfying~\eqref{eq:sympl_cond_lattice}. 

This shows that the set of symplectically integral lattices~$\cL$ (and thus GKP codes~$\mathsf{GKP}(\cL)$) can be partitioned into equivalence classes
of lattices that can be transformed into each other by symplectic Gaussian unitaries. Let us briefly argue that  there is in fact only a single equivalence class: Every symplectic integral lattice~$\cL$ can be obtained by applying a symplectic transformation~$S\in\mathsf{Sp}(2,\mathbb{R})$ to the square lattice~$\cL_\square$. 
Indeed, suppose that~$\cL$ is spanned by
$\xi_1,\xi_2\in\mathbb{R}^2$ satisfying the integrality condition
\begin{align}
    \innerprod{\xi_1}{\symplecticform\xi_2}&=4\pi\ ,\label{eq:xionetwo}
\end{align}
and let 
\begin{align}
    \xi_1^\square &:=\tvector{2\sqrt{\pi}}{0}\ ,\\
    \xi_2^\square &:=\tvector{0}{2\sqrt{\pi}}
\end{align}
be vectors generating~$\cL_\square$. Then it is easy to check (using the antisymmetry of the symplectic form and~\eqref{eq:xionetwo}) that the matrix
$S:=\frac{1}{2\sqrt{\pi}}(\xi_1|\xi_2)$ having
normalized versions of $\xi_1,\xi_2$ in its columns is symplectic and maps~$\cL_\square$ to~$\cL$.
We note that a generalization to $N>2$ of this argument can be obtained using a symplectic Gram-Schmidt procedure. Starting from the square lattice~$\cL_\square$, we can therefore obtain various deformed (asymmetric) GKP codes.
 
\subsubsection{Rectangular lattice GKP codes}\label{sec:rectangularGKP}

\begin{figure}
  	\subfloat[$r=1$\label{fig:GKPlatticer1}]{ \includegraphics[width=0.4\textwidth]{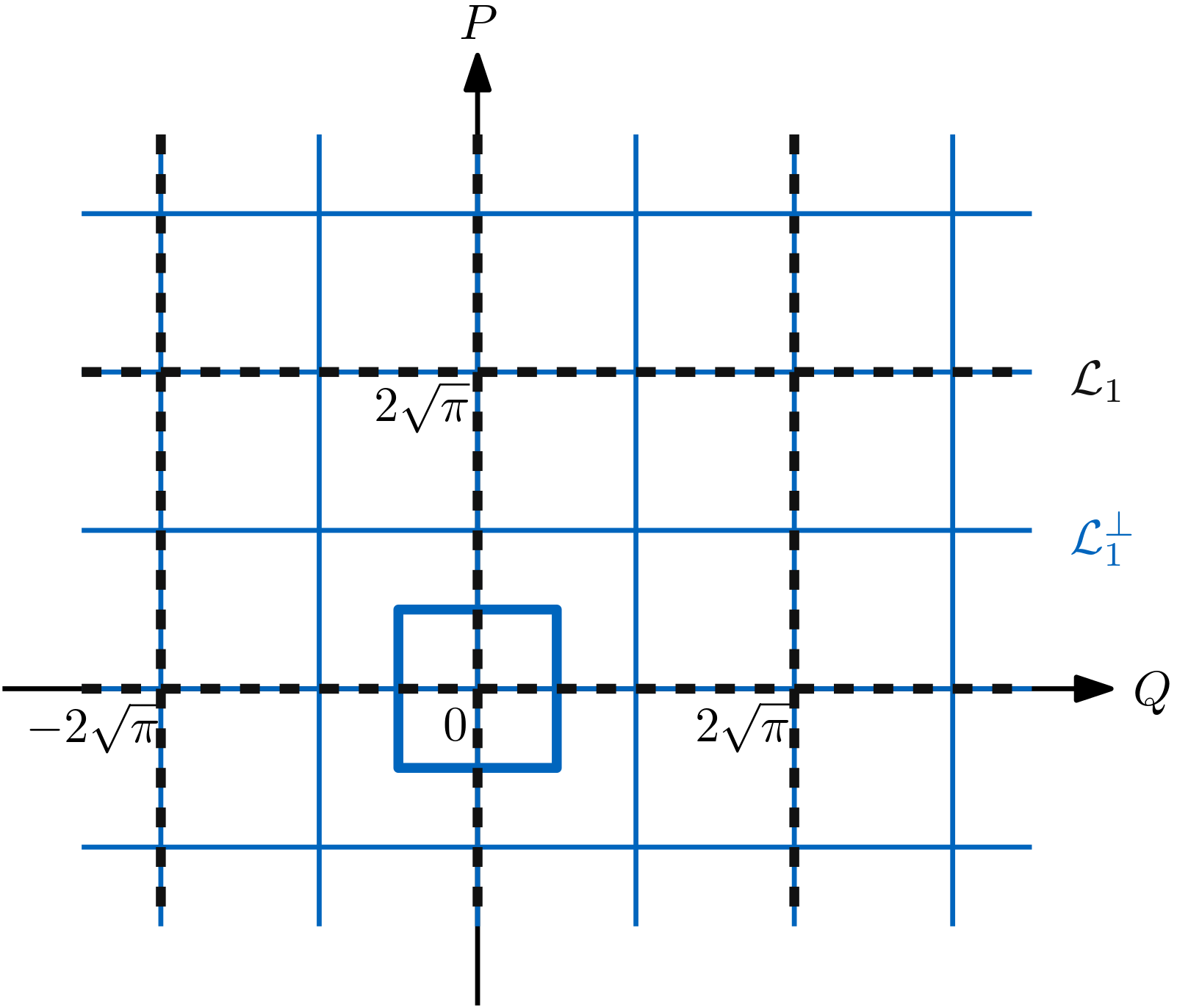}}\\ 
  	\subfloat[$r=4$\label{fig:GKPlatticer4}]{
  	\includegraphics[width=0.4\textwidth]{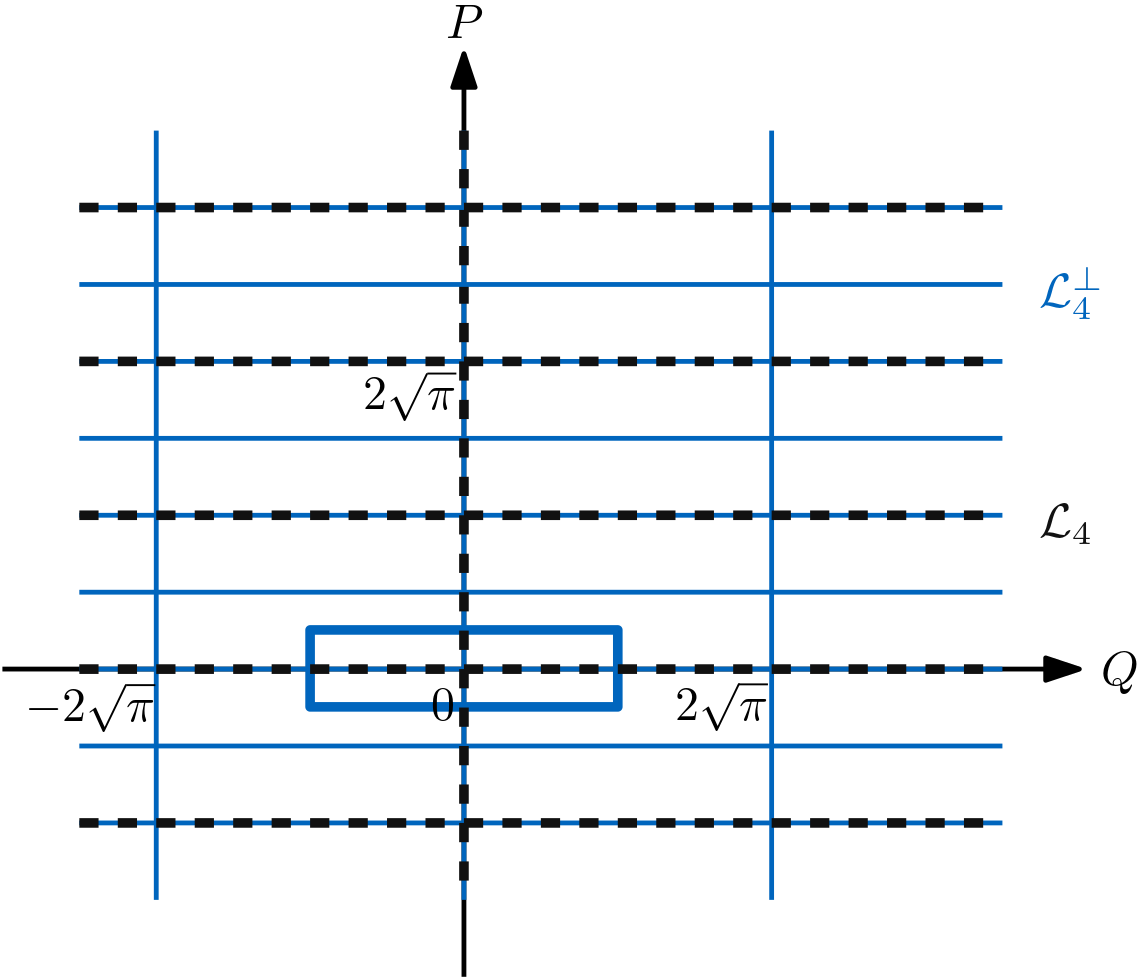}}
    \caption{The lattice $\cL_r$ (dashed black lines) and the dual lattice $\cL^\perp_r$ (solid blue lines) defining the GKP codes for two different ratio parameters $r=1$ (square lattice) and $r=4$. The Voronoi cell of the dual lattice $\cL^\perp_r$ is marked with blue edges. 
		\label{fig:GKPlattice}
		}
\end{figure}
Of primary interest is the following example, the rectangular lattice GKP code. Let $r\geq 1$ and consider the symplectic matrix
\begin{equation}
S_r\coloneqq \begin{pmatrix}
\sqrt{r} & 0\\
0& 1/\sqrt{r}
\end{pmatrix}\ \label{eq:squeezingsymplectic}
\end{equation}
associated with a one-mode squeezing unitary. 
The lattice generated by the corresponding symplectically transformed vectors $(S_r\xi_1^\square,S_r\xi_2^\square)$ and its dual lattice are rectangular:
\begin{align}\label{eq:rectangularlattice}
 \begin{matrix*}[l]
\cL_r \, = \left\{ \tvector{2\sqrt{\pi r}\ n_1}{2\sqrt{\pi/r}\ n_2 }\ \Big|\ n_1,n_2 \in \mathbb{Z} \right\} , & \\
\cL_r^\perp = \left\{ \tvector{\sqrt{\pi r} \ n_1}{\sqrt{\pi/r}\ n_2}\ \Big|\ n_1,n_2 \in \mathbb{Z} \right\} , &
\end{matrix*}
\end{align}
and the resulting code~$\mathsf{GKP}(\cL_r)$ is called a \emph{rectangular lattice GKP code}, see Fig.~\ref{fig:GKPlattice}. Here the parameter $r$ is the \emph{ratio} between the generating vectors of the lattice.
We denote the logical Pauli operators by 
\begin{align}
	\gkpreclogicalX &:=D\left(\tvector{\sqrt{\pi r}}{0}\right)\\
	\gkpreclogicalZ &:=D\left(\tvector{0}{\sqrt{\pi /r}}\right)\ .
\end{align}
The square lattice GKP code is a rectangular lattice GKP code with ratio $r=1$. For a discussion of the effect of a parameter~$r>1$ compared to $r=1$ in terms of error correction and state preparation see Section~\ref{sec:biasednoiserectGKP}.

We note that for a square or rectangular lattice~$\cL$ and its symplectically dual lattice~$\cL^\bot$, the Voronoi cells~$\cV\subset\mathbb{R}^2$ and~$\cV^\bot\subset\mathbb{R}^2$, respectively, are given by
\begin{align}\label{eq:voronoicell}
	 \begin{matrix*}[l]
	 \cV &\coloneqq\left\{\lambda_1 \xi_1+\lambda_2\xi_2\ |\ \lambda_1,\lambda_2\in [-\tfrac{1}{2},\tfrac{1}{2}] \right\} \ ,\\
	\cV^\perp &\coloneqq \left\{\lambda_1 \xi^\perp_1+\lambda_2\xi^\perp_2\ |\ \lambda_1,\lambda_2\in [-\tfrac{1}{2},\tfrac{1}{2}] \right\}\ ,
	\end{matrix*}
\end{align}
where we assume that the lattice~$\cL$ is spanned by $\xi_1,\xi_2$ and~$\cL^\bot$ is spanned by~$\xi_1^\bot,\xi_2^\bot$.

\subsubsection{Hexagonal lattice GKP codes}\label{sec:hexagonallatticeGKP}
We mainly focus on rectangular lattice GKP codes for concreteness and ease of illustration. Note, however, that one-mode squeezing is not the only symplectic transformation applicable to the square lattice. In particular, the transformed vectors need not be orthogonal. An example is the {\em  hexagonal lattice GKP code}~$\mathsf{GKP}(S_{\hexagon}\cL_\square)$  which is associated with the symplectic transformation 
\begin{equation}
S_{\hexagon} =\left(\frac{2}{\sqrt{3}}\right)^{1/2}\begin{pmatrix}
1 & 1/2 \\
0 & \sqrt{3}/2
\end{pmatrix}\ .
\end{equation} 
The lattice $S_{\hexagon}\cL_\square$ has
an angle of  $2\pi/3$ between the two lattice basis vectors -- this is the interior angle of a hexagon. Since the product of symplectic matrices is a symplectic matrix, one gets an {\em asymmetric hexagonal lattice GKP code} by applying~$S_{\hexagon}S_r$ to the square lattice code\footnote{Note that $S_{\hexagon}S_r\not=S_rS_{\hexagon}$ in general. Here we use the convention that the former order of transformations is associated with an asymmetric hexagonal lattice.}.

\subsection{Error recovery for the GKP code\label{sec:errorrecoveryGKP}}

Consider a logical qubit encoded in~$\GKP(\cL)$, i.e., described by a density operator~$\rho$ supported on~$\GKP(\cL)$. Assume that this state undergoes noise given by  a completely positive and trace preserving (CPTP) map $\cN\colon \mathcal{B}(L^2(\mathbb{R}))\to\mathcal{B}(L^2(\mathbb{R}))$ on the set $\mathcal{B}(L^2(\mathbb{R}))$ of bounded operators on the Hilbert space $L^2(\mathbb{R})$. Specifically, we are interested in random displacement channels $\cN_{f_Z}$ (cf.\ \eqref{eq:randomdisplacementnoisechannel}).
Error correction starting from the corrupted state~$\cN_{f_Z}(\rho)$ proceeds by
\begin{enumerate}[(i)]
\item\label{it:deconex}
Syndrome measurement:  This measures  the (eigenvalues of the) commuting stabilizer generators~$D(\xi_1)$, $D(\xi_2)$  (with the generating vectors~$\xi_1, \xi_2$ of the lattice~$\cL$). 
By~\eqref{eq:def_phase_space_translations} and~\eqref{eq:dual_lattice}, this measurement  amounts to a measurement of  the quadrature operators~$(Q,P)$ modulo the dual lattice~$\cL^\perp$. 
The measurement yields a {\em syndrome}~$s=\tvector{q}{p}\in\cV^\bot$ belonging to the Voronoi cell of the dual lattice~$\cL^\perp$.

For example, in the square lattice case, $Q$ and~$P$ are measured modulo $\sqrt{\pi}$, with outcomes $q,p\in (-\frac{\sqrt{\pi}}{2},\frac{\sqrt{\pi}}{2})$. This can be realized using a logical CNOT gate (realized by a beamsplitter) between the encoded GKP-qubit and an additional ancilla GKP-qubit, and subsequent homodyne measurements of~$Q$ and~$P$ of the ancilla mode~(cf.~\cite{gkp01}).
\item\label{it:dectwox}
Application of a unitary correction operation~$C(s)$ depending on the syndrome~$s$: The correction operation can be chosen to be a displacement, i.e., it is of the form $C(s)=D(c(s))$ 
for a function $c:\cV^\bot\rightarrow\mathbb{R}^2$ determined by the chosen recovery procedure (see below).
\end{enumerate}
 The sequence~\eqref{it:deconex} and~\eqref{it:dectwox} of operations 
 define a recovery CPTP map $\cR^c: \mathcal{B}(L^2(\mathbb{R}))\to\mathcal{B}(L^2(\mathbb{R}))$
 in terms of the function~$c:\cV^\bot\rightarrow\mathbb{R}^2$. We call this function the {\em recovery procedure} in the following.

To analyze the effect of the recovery map~$\cR^c$, first consider  a single (unitary) displacement error~$D(\nu)$ for some $\nu\in\mathbb{R}^2$ applied to a density operator~$\rho$ with support on~$\mathsf{GKP}(\cL)$. In the recovery scheme described above, the stabilizer measurement of the corrupted state~$D(\nu)\rho D(\nu)^\dagger$ yields the syndrome
\begin{align}
s(\nu)&= \nu\pmod{ \cL^\perp}\ ,\label{eq:syndromecomputation}
\end{align}
with certainty. Combined with the subsequent correction operator~$D\big(c(s(\nu))\big)$, this results in an overall (effective) displacement given by 
\begin{equation}\label{eq:overall_displacement}
D\big(c(s(\nu))\big)D(\nu)\propto D\big(c(s(\nu))+\nu\big)\ 
\end{equation}
where we omit the irrelevant global phase. This overall operation
acting on the initial state~$\rho$ therefore leaves the state invariant (i.e., recovery succeeds) if 
\begin{enumerate}[(a)]
\item
the operation~\eqref{eq:overall_displacement} is logical, i.e., maps the code space to itself, that is,
\begin{equation}\label{eq:cond_decoding_valid}
c(s(\nu))+\nu\in \cL^\perp\ ,
\end{equation}
and
\item
the  operation~\eqref{eq:overall_displacement} has trivial action on the code space, that is, the displacement vector belongs to the trivial coset: 
\begin{align}
c(s(\nu))+\nu\in \cL=[{\bf 0}]\ . \label{eq:nologicalerror}
\end{align}
\end{enumerate}
Here we have used the characterization of the effect of displacements on the code space discussed in Section~\ref{sec:gkpcodes}. We note that condition~\eqref{eq:cond_decoding_valid} is
typically guaranteed for all $\nu\in\mathbb{R}^2$ for reasonable choices of the recovery procedure~$c:\cV^\bot\rightarrow\mathbb{R}^2$ (as the ones discussed below). We call such 
recovery procedures~{\em valid}. The following analysis therefore focuses on~\eqref{eq:nologicalerror}.

For valid recovery procedures, condition~\eqref{eq:nologicalerror} guarantees that there is no residual logical error. More generally, the residual logical operator depends on the coset the vector~$c(s(\nu))+\nu$ belongs to and can be read off from the following table:
\begin{equation}\label{array:nontrivialerror}
\begin{array}{c|c}
c(s(\nu))+\nu\in & \text{logical Pauli operator applied}\\
\hline
[{\bf 0}] & \overline{I}\\
 \left[\xi_2^\bot \right] & \overline{X}\\
 \left[\xi_2^\bot-\xi_1^\bot\right] &\overline{Y}\\
 \left[-\xi_1^\bot\right] &\overline{Z}\ .
\end{array}
\end{equation}
(Here we write $\overline{I}$ for the case of a trivial operator leaving the code space  invariant.)

Returning to our error model~\eqref{eq:randomdisplacementnoisechannel}, an error $D(\nu)$ occurs with probability $f_Z(\nu)d^2\nu$. Table~\eqref{array:nontrivialerror} allows us to conclude that 
the combined CPTP map~$\cR^c\circ\cN_{f}$ obtained by applying the recovery operation after the noise -- when restricted to states supported on~$\mathsf{GKP}(\cL)$ -- is given by 
\begin{align}
 \begin{matrix*}[l]
\overline{\cN}_{p_{\overline{I}},p_{\overline{X}},p_{\overline{Y}},p_{\overline{Z}}}(\rho)=&p_{\overline{I}}\rho+
p_{\overline{X}}\overline{X}\rho\overline{X}^\dagger\\
&+p_{\overline{Y}}\overline{Y}\rho\overline{Y}^\dagger+p_{\overline{Z}}\overline{Z}\rho\overline{Z}^\dagger\ ,
\end{matrix*}\label{eq:averagedqubiterr}
\end{align}
where (cf.~\eqref{array:nontrivialerror})
\begin{align}
 \begin{matrix*}[l]
p_{\overline{P}}&=\Pr_\nu \Big[ c(s(\nu))+\nu \in  \left[\xi_{\overline{P}}^\bot \right]\Big]\\
&=\int_{\nu:\ c\left(s(\nu)\right)+\nu\ \in \left[\xi_{\overline{P}}^\bot \right]} f_Z\left(\nu \right) d^2\nu \ ,
\end{matrix*} \label{eq:cosetprobabilitiesgkp}
\end{align}
with $\overline{P}\in\{\overline{I},\overline{X},\overline{Y},\overline{Z}\}$ and
\begin{align}
    (\xi_{\overline{I}}^\bot,\xi_{\overline{X}}^\bot,\xi_{\overline{Y}}^\bot,\xi_{\overline{Z}}^\bot)\coloneqq ({\bf 0},\xi_2^\bot,\xi_2^\bot-\xi_1^\bot,-\xi_1^\bot)\label{eq:shiftlgmx}\ .
\end{align}
In particular, the success probability of a valid recovery procedure $c:\cV^\bot\rightarrow\mathbb{R}^2$ is thus given by
$P^c_{\textrm{success}}=p_{\overline{I}}$.

We note that the effective logical error channel~\eqref{eq:averagedqubiterr} is the result of averaging over displacement errors and syndrome measurement outcomes. The latter average is incorporated in the definition of the recovery map~$\cR^c$. In fact, a more detailed description  of this process treats the sequence~\eqref{it:deconex} and~\eqref{it:dectwox} as an instrument (rather than a CPTP map): this generates both syndrome information~$s\in\cV^\bot $ and a corresponding state~$\rho(s)$ (obtained by applying the correction to the post-measurement state).  Such a description is needed if the syndrome information~$s$ is used, e.g., in concatenated coding to update Bayesian priors. The distribution over syndromes~$s$ is 
\begin{equation}\label{eq:bayesiannorm}
P_s(s_0)=\Pr_\nu \left[s(\nu)=s_0\right]\qquad\textrm{ for }s_0\in\cV^\bot\ .
\end{equation}
Conditioned on the syndrome being~$s_0\in\cV^\bot$, the conditional probability $p_{\overline{P}}^{s_0}$ of finding an error~$\overline{P}\in\{\overline{I},\overline{X},\overline{Y},\overline{Z}\}$ is 
\begin{align}
 \begin{matrix*}[l]
p_{\overline{P}}^{s_0}&=\Pr_\nu \left[c(s(\nu))+\nu \in \left[\xi_{\overline{P}}^\bot \right]\big|\ s(\nu)=s_0\right]\\
&=\frac{\Pr_\nu \left[s(\nu)=s_0\textrm{ and }c(s(\nu))+\nu \in \left[\xi_{\overline{P}}^\bot \right]\right]}{P_s(s_0)}\ ,
\end{matrix*}\label{eq:bayesiandecodingprior}
\end{align}
by Bayes' rule, i.e., 
\begin{align}\label{eq:cosetprobfrombayesianerror}
p_{\overline{P}}&= \int_{\cV^\perp}P_s(s_0)\ p_{\overline{P}}^{s_0}d^2 s_0\\
=&\int_{\cV^\perp}{\Pr}_\nu \left[s(\nu)=s_0\textrm{ and }c(s(\nu))+\nu \in \left[\xi_{\overline{P}}^\bot \right]\right]d^2 s_0\ .
\end{align}
In particular, the  state after applying the correction operation is~$\rho(s_0)=
\overline{\cN}_{p^{s_0}_{\overline{I}},p^{s_0}_{\overline{X}},p^{s_0}_{\overline{Y}},p^{s_0}_{\overline{Z}}}(\rho)$. In other words, for a fixed syndrome measurement outcome~$s_0$, the residual error model is a (logical) Pauli-noise channel with syndrome-dependent error probabilities $(p^{s_0}_{\overline{I}},p^{s_0}_{\overline{X}},p^{s_0}_{\overline{Y}},p^{s_0}_{\overline{Z}})$.

\subsubsection{Nearest lattice point recovery procedure}\label{sec:gkpdecoding}

\begin{figure}
  	\subfloat[$r=1$\label{fig:GKPcosetsr1}]{\includegraphics[width=0.4\textwidth]{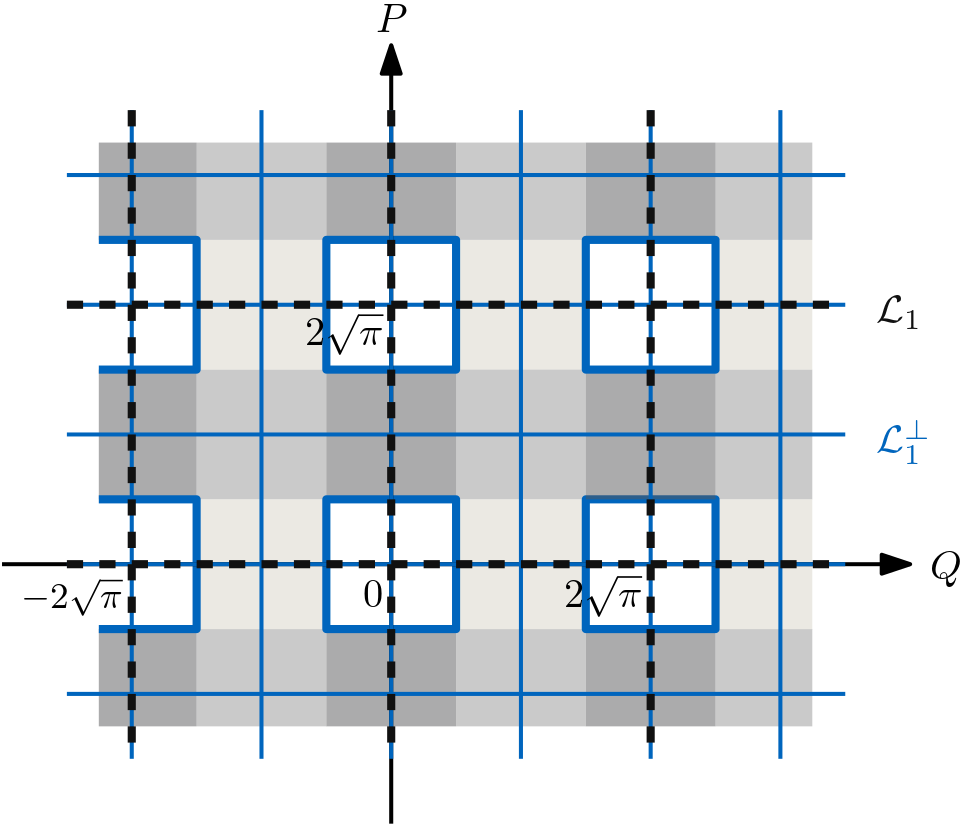}}\\
  	\subfloat[$r=4$\label{fig:GKPcosetsr4}]{\includegraphics[width=0.4\textwidth]{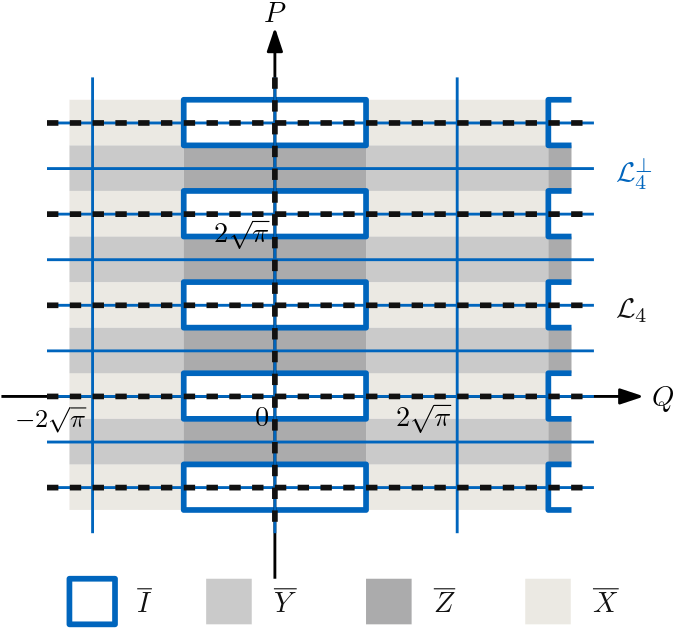}}
    \caption{Lattices for $\GKP(\cL_{r})$ for the ratio parameters $r=1,4$. In the phase space, the lattice $\cL_r$ (dashed black lines) and its dual $\cL^\perp_r$ (solid blue lines) are depicted, respectively. For the square lattice ($r=1$) GKP code undergoing a displacement $D(\nu)$, $\nu\in\mathbb{R}^2$, measuring the stabilizer generators $e^{i2\sqrt{\pi}Q}$ and $e^{i2\sqrt{\pi}P} $ corresponds to measuring the values $\eta=\nu \pmod{\cL^\perp}$, with $\eta=\tvector{\eta_1}{\eta_2}$, $\eta_1,\eta_2\in [-\frac{\sqrt{\pi}}{2},\frac{\sqrt{\pi}}{2}] $. Applying the correction $D(\eta)$ corresponds to a mapping to the nearest lattice point in $\cL^\perp_r$. The correction is successful if the actual error displacement $\nu=\tvector{\nu_1}{\nu_2}$ satisfies $\nu_1=2\sqrt{\pi}n_1 +\eta_1$, $\nu_2=2\sqrt{\pi}n_2 +\eta_2$, with $n_1,n_2\in\mathbb{Z}$. In other words, the correction is successful if the error vector $\nu$ lies in the areas marked by the blue edges. This correspond to shifts of the dual lattice Voronoi cell by a lattice vector~$ \xi \in \cL$. If the error vector~$\nu$ lies in one of the grey shaded areas, the correction results in a logical error $\bar{X},\bar{Y},\bar{Z}$ (cf.~table in~\eqref{array:nontrivialerror}).
		\label{fig:GKPcosets}}
\end{figure}

Let us now consider a specific simple  choice of recovery procedure $c\colon \cV^\perp\rightarrow\mathbb{R}^2$, namely 
\begin{equation}
c(s)=-s\qquad\textrm{ for } s\in  \cV^\perp\ .
\end{equation}
We call this the nearest lattice point recovery procedure:
it attempts to correct a given shift error by mapping to the nearest lattice point in $\cL^\perp$. 
While it is not equivalent to maximum likelihood decoding for general error distributions~$f_Z$, it is expected to work well for distributions concentrated around the origin (i.e., predominantly small displacement errors). Because of its simplicity, it is commonly used in the context of GKP codes.   Fig.~\ref{fig:GKPcosets} illustrates this recovery  procedure, as well as the phase space regions of
displacements associated with
errors that are successfully recovered from, respectively lead to a logical error.

To analyze its behavior, first observe that the recovery procedure $c$ is valid, i.e., satisfies~\eqref{eq:cond_decoding_valid} for all $\nu\in\mathbb{R}^2$: indeed, we have
\begin{align}
 \begin{matrix*}[l]
c(s(\nu))+\nu&=-s(\nu)+\nu\\
&=-\nu\pmod{\cL^\perp}+\nu\in\cL^\bot\
\end{matrix*}\label{eq:residualerrornlpd}
\end{align}
by definition. Thus  we can use the expression~\eqref{eq:cosetprobabilitiesgkp}  for the residual (i.e., post-correction) logical error (coset)  probabilities without syndrome information (cf.~\eqref{eq:shiftlgmx} and Appendix~\ref{app:compcosetprobnnd})
\begin{align}\label{eq:cosetprobabilitiesgkpnnd}
p_{\overline{P}}=\sum_{\xi \in \cL} \int_{\cV^\perp} f_Z\left(\nu+\xi+\xi^\bot_{\overline{P}} \right) &d^2\nu\ ,
\end{align}  for $\overline{P}\in \{\overline{I},\overline{X},\overline{Y},\overline{Z}\}$. 
Similarly, using the syndrome information $s_0$,   Eq.~\eqref{eq:bayesiandecodingprior} yields 
\begin{align}
p_{\overline{P}}^{s_0}=\frac{\sum_{\xi \in \cL} f_Z\left(s_0+\xi+\xi^\bot_{\overline{P}} \right)}{\sum_{\xi^\bot \in \cL^\perp}f_Z\left(s_0+\xi^\bot\right)} \ .\label{eq:bayesiandecodingpriornnd}
\end{align}

\subsubsection{Biased logical noise from isotropic physical noise in~$\mathsf{GKP}(\cL_r)$}\label{sec:biasednoiserectGKP}
As an example, consider the rectangular lattice GKP code $\mathsf{GKP}(\cL_r)$ with the physical mode  subject to the isotropic Gaussian displacement channel~$\cN_{f_{\sigma^2}}$ (cf.~\eqref{eq:Gaussnoisechannel}). Inserting the associated probability density~$f_{\sigma^2}$  
into~\eqref{eq:cosetprobabilitiesgkpnnd} and \eqref{eq:bayesiandecodingpriornnd} yields (see Appendix \ref{app:compcosetprobnndgauss})
\begin{align}\label{eq:Gaussianresidualerr}
\begin{matrix}
p_{\overline{I}}&=&(1-q_{\overline{X}})&\cdot&(1-q_{\overline{Z}})&\\
p_{\overline{X}}&=&q_{\overline{X}}& \cdot&(1-q_{\overline{Z}})&\\
p_{\overline{Y}}&=&q_{\overline{X}}& \cdot&q_{\overline{Z}}&\\
p_{\overline{Z}}&=&(1-q_{\overline{X}})& \cdot&q_{\overline{Z}}&\ 
\end{matrix}
\end{align}
where 
\begin{align}
    q_{\overline{P}}&\coloneqq \int_{\cV^\bot}n_{\overline{P}}(s_0)q_{\overline{P}}^{s_0} d^2 s_0\label{eq:logprobsq}
\end{align} for $ \overline{P}\in \{\overline{X},\overline{Z}\}$ with
\begin{align}
     &\begin{matrix}
    q_{\overline{X}}^{s_0}&\coloneqq &1-n_{\overline{X}}(s_0)^{-1}e(4r,x)\\
    q_{\overline{Z}}^{s_0}&\coloneqq& 1-n_{\overline{Z}}(s_0)^{-1}e(4/r,z)
    \end{matrix}\label{eq:condlogprobsq}
\end{align} and
\begin{align} 
    \begin{matrix}
        n_{\overline{X}}(s_0)&\coloneqq& e(r,x)\\
n_{\overline{Z}}(s_0)&\coloneqq&e(1/r,z)\ .
    \end{matrix}\label{eq:condlogprobsq2}
\end{align}
Here we write $s_0=\tvector{x}{z}$ for the syndrome and make use of the function
\begin{align}\label{eq:euw}
e(u,w)\coloneqq \frac{1}{\sqrt{2\pi \sigma^2}}\sum_{n\in\mathbb{Z}}e^{-\frac{(\sqrt{\pi u}n+w)^2}{2\sigma^2}}
\end{align} for $u,w\in \mathbb{R}, u\geq 0$.
More explicitly, \eqref{eq:logprobsq} can be written as
\begin{align}
& \begin{matrix*}[l]
1-q_{\overline{X}}=\frac{1}{2}\sum_{n\in\mathbb{Z}}\ & \mathrm{erf}\left(\sqrt{\frac{2\pi r}{\sigma^2}}\left(n+\frac{1}{4}\right)\right)\\
&-\mathrm{erf}\left(\sqrt{\frac{2\pi r}{\sigma^2}}\left(n-\frac{1}{4}\right)\right)\ ,
\end{matrix*}\label{eq:1minqx}\\
& \begin{matrix*}[l]
1-q_{\overline{Z}}=\frac{1}{2}\sum_{n\in\mathbb{Z}}\ & \mathrm{erf}\left(\sqrt{\frac{2\pi }{r\sigma^2}}\left(n+\frac{1}{4}\right)\right)\\
&-\mathrm{erf}\left(\sqrt{\frac{2\pi }{r\sigma^2}}\left(n-\frac{1}{4}\right)\right)\ ,
\end{matrix*}\label{eq:1minqz}
\end{align}
where $\operatorname{erf}(x)=\frac{2}{\sqrt{\pi}} \int_{0}^{x} e^{-\tau^{2}} \mathrm{d} \tau$. Note that \eqref{eq:1minqx} is a monotonically increasing function in $r$, whereas \eqref{eq:1minqz} is monotonically decreasing in $r$.

Similarly, we obtain the following expressions for the conditional logical error probabilities given syndrome~$s_0$:
\begin{align}\label{eq:GaussianresidualerrGKPinfo}
\begin{matrix}
p_{\overline{I}}^{s_0}&=&(1-q_{\overline{X}}^{s_0})&\cdot&(1-q_{\overline{Z}}^{s_0})&\\
p_{\overline{X}}^{s_0}&=&q_{\overline{X}}^{s_0}& \cdot&(1-q_{\overline{Z}}^{s_0})&\\
p_{\overline{Y}}^{s_0}&=&q_{\overline{X}}^{s_0}& \cdot& q_{\overline{Z}}^{s_0}&\\
p_{\overline{Z}}^{s_0}&=&(1-q_{\overline{X}}^{s_0})& \cdot& q_{\overline{Z}}^{s_0}&\ .
\end{matrix}
\end{align}
As discussed, the rectangular lattice GKP code~$\mathsf{GKP}(\cL_r)$ 
with squeezing parameter~$r>1$   is the result of applying the one-mode squeezing operation with parameter $r>1$ to the square lattice GKP code. Thus, using identity~\eqref{eq:sympl_trafo_displacements} and a variable substitution, we get
 $\cN_{f_{\sigma^2}}(U_{S_r^{-1}}\rho U_{S_r^{-1}}^\dagger)=U_{S_r^{-1}}\cN_{f_{\tilde{Z}}}(\rho) U_{S_r^{-1}}^\dagger$ where $\tilde{Z}\sim \mathsf{N}({\bf 0},\Sigma_r)$, with the diagonal matrix~$\Sigma_r$ defined in~\eqref{eq:tildeSigmaimprov}.  That is, the noise is effectively transformed to one
where the two quadratures are displaced independently
with variances $(\tilde{\sigma}_Q^2,\tilde{\sigma}_P^2):=(\sigma^2/r,r\sigma^2)$ (cf.~\eqref{eq:tildeSigmaimprov}).  Note that $\tilde{\sigma}_Q^2< \tilde{\sigma}_P^2$ for $r> 1$. This results in biased noise on the level of GKP-qubits  as can be seen in \eqref{eq:Gaussianresidualerr}: for $r>1$, we have $p_{\overline{X}}< p_{\overline{Z}}$.

Since we use the code~$\mathsf{GKP}(\cL_r)$ extensively below, let us briefly discuss the degree of squeezing necessary to produce associated code states from
standard square lattice GKP code states, i.e., 
the parameter~$r$ in~\eqref{eq:squeezingsymplectic}. As common in quantum optics, we quantify squeezing in terms of the \emph{squeezing factor} $s$, which is given in units of dB and defined as
\begin{align}
s(\sigma_Q):=-10\log_{10}\left(\frac{\sigma_Q^2}{\sigma_0^2}\right)\ ,
\end{align}
where $\sigma_0^2=1/2$ is the variance in the $Q$-quadrature in the vacuum state. As just derived, the effective variance in the $Q$-quadrature of a rectangular ($r>1$) lattice code is $\tilde{\sigma}_Q^2=\sigma^2/r$, where~$\sigma^2$ is the corresponding variance of a square ($r=1$) lattice code. We therefore have $s(\tilde{\sigma}_Q)=s(\sigma^2)+10\log_{10}(r)$, i.e., the introduction of asymmetry $(r>1)$ yields an increase of the squeezing factor by  an additive term~$10\log_{10}(r)$. 

We note that for  more general Gaussian encodings corresponding to some Gaussian unitary~$U_S$, the 
resulting modified density $f_{\tilde{Z}}$ 
over displacements may not be a product distribution with respect to the quadratures. This generally results in correlations between the $\overline{X}$ and $\overline{Z}$ errors. One such example is the asymmetric hexagonal lattice GKP code (cf.\ Section~\ref{sec:hexagonallatticeGKP}). Expressions for the associated logical error probabilities under the noise channel \eqref{eq:Gaussnoisechannel} are given in Appendix~\ref{app:residuallogerrmixing}.

\section{The (modified) surface-GKP code\label{sec:surfacegkp}}
In this section, we introduce our main idea, namely the effective generation of biased noise in modified surface-GKP codes by the introduction of asymmetry. The  modified surface-GKP codes we study  are obtained by concatenating an asymmetric GKP code at the base (inner) level with a qubit (surface) code at the top (outer) level. They benefit from the fact that surface codes are resilient to certain forms of biased noise.

In Section~\ref{sec:priorbiased} we briefly review prior uses of biased noise in the discrete- and continuous-variable settings. 
In Section~\ref{sec:engineered}, we give the details of how 
the concatenation has to be done in order to yield an improvement over standard surface-GKP codes. Then, in Section~\ref{sec:decodingasymmetric}, we explain how to use the BSV decoder to decode asymmetric surface-GKP codes with and without GKP side information.

\subsection{Prior work on biased noise in codes\label{sec:priorbiased}}
For certain physical systems, physically relevant processes naturally lead to biased noise. There is a long history of considering such scenarios in quantum error correction:  it was shown that biased noise is typically less detrimental than more general noise if suitable encodings are used. For example, early work~\cite{gourlaysnowdon00} on qubit codes considered the extreme case of pure dephasing noise. Fault-tolerance schemes for asymmetric (i.e., predominantly dephasing) qubit noise models were introduced and analyzed in~\cite{stephensetal08,evansetal07,aliferispreskill08}, giving improved estimates for error rates and error thresholds for biased noise. 

More recently, and  directly relevant to our work, a significant degree of resilience to  biased noise of surface codes was observed numerically by Tuckett, Bartlett and Flammia~\cite{tuckettetal18} when using the
 approximate maximum likelihood decoder of Bravyi, Suchara and Vargo~\cite{bsv14}. That is, recovery from Pauli noise
\begin{align}
 \begin{matrix*}[l]
\cN_{(p_I,p_X,p_Y,p_Z)}(\rho)=&p_I\rho+ p_X  X\rho X^\dagger\\
&+p_Y Y\rho Y^\dagger +p_Z Z\rho Z^\dagger
\end{matrix*}\label{eq:qubitnoisechanneln}
\end{align}
(i.i.d.~on each qubit) is 
successful (in the limit of large code sizes) even for values of $p_Y$  close to~$1/2$, in the case where $p_X=p_Z$ and the bias
\begin{align}
\eta:=p_Y/(p_X+p_Z)\label{eq:biasednoiseoriginallystudied}
\end{align}
is sufficiently large. This is in sharp contrast to the case of independent $X$ and $Z$ noise, where 
\begin{align}\label{eq:independentXZnoise}
\begin{matrix}
p_X&=&q_X&\cdot &(1-q_Z)\\
p_Y&=&q_X&\cdot &q_Z\\
p_Z&=&(1-q_X)&\cdot &q_Z\ 
\end{matrix}
\end{align}
and where the tolerable noise thresholds for $q_X$ and $q_Z$ are smaller than~$\approx 11\%$~\cite{kitaevetal02}.

 Subsequent work by Tuckett, Darmawan et al.~\cite{tuckettetal19} extended these findings significantly in several directions: on the one hand, new numerical results for (rotated), non-square, i.e., $r\times s$-surface codes were established. On the other hand, additional analytical arguments were given to support the numerical evidence: the minimum weight of a   $Y$-type logical operator was computed (as a function of $r,s$), and in the limiting case of pure $Y$-noise, a threshold of~$50\%$ error probability was established by analyzing a corresponding decoder.

Note, however, that not all qubit codes show such improvements: in fact, it was shown numerically in~\cite{tuckettetal19} that the threshold of color codes~\cite{bombindelgado06} actually decreases with stronger bias. It should also be emphasized that improvements can only be obtained by suitably aligning the code's stabilizers
with the asymmetry in the noise as discussed below.

While these results apply to qubit codes, more recent work has 
shown that biasedness of noise can naturally emerge when using continuous variable (CV) systems to encode logical qubits.
Specifically, Guillaud and Mirrahimi~\cite{guillaudmirrahimi19} and Puri et al.~\cite{purietal19} argued that bosonic cat-codes exhibit logical-level  noise biased towards dephasing (under certain bosonic noise processes affecting the physical modes), where the bias is tunable by adjusting the ``cat size''   parametrizing the code. They show how to exploit this by concatenating the cat code with the (qubit) repetition code, which has some resilience to biased noise. In addition, methods for achieving bias-preserving logical gates are proposed and analyzed in detail.

\subsection{Exploiting engineered bias in surface-GKP codes\label{sec:engineered}}
To tap into the potential of the surface code to correct biased noise, we  require the following modifications of what are typically considered to be surface-GKP codes:
\begin{enumerate}[(i)]
\item
First, we use a Gaussian unitary~$U_{S}$ applied to each mode to turn 
isotropic Gaussian noise into biased noise at the GKP-qubit level, that is, we work with
modified GKP code states~$U_{S}\gkpsquarezero$ and $U_{S}\gkpsquareone$, where $\gkpsquarezero$ and $\gkpsquareone$ are the standard square lattice GKP basis states.   
\item \label{surfaceGKPmodification2}
Second, we change the way the GKP code is concatenated with the surface code.
More precisely, we map   Pauli-$Y$ (instead of Pauli-$Z$) eigenstates of a (GKP) qubit to the modified GKP basis states~$U_{S}\gkpsquarezero$ and $U_{S}\gkpsquareone$ according to 
\begin{align}
\begin{matrix}
\ket{+i}&=\frac{1}{\sqrt{2}}(\ket{0}+i\ket{1}) &\quad \mapsto & \quad U_{S}\gkpsquarezero&\\
\ket{-i}&=\frac{1}{\sqrt{2}}(\ket{0}-i\ket{1}) &\quad \mapsto & \quad U_{S}\gkpsquareone& ,
\end{matrix}\label{eq:Yencodedi}
\end{align}
and linearly extend this to a definition of a (modified) isometric GKP-qubit encoding~$\mathbb{C}^2\rightarrow L^2(\mathbb{R})$.
\end{enumerate}
We note that the definition~\eqref{eq:Yencodedi} ensures that qubit operators are identified with logical GKP operators according to
\begin{align}
\begin{matrix}
Y&\leftrightarrow &U_{S} \gkpsquarelogicalZ U_{S}^\dagger&\\
Z&\leftrightarrow &U_{S} \gkpsquarelogicalX U_{S}^\dagger&\ .
\end{matrix}\label{eq:mappingmodifiedconseq}
\end{align}
In the case of a rectangular lattice GKP code, the correspondence~\eqref{eq:mappingmodifiedconseq} leads to effective (qubit level) noise channels with independent $Y$ and $Z$ noise, i.e.,
\begin{align}\label{eq:ourpxpypzdependence}
\begin{matrix}
p_X&=&q_Z&\cdot & q_Y&\\
p_Y&=&(1-q_Z)&\cdot & q_Y&\\
p_Z&=&q_Z&\cdot & (1-q_Y)&\ ,
\end{matrix}
\end{align}
in the resulting qubit noise channel~\eqref{eq:qubitnoisechanneln}, where 
\begin{align}\label{eq:relqmod}
(q_Y,q_Z)=(q_{\overline{Z}},q_{\overline{X}})\ ,
\end{align}
with $q_{\overline{X}},q_{\overline{Z}}$ given in~\eqref{eq:logprobsq}.
In particular, $p_Z$ is biased compared to $p_Y$ for $r$ large enough (i.e., $p_Z\ll p_Y$), since $q_Z$ is monotonically decreasing in $r$ whereas $q_Y$ is monotonically increasing in $r$ by the relation \eqref{eq:relqmod} and Eqs.~\eqref{eq:1minqx}, \eqref{eq:1minqz}. We note that
the triple $(p_X,p_Y,p_Z)$ thus is a two-parameter family parameterized by $(q_Y,q_Z)$, and the set of such triples differs from the one studied
in~\cite{tuckettetal18,tuckettetal19}, which consists of tuples of the form $(p_X,p_Y,p_Z)=(p_X,p_Y,p_X)$, cf.~\eqref{eq:biasednoiseoriginallystudied}. In particular, this means that the threshold estimates obtained in~\cite{tuckettetal18,tuckettetal19} cannot directly be lifted to our setting. Rather, we need to separately study the threshold behavior of biased noise of the form~\eqref{eq:ourpxpypzdependence}. 

We find (see Section~\ref{sec:numerical}) that the surface code is well-equipped against displacement noise specified by~\eqref{eq:ourpxpypzdependence} if $p_Z\ll p_Y$. Specifically, we find threshold standard deviations $\sigma$ of the noise corresponding to -- in a standard (non-modified) surface-GKP code -- Pauli-$X$ (equivalently: Pauli-$Z$) error probabilities strictly above the $\approx 11\%$~of the plain surface code, similar to the results in~\cite{tuckettetal18,tuckettetal19}.

\subsection{Decoding modified surface-GKP codes with the BSV decoder\label{sec:decodingasymmetric}}
In the surface-GKP code, each of the $n$~qubits of the surface code is encoded
 in a GKP-qubit (i.e., in a single mode). As with any concatenated code, the surface-GKP code  provides natural families of decoders obtained by combining a decoder for the GKP code with a decoder for the surface code. Such a decoding procedure first decodes the GKP code (the inner code), and subsequently the surface code (the outer code). Here we primarily use  the nearest lattice point recovery procedure discussed in Section~\ref{sec:gkpdecoding} to decode the GKP code, and the BSV decoder discussed in Section~\ref{sec:bsvdecoder} for the surface code.
 
\begin{figure}
\vspace{5ex}
 	\subfloat[Decoding without GKP side information.\label{fig:GKPwithout}]{  \includegraphics[width=.4\textwidth]{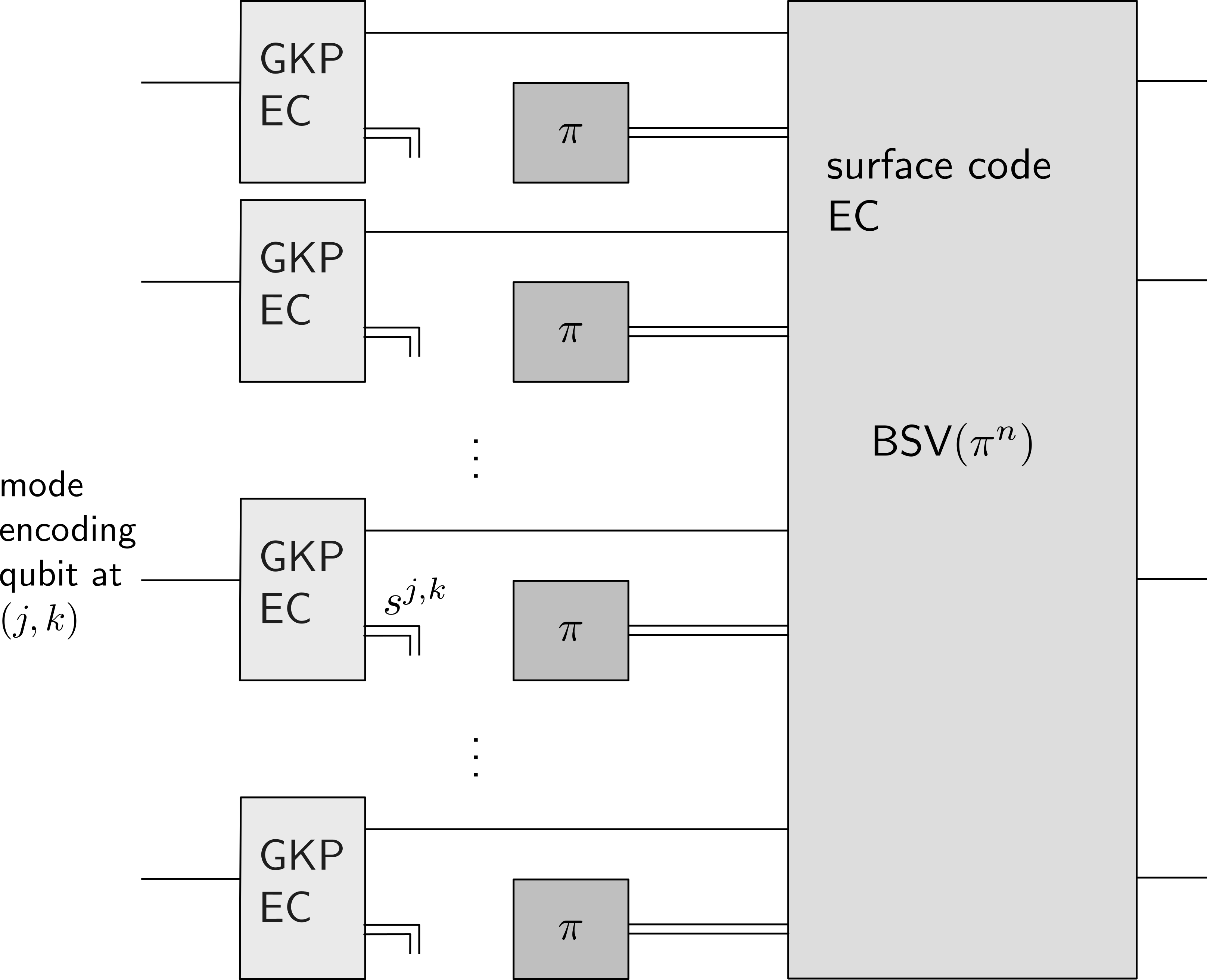}}\\ 
  	\subfloat[Decoding with GKP side information.\label{fig:GKPwith}]{  \includegraphics[width=.4\textwidth]{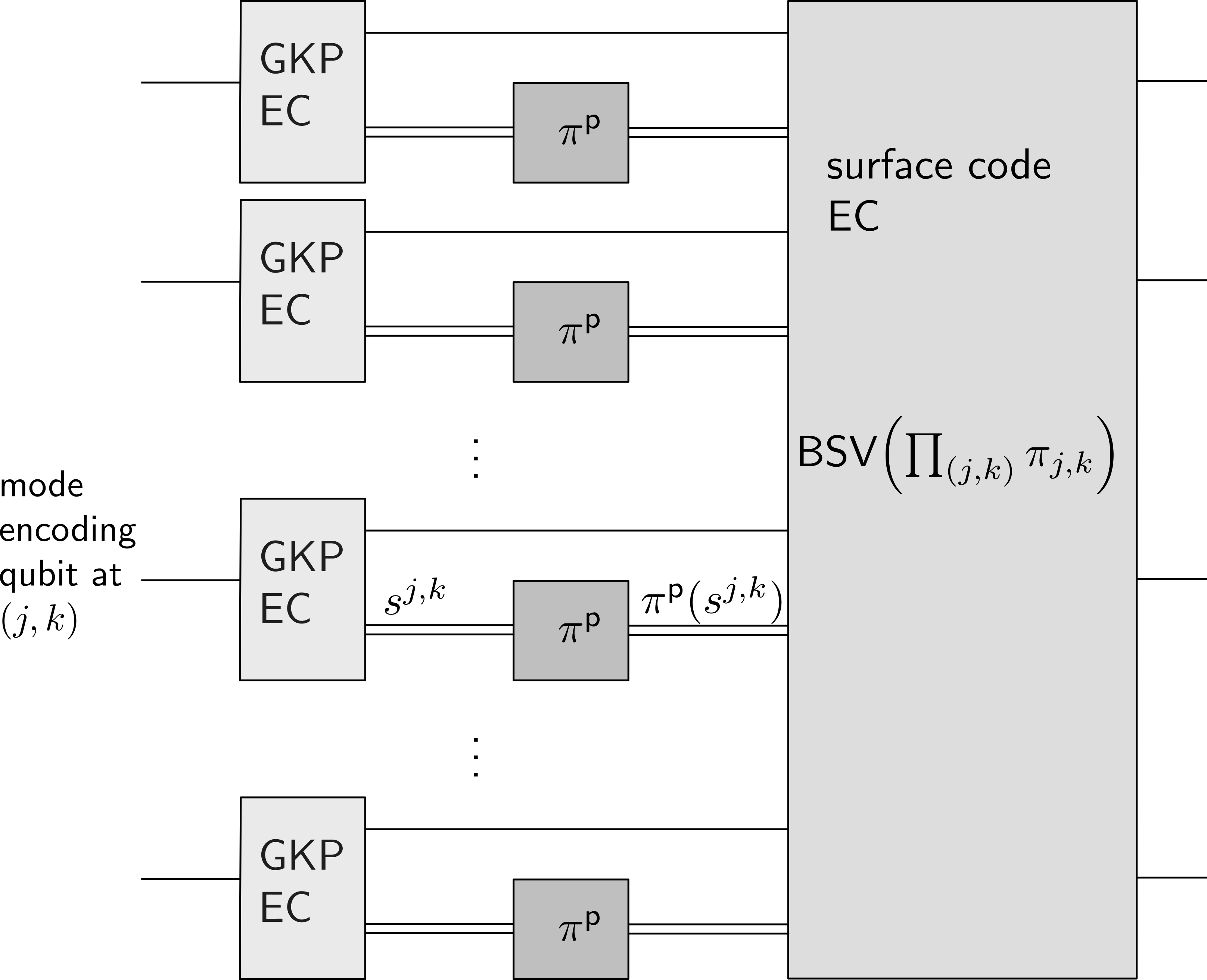}}
    \caption{Two ways of decoding the surface-GKP code. In both cases the GKP error correction circuit produces syndrome information $s^{j,k}$. When decoding without  side information (Fig.~\ref{fig:GKPwithout}), this  information is ignored and the same prior distribution~$\pi$ over logical Pauli errors $\{\overline{I},\overline{X},\overline{Y},\overline{Z}\}$ is used for every site. The surface-code BSV decoder (which also involves syndrome measurement but whose syndromes are not shown here) is run with the associated product (i.i.d.)~distribution~$\pi^n$. Fig.~\ref{fig:GKPwith} shows how to exploit the GKP side information: For each qubit~$(j,k)$ a prior distribution $\pi_{j,k}=\pi^{\mathrm{p}}(s^{j,k})$ over logical Pauli errors is computed from the  GKP syndrome~$s^{j,k}$, and this is then used in the BSV decoder as the prior distribution.\label{fig:decodingwithorwithout}}
\end{figure}

 Recall that the GKP decoder produces syndrome information~$s^{j,k}$ for every GKP-encoded qubit~$(j,k)$ (see Section~\ref{sec:errorrecoveryGKP}).  Here we use $j$ and $k$ to index the row and column in the square lattice geometry (see Fig.~\ref{surfacecodedfour}).  This syndrome information can either be ignored or passed on to the outer decoder. The difference is illustrated in Fig.~\ref{fig:decodingwithorwithout}. In the former case,  i.e., decoding without side information, the prior distribution $\pi_{j,k}=(p^{j,k}_{\overline{I}},p^{j,k}_{\overline{X}},p^{j,k}_{\overline{Y}},p^{j,k}_{\overline{Z}})$ over logical Pauli-error probabilities  (seen by the outer decoder) are identical for each qubit $(j,k)$, i.e., $\pi_{j,k}=\pi$, and equal to
 the averaged  GKP-qubit probabilities $\pi=(p_{\overline{I}},p_{\overline{X}},p_{\overline{Y}},p_{\overline{Z}})$  given in expression~\eqref{eq:cosetprobabilitiesgkp} (that is, \eqref{eq:Gaussianresidualerr} for rectangular lattice GKP-qubits subject to isotropic Gaussian noise). Thus the BSV decoder is provided with the i.i.d.~distribution $\pi^n$, that is, the tensors in Fig.~\ref{bulktensors} are qubit-independent.

 Alternatively, one may use the GKP syndrome information in the surface code decoding step. In this case, the prior probabilities $\pi_{j,k}$ passed to the surface code decoder are potentially different for each GKP-encoded qubit~$(j,k)$. They are computed from the associated GKP syndrome~$s^{j,k}$ according to expression~\eqref{eq:bayesiandecodingprior}, which involves the distribution over the bosonic displacement errors, and we write $\pi_{j,k}=\pi^{\mathrm{p}}(s^{j,k})$. The BSV decoder can then be run with the (non-identical) product distribution $\prod_{(j,k)}\pi_{j,k}$, where each qubit experiences independent noise with a distribution depending on the outcome of the associated GKP measurement.
 
 We note that decoding with and without side information
 has been studied previously in~\cite{fukuietal18,vuillotetal}, for the surface-GKP code and the toric-GKP code respectively, based on square lattices. 
 These authors used a minimum weight matching decoder for the toric/surface code. Syndrome information from the GKP measurements was  translated into different edge weights (where the two papers use different heuristic expressions) to be passed to the minimum weight matching decoder.
 In contrast, we consider non-square lattice GKP codes, and use the BSV decoder: here the way syndrome information is passed is completely determined. Indeed, for bond dimension $\chi\rightarrow\infty$, this results in maximum likelihood decoding based on the available syndrome information.
 
\section{Threshold estimates for the modified surface-GKP code\label{sec:numerical}}
In this section, we present numerical results obtained by applying the decoding procedure discussed in Section~\ref{sec:decodingasymmetric} to study the effect of asymmetry in surface-GKP codes.  We primarily study if asymmetry increases thresholds, even when no GKP side information is used, in rectangular lattice GKP codes. Additionally, we provide numerical data for rectangular and asymmetric hexagonal lattice GKP codes using GKP side information.

\subsection{Simulation method}\label{sec:simulationmethods}
The results are based on Monte-Carlo simulation of the error correction processes illustrated in Fig.~\ref{fig:decodingwithorwithout}. A pseudocode of the corresponding routine in the case where the GKP side information is ignored is presented in Fig.~\ref{fig:montecarlowithoutsideinfo} in Appendix~\ref{sec:simulationpseudocode}, whereas Fig~\ref{fig:montecarlowithsideinfo} in Appendix~\ref{sec:simulationpseudocode} gives pseudocode in the case where the side information is used in the BSV decoder. Given a standard deviation~$\sigma$, both routines simulate the process of applying a displacement error distributed according to  $\mathsf{N}({\bf 0},\sigma^2 I_{2})$  independently to each GKP-qubit of a distance-$d$ surface code,
 subsequent GKP syndrome extraction and error correction, and surface-code error correction with or without this syndrome information. The output of both routines is the residual logical Pauli error that this process yields on the logical surface-GKP-qubit. Thus the routines permit to empirically estimate the averaged logical error channel. More coarsely, we use these procedures to study the logical error probability~$P_{\textrm{err}}=P_{\textrm{err}}(\sigma,d,r)$ as a function of the standard deviation~$\sigma$ of the noise (cf.~\eqref{eq:Gaussnoisechannel}), the code distance~$d$, and the asymmetry ratio~$r$ defining the GKP code lattice.

\onecolumngrid  
\begin{figure*}
\begin{minipage}{0.98\linewidth}
\subfloat[Ratio $r=1.0$, threshold $\sigma_c\approx 0.540$ \label{Ratio1}]{  \includegraphics[width=.47\textwidth]{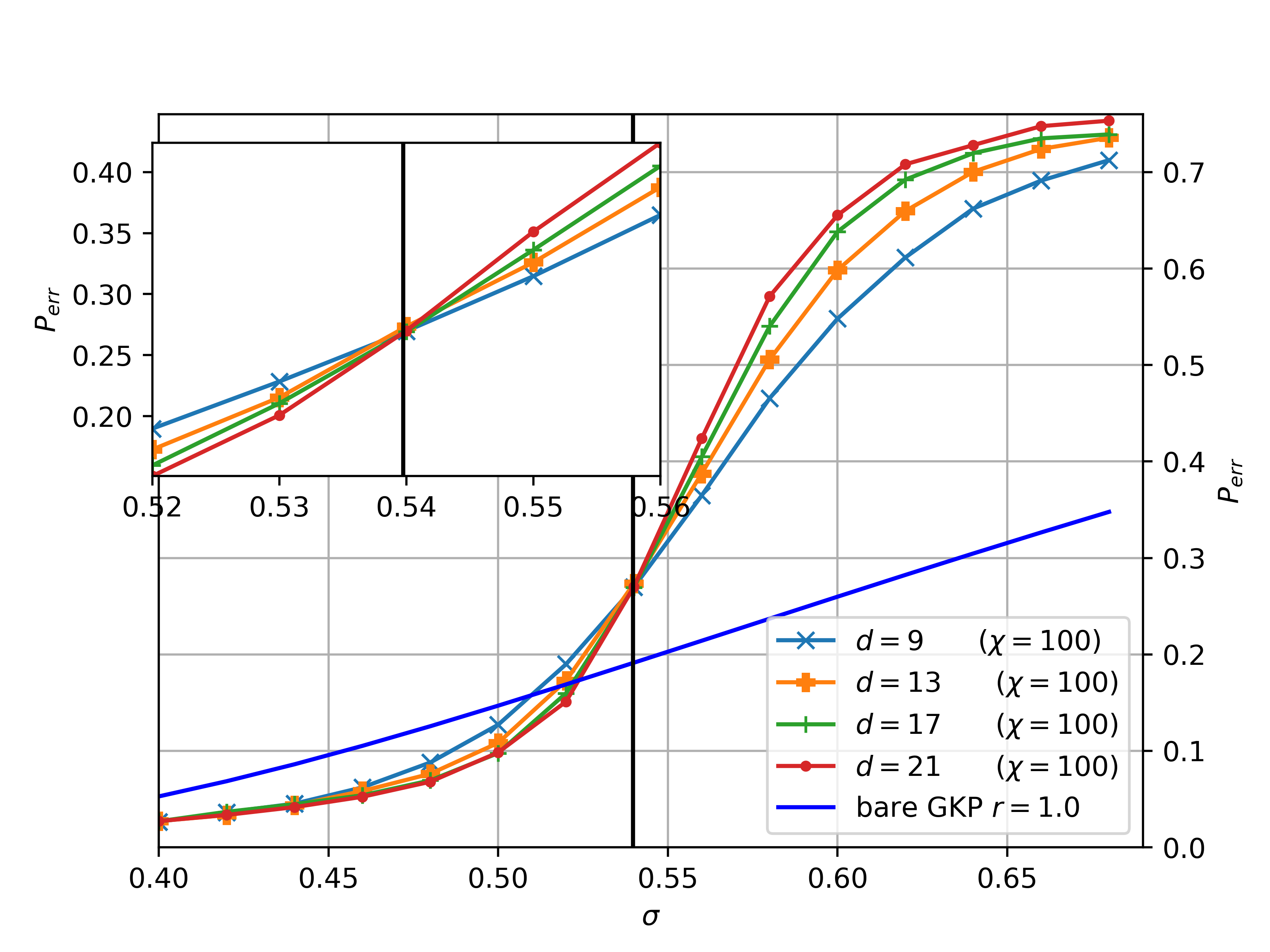} 
\includegraphics[width=.47\textwidth]{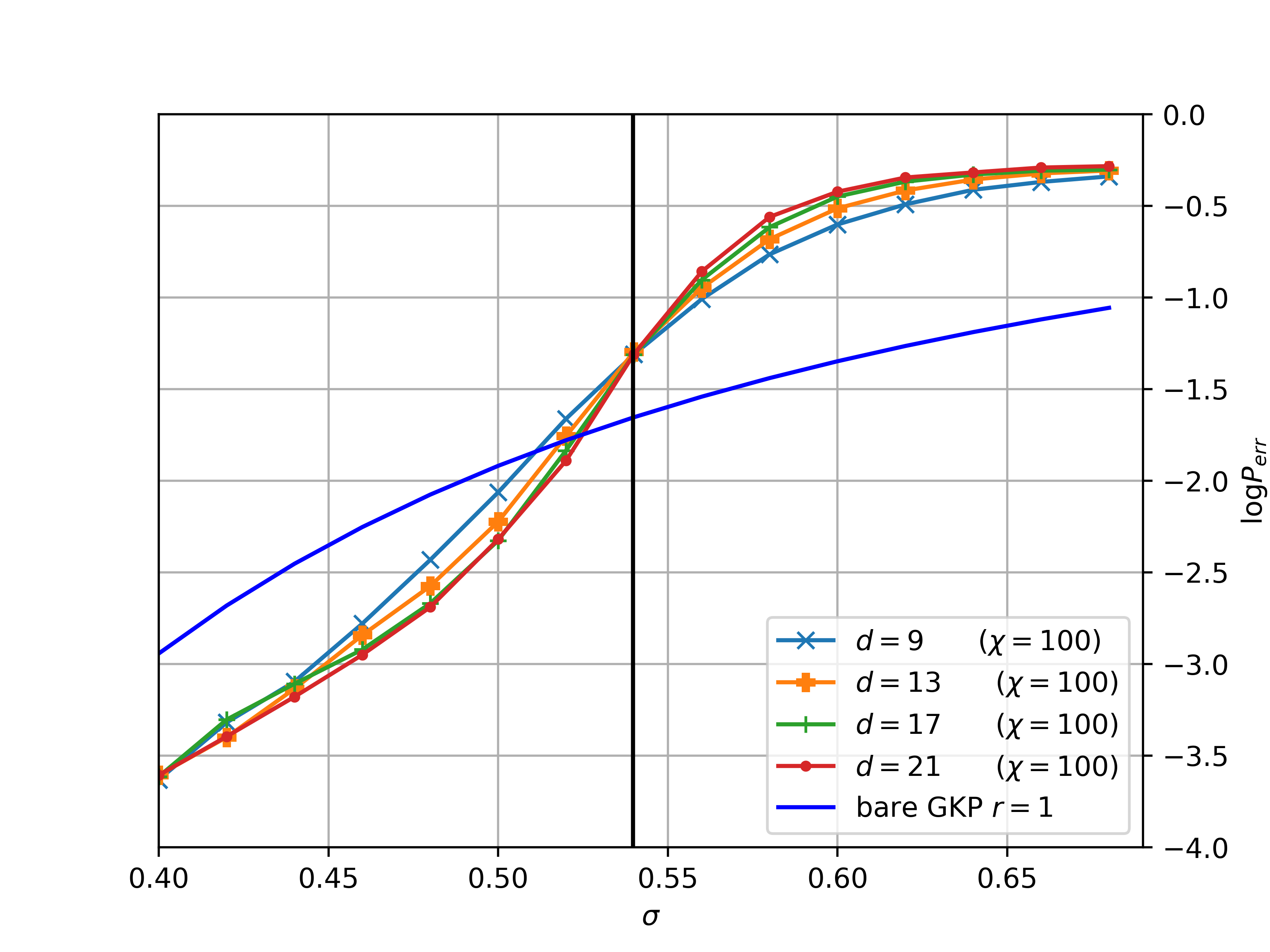}}\\
\end{minipage}
\begin{minipage}{0.98\linewidth}
\subfloat[Ratio $r=2.0$,  threshold $\sigma_c\approx 0.562$ \label{Ratio2}]{  \includegraphics[width=.47\textwidth]{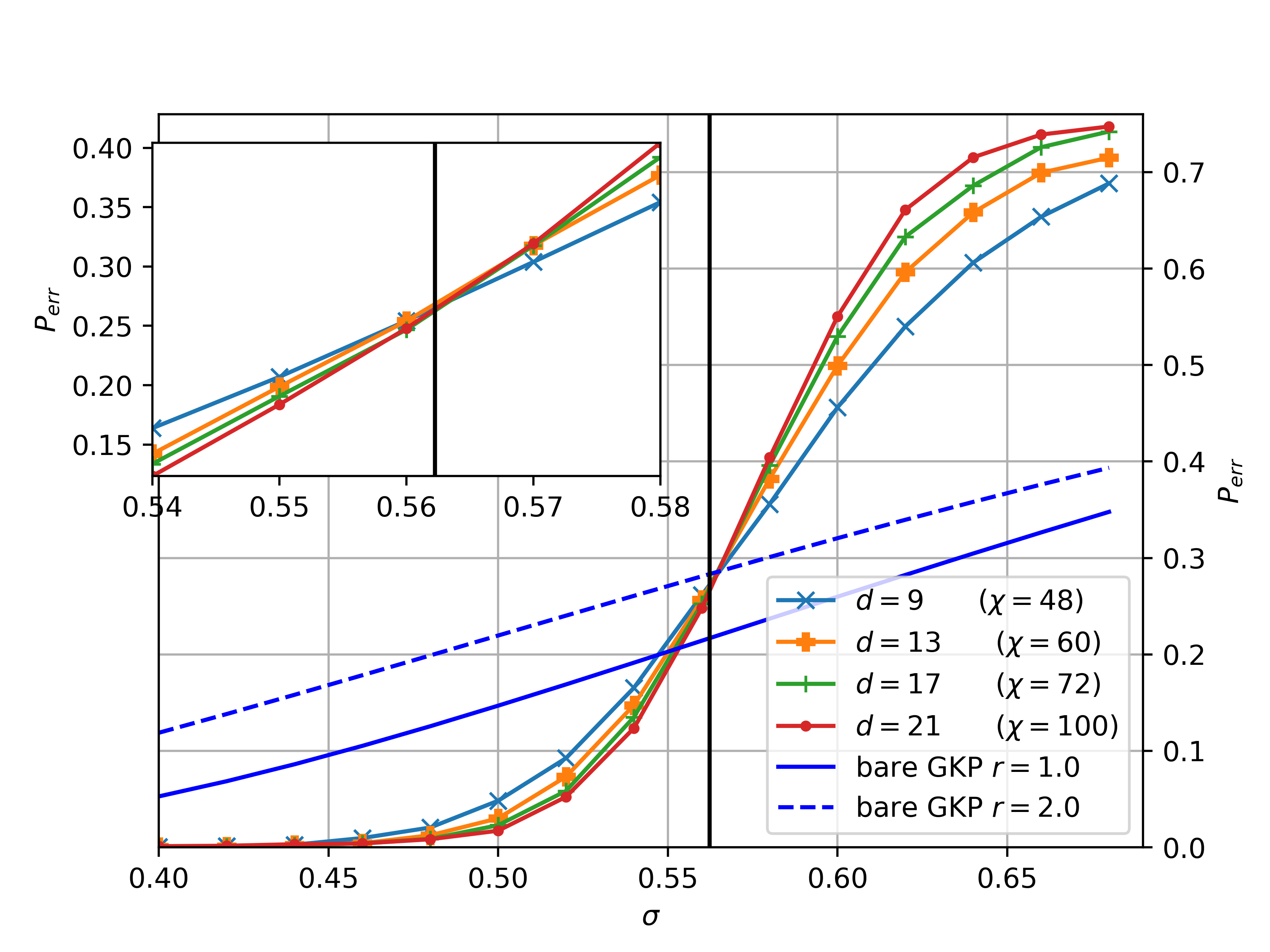}
\includegraphics[width=.47\textwidth]{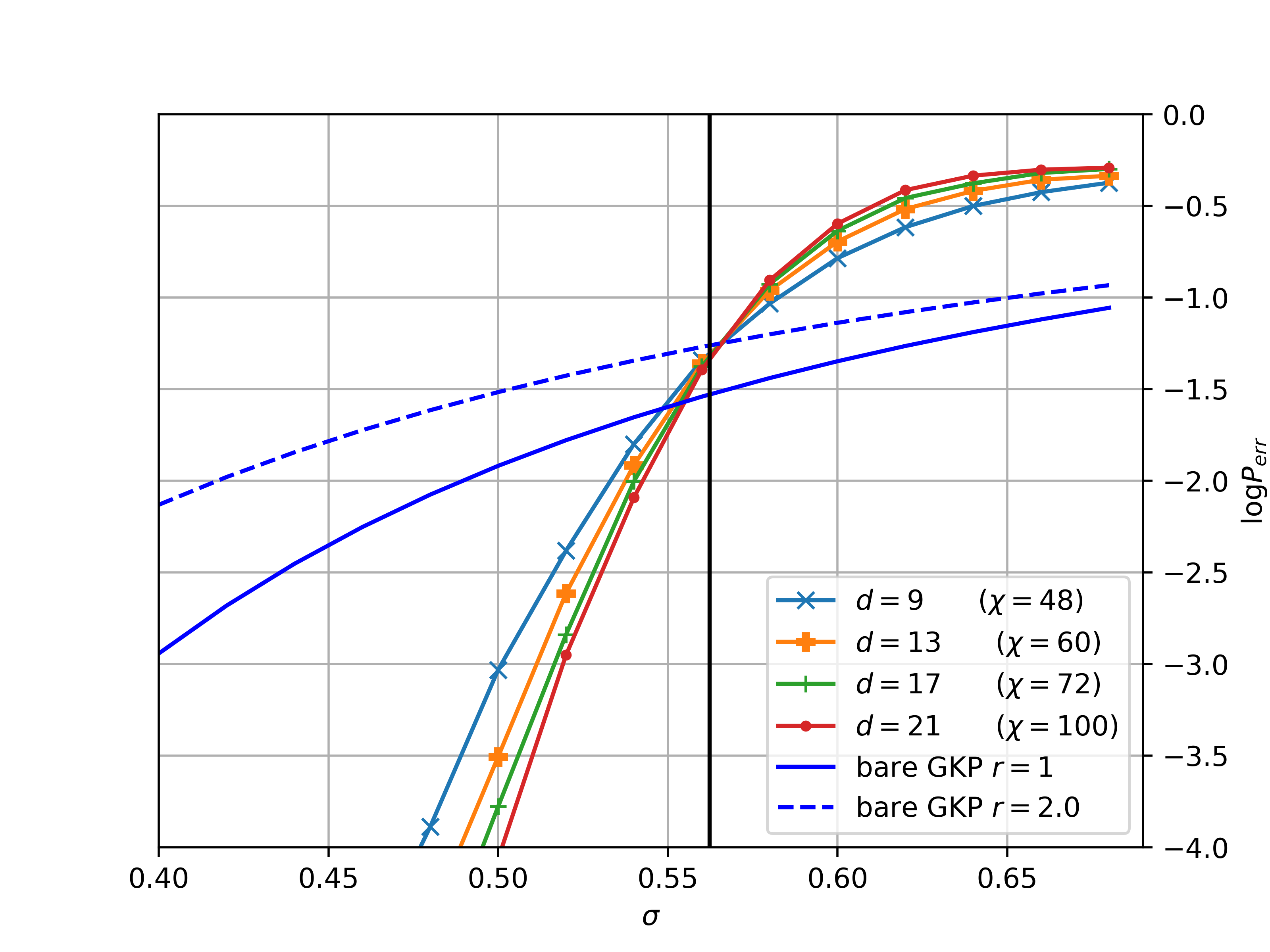}}\\
\end{minipage}
\begin{minipage}{0.98\linewidth}
\subfloat[Ratio $r=3.0$, threshold $\sigma_c\approx 0.581$ \label{Ratio3}]{  \includegraphics[width=.47\textwidth]{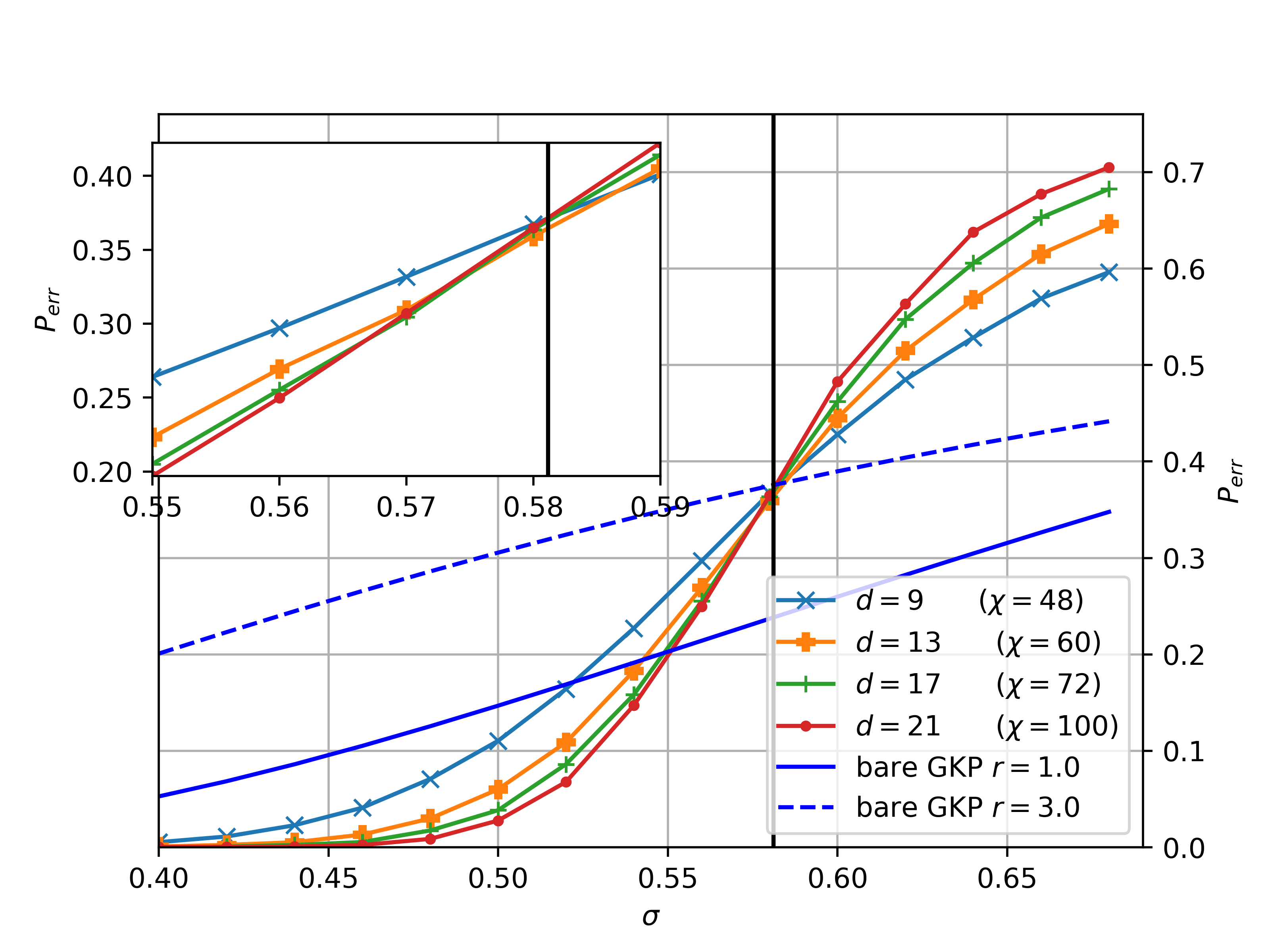}
\includegraphics[width=.47\textwidth]{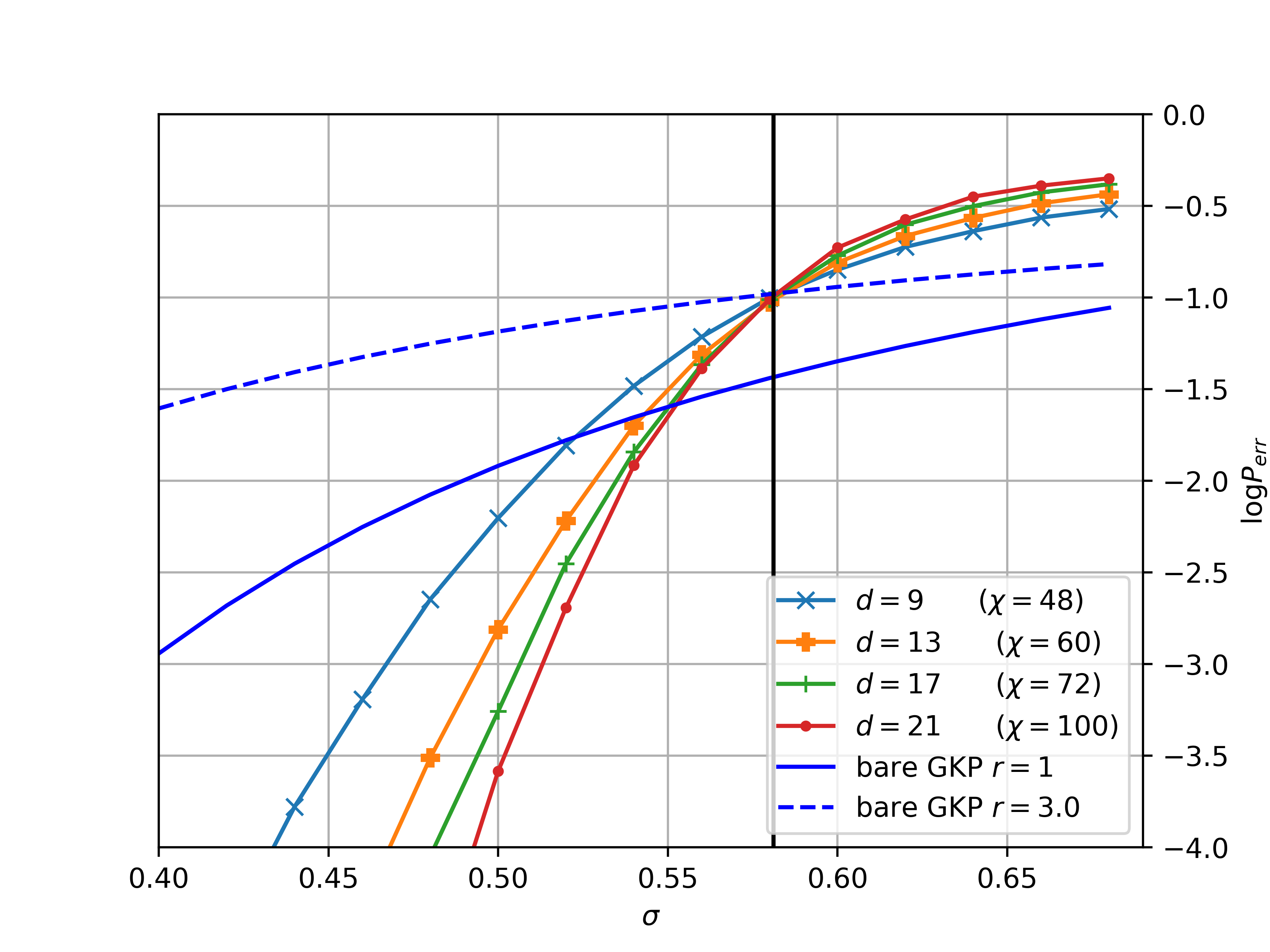}}
\end{minipage}
\caption{The error probability~$P_{\textrm{err}}(\sigma,d,r)$ (left)  for decoding without GKP side information for different asymmetry ratios~$r\in \{1.0,2.0,3.0\}$. Insets give    higher-resolution data around the observed threshold estimate~$\sigma_c$ for the critical noise variance. The latter is indicated by a vertical line. The right hand side shows a log-plot of the error probability~$P_{\textrm{err}}(\sigma,d,r)$.
For comparison, the error probability of the bare GKP code, i.e., without concatenation with the surface code, is shown in blue with the solid line corresponding to ratio $r=1.0$ and the dashed lines corresponding to the respective ratio of the figure.  
}\label{fig:GKPwithoutsideinfo}
\end{figure*}
\clearpage
\twocolumngrid  
 
For a fixed ratio~$r$ we are interested in the {\em threshold}, which is the critical noise level, i.e., the critical value of~$\sigma$, below which the error probability~$P_{\textrm{err}}$ can be made arbitrarily small by choosing a sufficiently large code distance~$d$. Following standard reasoning, an estimate~$\sigma_c$ for this quantity can be obtained by studying the intersection points of a family~$\{\sigma\mapsto P_{err}(\sigma,d,r)\}_{d\in D}$ of curves parameterized by a (finite) set of distances~$D\subset\mathbb{N}$. More precisely, we use the critical exponent method of~\cite{WANG200331}: define a correlation length $\xi=(\sigma-\sigma_c)^{-\mu}$ for some critical exponent $\mu$, and assume that for $d\gg \xi$, the failure probability only depends on the dimensionless ratio $d/\xi$, i.e., it is a function of the variable $x\coloneqq (\sigma-\sigma_c)d^{1/\mu}$. Our thresholds are obtained by numerically fitting the data to a power series up to quadratic terms in $x$, i.e., using the same fitting formula as in~\cite{tuckettetal18}.

For our simulations, we choose physical parameters as follows. We consider code sizes~$d$ satisfying $9\leq d\leq 21$, Gaussian displacement noise standard deviations~$\sigma$ in the interval $[0.4,0.7]$, and asymmetry ratios $r\in [1,4]$. The parameters for the simulation are as follows: the number of Monte Carlo samples for an  estimate of $P_{err}$ for fixed physical parameters~$(\sigma,d,r)$ is chosen  between~$10000$  and $50000$ depending on the distance~$d$. For the BSV decoder, we use bond dimensions~$\chi$ in the interval~$[48,100]$. We provide a more detailed description and justification of these choices in  Appendix~\ref{sec:choiceofparams}.
  
\subsection{Numerical results for surface-$\GKP(\cL_r)$ codes}\label{sec:results}
 We primarily focus on decoding surface-GKP codes  without making use of GKP side information (other scenarios will be discussed below). We present (selected) simulation results for this scenario: in Figs.~\ref{Ratio1}-\ref{Ratio3}, the curves
$\{\sigma\mapsto P_{err}(\sigma,d,r)\}_{d\in \{9,13,17,21\}}$ are shown for asymmetry ratios~$r\in \{1,2,3\}$,  respectively. These curves are used to extract a threshold value~$\sigma_c$ as described above. Fig.~\ref{fig:distances} shows the curves 
$\{\sigma\mapsto P_{err}(\sigma,d,r)\}_{r\in \{1,1.5,2,2.5,3,3.5,4\}}$ indicating how different asymmetry ratios~$r$ change the logical error probability for a fixed distance $d\in \{9,13,17,21\}$.

\subsubsection{Standard (symmetric) surface-GKP code}
The standard (symmetric, i.e., $r=1$) GKP-code corresponding to the square lattice provides our reference point for comparison when studying the effect of asymmetry. For this reason, we work with a relatively high bond dimension of $\chi=100$ for the BSV decoder with the goal of obtaining accurate estimates of the maximum likelihood decoding probability (see Appendix~\ref{sec:choiceofparams} for a detailed discussion). We  obtain a threshold value~$\sigma_c\approx 0.540$, see Fig.~\ref{Ratio1}. 

We  note that this value is comparable to the threshold estimate for $\sigma_c$ between~$0.54$ and~$0.55$ obtained in~\cite{vuillotetal} for independent $X$- and $Z$-noise using the minimum weight matching decoder.
We emphasize, however, that while these results  deal with the same physical error model, there is no a priori reason the obtained thresholds should coincide. This is because the mapping to qubit-level noise used here (cf.\ Section~\ref{sec:engineered},  in particular modification \eqref{surfaceGKPmodification2}) differs from that used in~\cite{vuillotetal}. 

Note also that the data presented in~\cite{bsv14} suggests that the difference between using the minimum weight matching decoder and the BSV decoder may play a minor role for the symmetric case~$r=1$. In fact, for pure Pauli-$X$ (equivalently: Pauli-$Z$) errors even the discrepancy between the threshold values obtained by using the minimum weight matching decoder~\cite{WANG200331} and the maximum likelihood decoder (for which a threshold can be derived from the  numerical  estimates for the
random-bond Ising model obtained in~\cite{MerzChalker02}) is known to be less than~$0.7\%$, see~\cite{bsv14}. This is within the accuracy regime of the results found in~\cite{vuillotetal}.

\subsubsection{Asymmetric surface-GKP codes}
Consider now the asymmetric case, i.e., $r>1$. 
 A subset of corresponding simulation results is  shown in Fig.~\ref{Ratio2} and Fig.~\ref{Ratio3}. These curves  are used to extract threshold estimates. We note that the choice of bond dimension (discussed in Appendix~\ref{sec:choiceofparams}) is less crucial here as our goal is mainly to demonstrate the advantage of asymmetry. Indeed, for any chosen value of~$\chi$, the associated curve in Fig.~\ref{fig:GKPwithoutsideinfo} represents the actual error probability of some (albeit approximate and hence possibly non-optimal)  decoder.

\begin{figure*}
    \subfloat[Distance $d=9$ \label{Dist9}]{  \includegraphics[width=.47\textwidth]{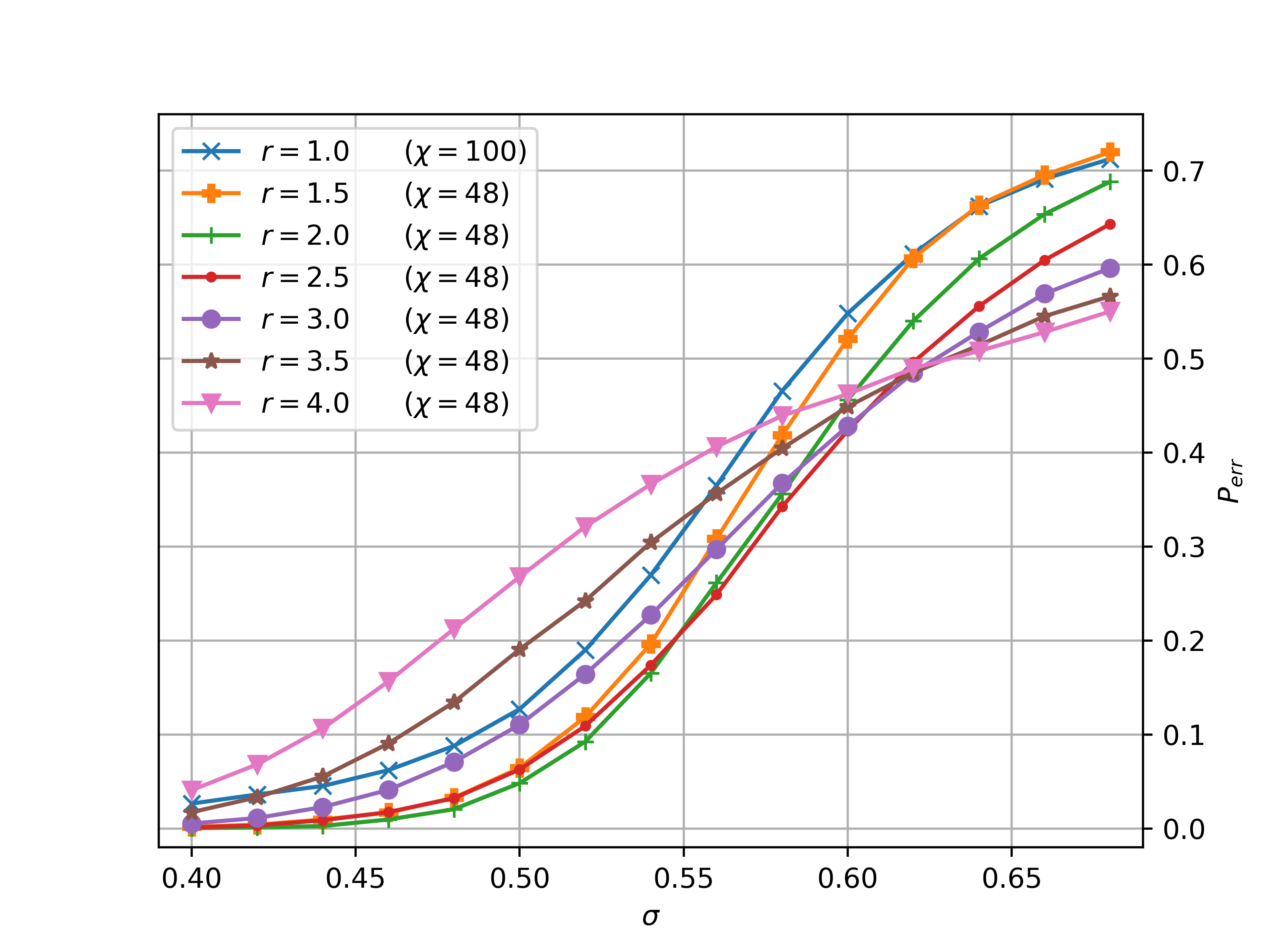}}
    \subfloat[Distance $d=13$ \label{Dist13}]{  \includegraphics[width=.47\textwidth]{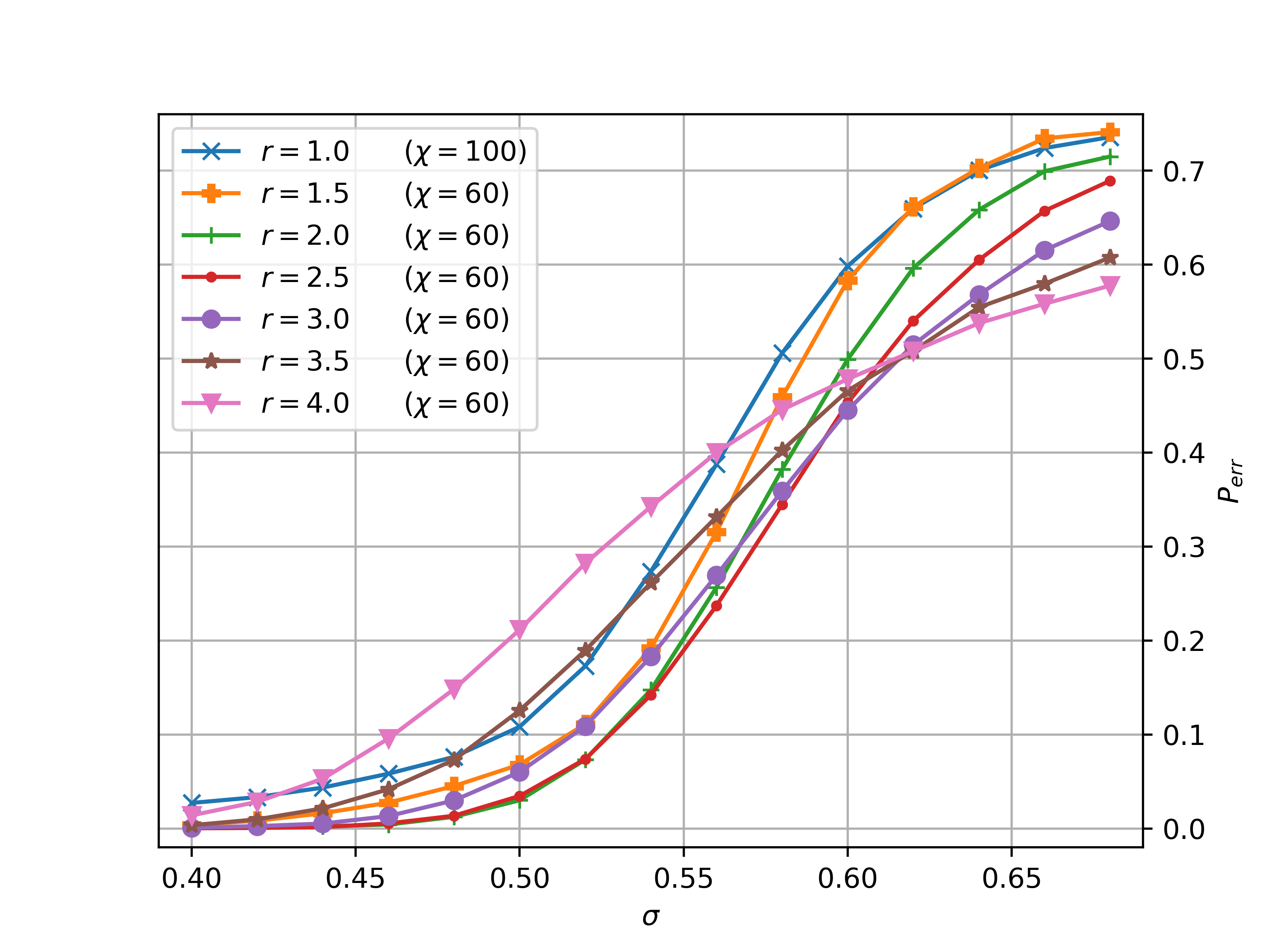}}\\
    \subfloat[Distance $d=17$ \label{Dist17}]{  \includegraphics[width=.47\textwidth]{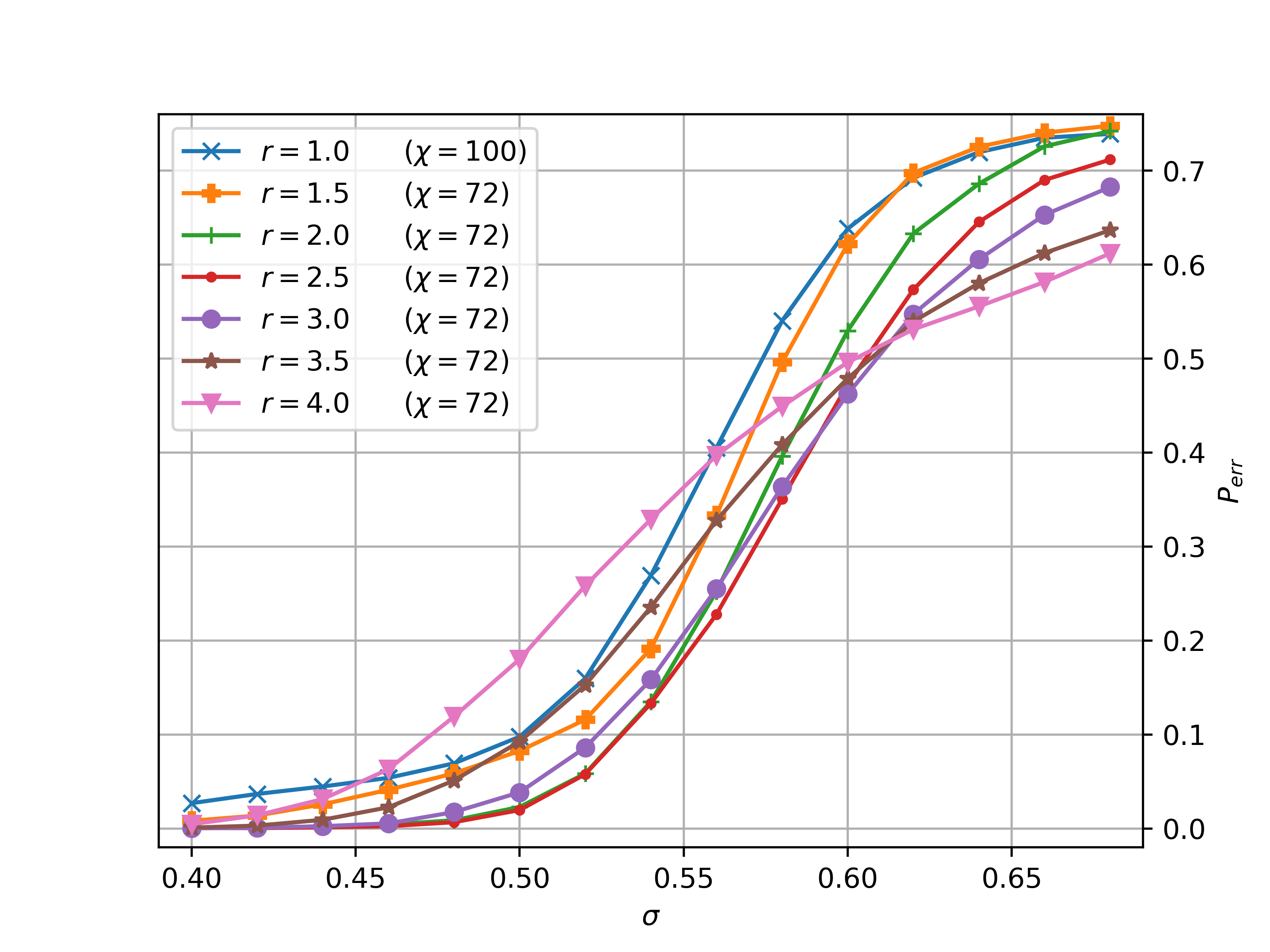}}
    \subfloat[Distance $d=21$ \label{Dist21}]{  \includegraphics[width=.47\textwidth]{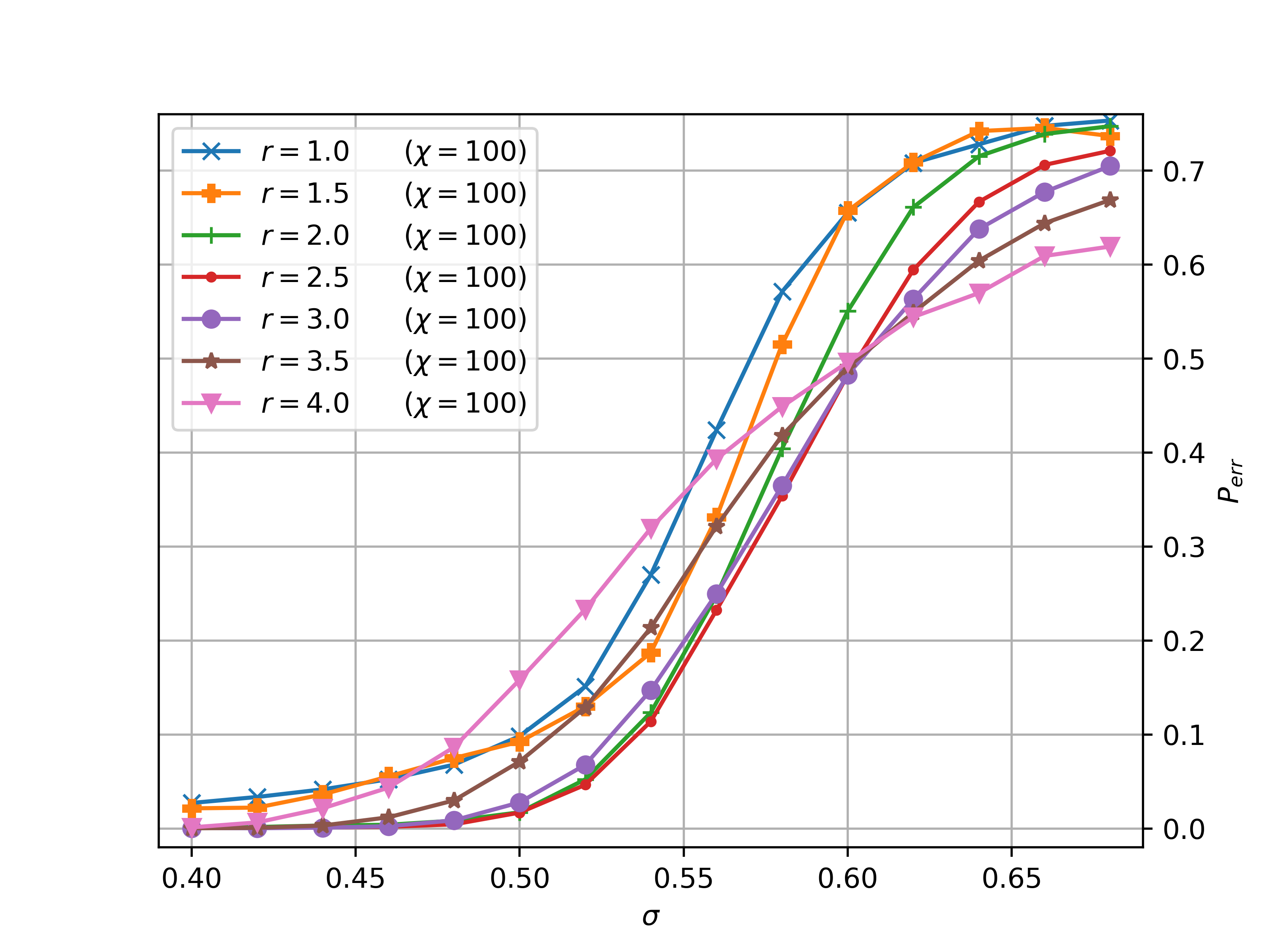}}
    \caption{The dependence of the error probability~$P_{\textrm{err}}(\sigma,d,r)$ on the noise variance~$\sigma$    for decoding without GKP side information. In each figure, the curves $\sigma\mapsto P_{\textrm{err}}(\sigma,d,r)$ for asymmetry ratios~$r$ between $1.0$ and~$4.0$ are given. The individual figures treat code distances~$d\in \{9,13,17,21\}$.
    }\label{fig:distances}
\end{figure*}

 It is also instructive and experimentally  relevant in the near future to consider the error probability for a fixed code distance~$d$. Corresponding curves for a collection of asymmetry ratios are shown in Fig.~\ref{fig:distances}. Figs.~\ref{Dist9}-\ref{Dist21} give the curves $\{\sigma\mapsto P_{err}(\sigma,d,r)\}_{r\in \{1,2,2.5,3,3.5,4\}}$
for different ratios in separate graphs for the same distance~$d\in \{9,13,17,21\}$.

We observe a qualitative difference between the regime of small distances and small asymmetry ratios, and the regime of large  distances  and large asymmetry ratios.  More precisely, we may ask where the error probability decreases monotonically with increasing asymmetry ratio~$r$, i.e.,  $P_{err}(\sigma,d,r')<P_{err}(\sigma,d,r)$ for $r'>r$ (for all $\sigma$). We find that this property holds for all $r,r'\leq 2.0$ for all distances~$d$. For higher distances, e.g., $d=17$ and $d=21$, this property extends up to all asymmetry ratios $r,r'\leq 2.5$. In other words, the monotonicity in~$r$ appears to be a function of the distance~$d$: For example, for $d=9$ and ratios $r'=3.5$ and $r=1$, there are values $\sigma$ such that  $P_{err}(\sigma,9,3.5)>P_{err}(\sigma,9,1)$, whereas for $d=21$ we find that $P_{err}(\sigma,21,3.5)<P_{err}(\sigma,21,1)$ for all $\sigma$. 

It is conceivable  that this non-monotonic behavior of~$P_{err}$ is merely an artefact of the finite size, i.e., the limited code distances~$d$ that can be explored by simulation. For large distances~$d$, the logical error probability may in fact decrease monotonically for increasing asymmetry ratios~$r$. Our current numerical data does not permit us to draw a definite conclusion in this regard.

\subsubsection{Thresholds}
\begin{figure}
\begin{center}
\includegraphics[width=0.48\textwidth]{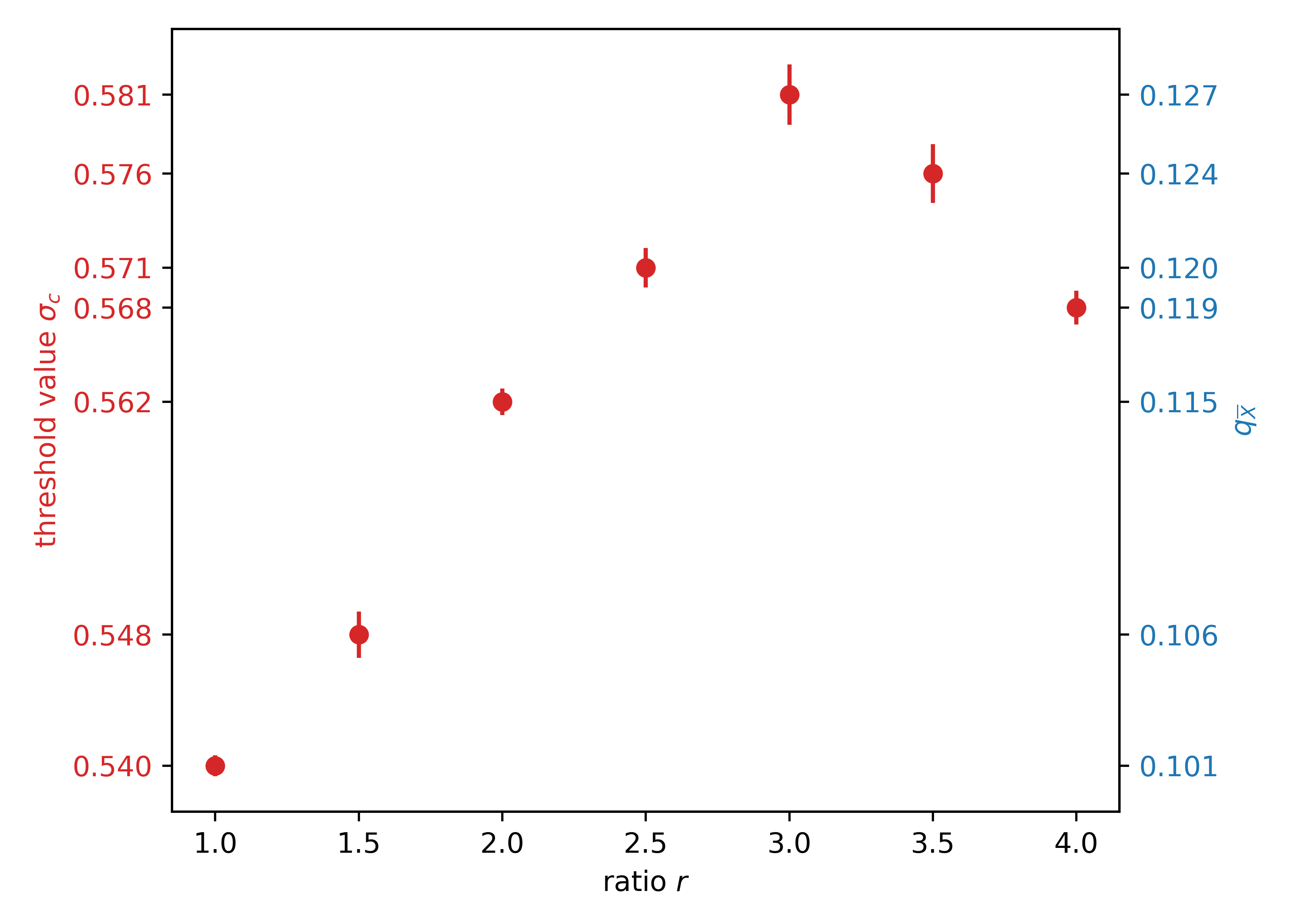}
\end{center}
        \caption{Empirically computed thresholds for different asymmetry ratios~$r$. The error bars depict the standard deviation of the fitted threshold value $\sigma_c$ (left hand side axis) and range from 0.0006 (for $r=1)$ to 0.0019 (for $r=3$). The values on the right hand side axis correspond to $q_{\bar{X}}$ (Eq.~\eqref{eq:1minqx}) for the parameters $\sigma=\sigma_c$ and $r=1$. \label{fig:thresholds} }
\end{figure}

Fig.~\ref{fig:thresholds} summarizes our main numerical findings for the effect of asymmetry on the threshold. It gives the observed threshold values $\sigma_c$ for all considered asymmetry ratios~$r$. For comparison, the vertical axis on the right hand side gives the value of $q_{\overline{X}}$ (cf.~\eqref{eq:1minqx}) for $\sigma=\sigma_c$ and $r=1$, i.e., the Pauli-$X$ (equivalently: Pauli-$Z$) error probability of 
the individual qubits of the surface code. This probability corresponds to the scenario where -- under the same displacement error noise model -- the standard GKP code and GKP error correction without syndrome information is used to encode each qubit.

The numerical results gathered in Fig.~\ref{fig:thresholds} show that every asymmetry ratio~$r>1$ considered here yields an improved threshold compared to the symmetric ($r=1$) case: for example, the threshold (tolerable error probability) improves to $\sigma_c=0.581$ ($0.127$) for $r=3$ from $\sigma_c=0.540$ ($0.101$) for the symmetric case~$r=1$. That is, even a moderate, experimentally achievable amount (cf.\ Appendix~\ref{sec:asymmetryratioused}) of squeezing results in more noise-resilience.

Theoretically it is also interesting to examine the limit of large asymmetry~$r$.  Fig.~\ref{fig:thresholds} shows that the estimated threshold value $\sigma_c$  obtained by simulation is non-monotonic, exhibiting a maximum around~$r=3$. As explained above, this numerical finding may be due to finite-size effects, i.e., the limited system sizes considered in our simulations. We cannot conclusively deduce   that the actual threshold increases monotonically with~$r$ (for large distances~$d$). We note that the limiting effective surface-code qubit-level noise for $r\rightarrow\infty$ is pure $Y$-noise with $50\%$ probability. For the latter, an analytical threshold result is known~\cite{tuckettetal19}. However, this result does not allow us  to draw conclusions about the threshold for any fixed asymmetry ratio~$r$ because of the different order of limits with respect to $d$ and $r$, respectively.

 Note that the effective qubit error distribution resulting from the use of asymmetric GKP codes depends on both the asymmetry ratio~$r$ (a parameter which may in principle be chosen arbitrarily subject to experimental capabilities) and the physical error strength~$\sigma$. In particular, the parameter~$r$ determines the relative weight of $X$, $Y$ and $Z$-Pauli errors similar to the bias parameter~$\eta$ in the work of~\cite{tuckettetal18,tuckettetal19}. We emphasize however that our effective noise model does not have the form of the error distributions considered in those works. In particular, their bias parameter~$\eta$ is not in one-to-one correspondence with our asymmetry ratio~$r$.

\begin{figure}
\vspace{-0.5cm}
    \subfloat[Rectangular lattice GKP without side information, threshold $\sigma_c\approx 0.562$\label{Ratio2without}]{  \includegraphics[width=.47\textwidth]{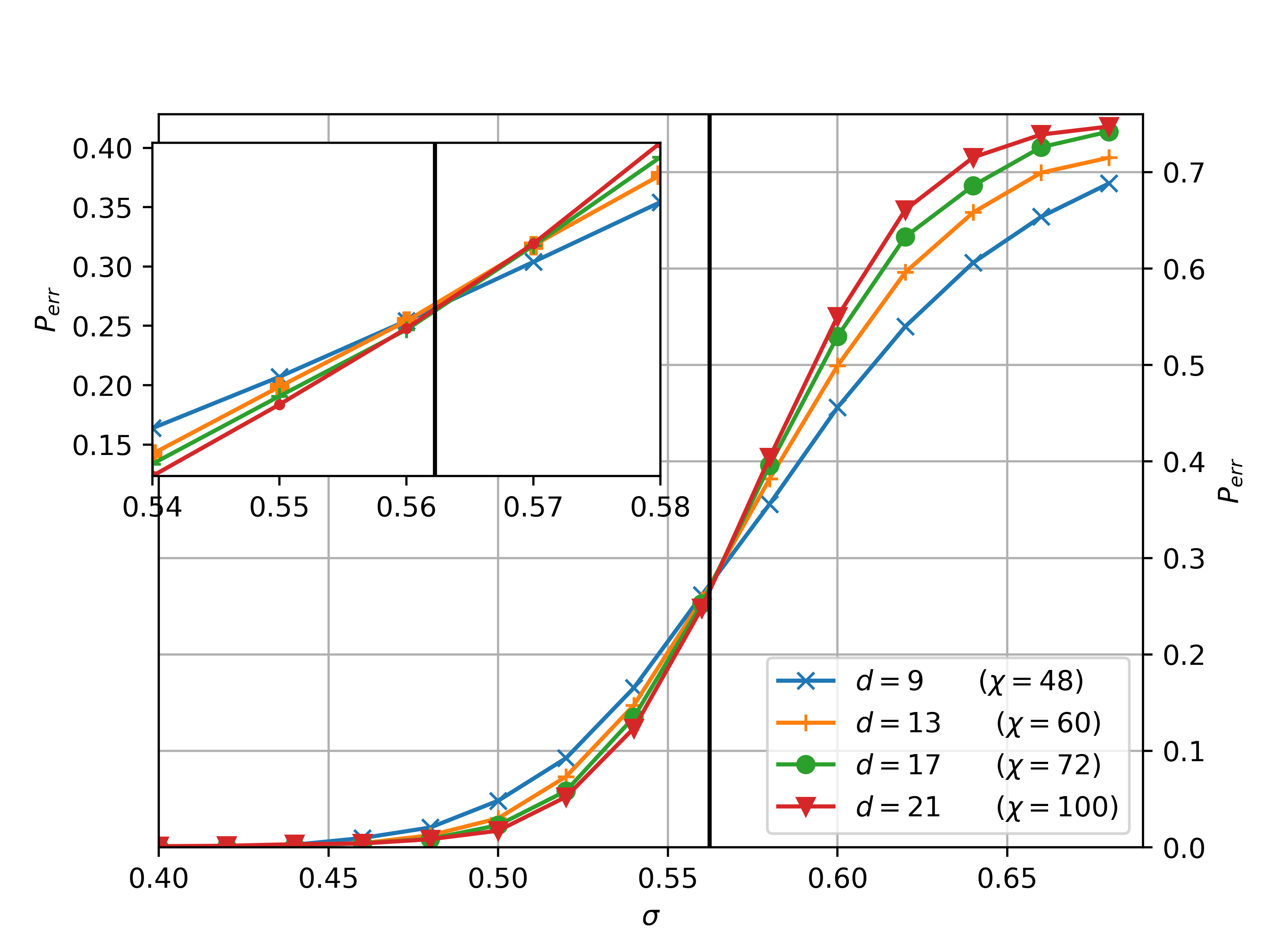}}\\
    \subfloat[Rectangular lattice GKP with side information, threshold $\sigma_c\approx0.6062$ (with standard deviation $ 0.0007$) \label{Ratio2withInfo}]{  \includegraphics[width=.47\textwidth]{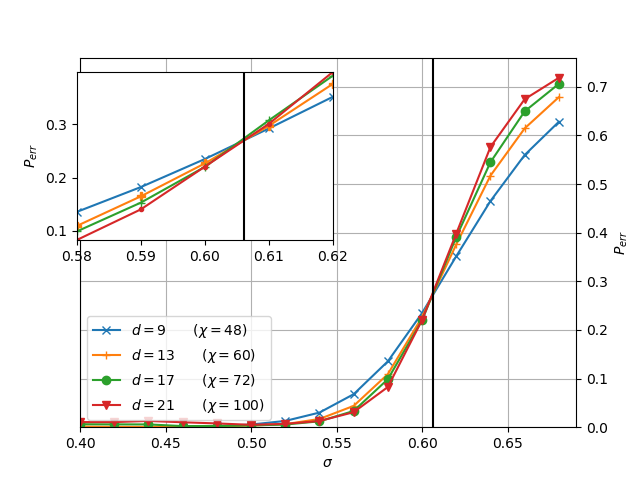}}\\
    \subfloat[Asymmetric hexagonal lattice GKP with side information, threshold $\sigma_c\approx 0.6045$ (with standard deviation $0.0009$) \label{Ratio2withInfohex}]{  \includegraphics[width=.47\textwidth]{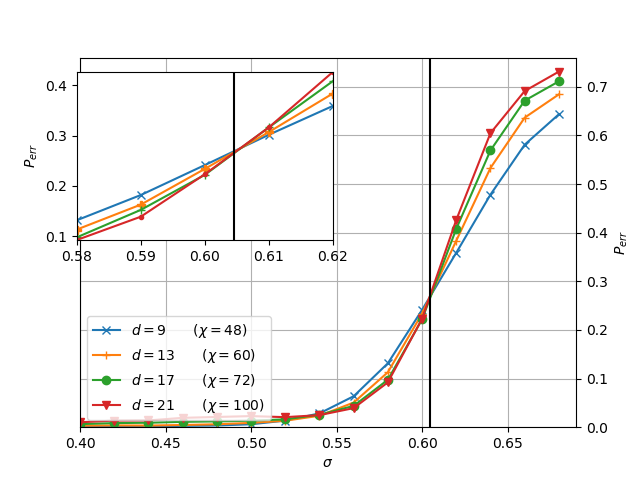}}
    \caption{Other improvements for ratio $r=2.0$: Here the plots show the error probability~$P_{\textrm{err}}(\sigma,d,r)$ as a function of~$\sigma$ for decoding without side information (Fig.~\ref{Ratio2without}) and  with side information (Fig.~\ref{Ratio2withInfo}) for the rectangular lattice GKP code. Fig.~\ref{Ratio2withInfohex} shows the error probability for the asymmetric hexagonal lattice GKP code decoded with side information. Observed threshold estimates~$\sigma_c$ are indicated with vertical lines.  }\label{fig:hexGKPinfo}
\end{figure}

 \subsection{Numerical results for further scenarios\label{sec:resultsasymmetric}}
 Up to this point, we have not considered all possible optimizations over different choices of codes and decoders in our simulations. For example, we restricted our attention to rectangular GKP codes and  did not attempt to optimize the underlying lattice by varying its axial angle. It can be expected that such an optimization yields additional benefits. Indeed, as already pointed out in the seminal paper~\cite{gkp01}, using 
 the hexagonal lattice~$\cL_{\hexagon}$ instead of $\cL_{\square}$ results in  improved error correcting properties because of the difference in volume of the associated Voronoi cells.
 
 A further significant improvement should result from modifying the decoder: so far,  we did not use the GKP syndrome information in the surface code decoding step. For the standard (symmetric) surface-GKP code, it was shown in~\cite{fukuietal18,vuillotetal} that using this side information improves  threshold estimates.
 
 To examine if such modifications lead to improvements as expected, we numerically estimate logical error probabilities for  surface-GKP codes based on the rectangular lattice~$\cL_r$ with $r=2$,  decoded without  and with side information (Fig.~\ref{Ratio2without} and Fig.~\ref{Ratio2withInfo}, respectively). Analogously to the results of~\cite{fukuietal18,vuillotetal} for the symmetric case where a threshold of $\sigma_c\approx 0.61$ was observed, we find that the use of side information significantly increases the threshold  in the asymmetric case also.

As a paradigmatic test case, we additionally consider the use of the asymmetric hexagonal lattice surface GKP code $\mathsf{GKP}(S_{\hexagon}S_r\cL_\square)$ introduced in Section~\ref{sec:hexagonallatticeGKP} for asymmetry ratio~$r=2$, see Fig.~\ref{Ratio2withInfohex}. Again, this  simulation data is for a decoder using side information.  Here the observed threshold is comparable to that obtained for the square lattice surface GKP code.
 
These numerical results indicate that such  additional modifications aimed to improve fault-tolerance properties do not   negatively interfere with the use of asymmetry.  In particular, combining these strategies and additionally optimizing
over lattices (rather than simply the hexagonal and rectangular ones) 
has the  potential to yield  further increases of the threshold estimates.

\section{Conclusions\label{sec:conclusions}} 
Our results show that no-go theorems for Gaussian CV-into-CV encodings (see Section \ref{sec:introduction}) no longer apply when considering concatenated codes: Our numerical results for the (modified) surface-GKP code indicate that suitably chosen Gaussian CV-into-CV encodings (such as the one presented in Section~\ref{sec:engineered}) effectively lead to a deformation of the effective error distribution for the GKP-qubits which   known decoders for the surface code can benefit from. We expect that similar strategies may improve other CV codes constructed by concatenation. Due to the minimal additional experimental requirements, this kind of modification of bosonic fault-tolerance schemes may be particularly attractive once  the basic components of an experimental setup are in place.

Our work provides a proof of principle and shows that artificially engineered asymmetry not only benefits surface-GKP codes, but is also compatible with other improvements to the GKP encoding: this includes e.g., the use of non-rectangular lattices having larger Voronoi cells,  and the use of GKP side information at the 
surface-code decoding stage. 

Future work may seek to establish monotonicity of the logical error probability as a function of the degree of asymmetry. Furthermore, the effect of non-ideal (finitely squeezed) GKP states as well as noise in the syndrome information should be examined. Finally, methods for fault-tolerant computation with asymmetric surface-GKP codes, e.g., using bias-preserving gates  should be further developed.

\section*{Acknowledgements}
RK acknowledges support by the Technical University of Munich – Institute of Advanced Study, funded by the
German Excellence Initiative and the European Union Seventh Framework Programme under grant
agreement no.~291763 and by the DFG cluster of excellence 2111 (Munich Center for Quantum Science
and Technology). 
RK and LH acknowledge support by the German Federal Ministry of Education through the  program
Photonics Research Germany, contract no. 13N14776 (QCDA-QuantERA). 
MH is supported by
the International Max Planck Research School for Quantum Science and Technology at the Max-Planck-Institut
f\"ur Quantenoptik. The authors gratefully acknowledge computing resources  of  the Leibniz-Rechenzentrum.

\newpage
\onecolumngrid
\appendix
\newpage

\section{Computation of coset probabilities}\label{app:compcosetprob}
In this appendix we derive the logical error (coset) probabilities for nearest lattice point decoding (NLPD, cf.\ Section \ref{sec:gkpdecoding}) given a classical isotropic Gaussian displacement noise channel $\cN_{f_{\sigma^2}}$ (cf.\ \eqref{eq:Gaussnoisechannel}); the latter channel is characterized by its density
\begin{align}\label{eq:Gaussiandensity2}
f_{\sigma^2}(\nu)=\frac{1}{2\pi\sigma^2}e^{-\frac{\|\nu\|^2}{2\sigma^2}}\ ,
\end{align} 
where $\|\nu\|^2=\innerprod{\nu}{\nu}=\nu_1^2+\nu_2^2$. We start by proving the formulas \eqref{eq:cosetprobabilitiesgkpnnd}, \eqref{eq:bayesiandecodingpriornnd} for the coset probabilities without and with GKP side information for NLPD for a general random displacement error channel \eqref{eq:randomdisplacementnoisechannel} in Section \ref{app:compcosetprobnnd}. Subsequently, the corresponding explicit expressions \eqref{eq:Gaussianresidualerr}, \eqref{eq:GaussianresidualerrGKPinfo} for a rectangular lattice and the error channel corresponding to~\eqref{eq:Gaussiandensity2} are derived in Section~\ref{app:compcosetprobnndgauss}. Finally, in Section~\ref{app:residuallogerrmixing}, we give an explicit expression for the coset probabilities with GKP side information in the case of an asymmetric hexagonal lattice and density \eqref{eq:Gaussiandensity2}. The latter expression is not presented in the main text but added here for completeness as it is used in the simulations for the asymmetric hexagonal case.

\subsection[A.1]{Nearest lattice point decoding (NLPD)}\label{app:compcosetprobnnd}
Recall that for NLPD (cf.\ Section \ref{sec:gkpdecoding}) we have $c(s(\nu))=-s(\nu)\in \cV^\perp$ and thus 
\begin{align}
\{\nu\in\mathbb{R}^2\ |\ c(s(\nu))+\nu \in  \left[\xi_{\overline{P}}^\bot \right]\}&=\{\nu\in\mathbb{R}^2\ |\ \nu-s(\nu) \in \xi_{\overline{P}}^\bot+ \cL \}\\
&=\{\nu\in\mathbb{R}^2\ |\ \nu=\omega+\xi+\xi_{\overline{P}}^\bot,\ \xi \in  \cL\text{ and }\omega\in \cV^\perp \}\ .
\end{align}
Hence by \eqref{eq:cosetprobabilitiesgkp} and variable substitution we have
\begin{align}
p_{\overline{P}}=\Pr_\nu \left[ c(s(\nu))+\nu \in  \left[\xi_{\overline{P}}^\bot \right]\right]=\sum_{\xi \in \cL} \int_{\cV^\perp} f_Z\left(\nu+\xi+\xi^\bot_{\overline{P}} \right) &d^2\nu\ ,
\end{align}
proving \eqref{eq:cosetprobabilitiesgkpnnd}.
Furthermore, since $s(\nu)=\nu \pmod{ \cL^\perp}$, we have 
\begin{align}
\{\nu\in\mathbb{R}^2\ |\ s(\nu)=s_0\}&=\{\nu\in\mathbb{R}^2\ |\ \nu=\xi^\bot+s_0,\ \xi^\bot\in \cL^\bot\}\ .
\end{align}
By~\eqref{eq:bayesiannorm} and variable substitution it thus follows that
\begin{align}\label{eq:bayesiannormapp}
P_s(s_0)
=\Pr_\nu \left[s(\nu)=s_0\right]
=\sum_{\xi^\bot\in \cL^\bot} f_Z\left(\xi^\bot+s_0 \right)\ .
\end{align}
This together with
\begin{align}
\{\nu\in\mathbb{R}^2\ |\ c(s_0)+\nu \in \left[\xi_{\overline{P}}^\bot \right]\}&=\{\nu\in\mathbb{R}^2\ |\ -s_0+\nu\in \left[\xi_{\overline{P}}^\bot \right]\}\\
&=\{\nu\in\mathbb{R}^2\ |\ \nu=s_0+\xi+\xi_{\overline{P}}^\bot,\ \xi \in  \cL \}\ 
\end{align}
and an additional variable substitution implies, by \eqref{eq:bayesiandecodingprior}, that
\begin{align}\label{eq:bayesiandecodingpriornndapp}
p_{\overline{P}}^{s_0}=&\Pr_\nu \left[c(s(\nu))+\nu \in \left[\xi_{\overline{P}}^\bot \right]|\ s(\nu)=s_0\right]\\
=&P_s(s_0)^{-1}\Pr_\nu \left[s(\nu)=s_0\textrm{ and }c(s(\nu))+\nu \in \left[\xi_{\overline{P}}^\bot \right]\right]\\
=&\left(\sum_{\xi^\bot \in \cL^\perp}f_Z(s_0+\xi^\bot)\right)^{-1} \sum_{\xi \in \cL} f_Z\left(s_0+\xi+\xi^\bot_{\overline{P}} \right) \ .
\end{align}
This yields~\eqref{eq:bayesiandecodingpriornnd}.

\subsection{NLPD with Gaussian noise channel: rectangular lattice}\label{app:compcosetprobnndgauss}
Let now $\cL\equiv \cL_r$ and $f_Z\equiv f_{\sigma^2}$ be given by \eqref{eq:Gaussiandensity2}. Furthermore, let us write $s_0=\tvector{x}{z}$. Then by the definition of the dual lattice $\cL_r^\bot$ (cf.~\eqref{eq:rectangularlattice}) we may write Eq.~\eqref{eq:bayesiannormapp} as
\begin{align}
P_s(s_0)
=&\frac{1}{2\pi\sigma^2}\sum_{n_1,n_2\in\mathbb{Z}} e^{-\frac{(\sqrt{\pi r}n_1+x)^2}{2\sigma^2}}e^{-\frac{(\sqrt{\pi/r}n_2+z)^2}{2\sigma^2}}
=e(r,x)\ e(1/r,z)\ ,\label{eq:comppsgauss}
\end{align}
cf.\ \eqref{eq:euw}. Recall that $(\xi_{\overline{I}}^\bot,\xi_{\overline{X}}^\bot,\xi_{\overline{Y}}^\bot,\xi_{\overline{Z}}^\bot)\coloneqq (0,\xi_2^\bot,\xi_2^\bot-\xi_1^\bot,-\xi_1^\bot)$. If we interpret $\xi_{\overline{P}}^\bot$ as a function of $\overline{P}$ we may thus write 
\begin{align}\xi_{\overline{P}}^\bot=\tvector{\sqrt{\pi r}\cdot\ 1_{\overline{P}\in\{\overline{X},\overline{Y}\}}}{\sqrt{\pi/ r}\cdot 1_{\overline{P}\in\{\overline{Z},\overline{Y}\}}}\ ,
\end{align}
where $1_{\overline{P}\in\{\overline{X},\overline{Y}\}}$, $1_{\overline{P}\in\{\overline{Z},\overline{Y}\}}$ are indicator functions. Then, by the definition of the lattice $\cL_r$ (cf.\ \eqref{eq:rectangularlattice}),
\begin{align}\label{eq:probsomega}
\sum_{\xi \in \cL_r} f_Z\left(s_0+\xi+\xi^\bot_{\overline{P}} \right)
=&\frac{1}{\sqrt{2\pi\sigma^2}}\sum_{n_1\in\mathbb{Z}} 
e^{-\frac{\left(\sqrt{\pi r}\left(2n_1+1_{\overline{P}\in\{\overline{X},\overline{Y}\}}\right)+x\right)^2}{2\sigma^2}}
\frac{1}{\sqrt{2\pi\sigma^2}}\sum_{n_2\in\mathbb{Z}} e^{-\frac{\left(\sqrt{\pi/r}\left(2n_2+1_{\overline{P}\in\{\overline{Z},\overline{Y}\}}\right)+z\right)^2}{2\sigma^2}}\\
=&e\left(4r,x+\sqrt{\pi r}\ 1_{\overline{P}\in\{\overline{X},\overline{Y}\}}\right)\ e\left(4/r,z+\sqrt{\pi/r}\ 1_{\overline{P}\in\{\overline{Z},\overline{Y}\}}\right)\ ,
\end{align}
whence the coset probabilities with GKP information \eqref{eq:bayesiandecodingpriornndapp} can be written as 
\begin{align}
p_{\overline{P}}^{s_0}
=&P_s(s_0)^{-1}\sum_{\xi \in \cL_r} f_Z\left(s_0+\xi+\xi^\bot_{\overline{P}} \right)\\
=&\frac{e\left(4r,x+\sqrt{\pi r}\ 1_{\overline{P}\in\{\overline{X},\overline{Y}\}}\right)}{e(r,x)}\frac{e\left(4/r,z+\sqrt{\pi/r}\ 1_{\overline{P}\in\{\overline{Z},\overline{Y}\}}\right)}{e(1/r,z)}\ .
\end{align}
This together with the identities 
\begin{align}
e(r,x)-e(4r,x+\sqrt{\pi r})=e(4r,x)\ ,\quad e(1/r,z)-e(4/r,z+\sqrt{\pi/ r})=e(4/r,z)
\end{align}
yields \eqref{eq:condlogprobsq}, \eqref{eq:condlogprobsq2} and \eqref{eq:GaussianresidualerrGKPinfo}.

By \eqref{eq:cosetprobfrombayesianerror}, the coset probabilities \eqref{eq:Gaussianresidualerr} without GKP information follow from \eqref{eq:GaussianresidualerrGKPinfo} by integration, cf.\ \eqref{eq:logprobsq}.
Finally, \eqref{eq:1minqx} is verified as
\begin{align}
(1-q_{\overline{X}})=&\int_{\cV_r^\perp}n_{\overline{X}}(s_0)(1-q_{\overline{X}}^{s_0}) d^2 s_0\\
=&\int_{\left\{\left.x=\lambda_1\sqrt{\pi r}\ \right|\ \lambda_1\in[-1/2,1/2]\right\}}e(4r,x) d x\\
=& \sum_{n_1\in\mathbb{Z}} \frac{1}{\sqrt{\pi}}\int_{\sqrt{\frac{2\pi r}{\sigma^2}}\left(n_1-1/4\right)}^{\sqrt{\frac{2\pi r}{\sigma^2}}\left(n_1+1/4\right)}
e^{-\tau_1^2} d \tau_1\\
=&\frac{1}{2}\sum_{n\in\mathbb{Z}} \mathrm{erf}\left(\sqrt{\frac{2\pi r}{\sigma^2}}\left(n+\frac{1}{4}\right)\right)-\mathrm{erf}\left(\sqrt{\frac{2\pi r}{\sigma^2}}\left(n-\frac{1}{4}\right)\right)\ ,
\end{align}
where we used (in this order) the identity
\begin{align}
\int_{\cV_r^\perp}n_{\overline{P}}(s_0) d^2 s_0=1\ , 
\end{align}
the definition of the dual Voronoi cell $\cV_r^\perp$ (cf.\ \eqref{eq:voronoicell}), variable substitution, and the definition of the error function. The identity \eqref{eq:1minqz} is verified analogously.

\subsection{NLPD with Gaussian noise channel: hexagonal lattice}\label{app:residuallogerrmixing}
Here we derive the coset probabilities with GKP side information for an asymmetric hexagonal lattice $\cL_{\hexagon,r}$ (cf.\ Section \ref{sec:hexagonallatticeGKP}), with $f_Z\equiv f_{\sigma^2}$ given in \eqref{eq:Gaussiandensity2}. We have 
\begin{align}
\cL_{\hexagon,r} &= \left\{\left.c\tvector{2\sqrt{\pi r}n_1+\sqrt{\pi/ r}n_2}{ \sqrt{3\pi/ r} n_2 }\ \right|\ n_1,n_2 \in \mathbb{Z} \right\}\ , \\
\cL_{\hexagon,r}^\perp &= \left\{\left. \frac{c}{2}\tvector{2\sqrt{\pi r}n_1+\sqrt{\pi/ r}n_2}{ \sqrt{3\pi/ r} n_2 }\ \right|\ n_1,n_2 \in \mathbb{Z} \right\} \ ,
\end{align}
and 
\begin{align}\label{eq:defvoronoicellhex}
	\cV_{\hexagon,r}^\perp &= \left\{\left. \frac{c}{2}\tvector{2\sqrt{\pi r}\lambda_1+\sqrt{\pi/ r}\lambda_2}{ \sqrt{3\pi/ r} \lambda_2 }\ \right|\ \lambda_1,\lambda_2\in [-\tfrac{1}{2},\tfrac{1}{2}] \right\}\ ,
\end{align}
where $c\coloneqq \left(2/\sqrt{3}\right)^{1/2}$. For $u,w_q,w_p\in \mathbb{R}, u\geq 0$, define
\begin{align}\label{eq:ehuwqwp}
\ehex(u,w_q,w_p)\coloneqq \frac{1}{{2\pi\sigma^2}} \sum_{n_1,n_2\in\mathbb{Z}} e^{-\frac{\left(c\sqrt{\pi u}n_1+(c/2)\sqrt{\pi/ u}n_2+w_q\right)^2+\left((c/2)\sqrt{3\pi/u} n_2+w_p\right)^2}{2\sigma^2}}\ .
\end{align}
Then, with $s_0=\tvector{x}{z}$, Eq.~\eqref{eq:bayesiannormapp} can be written as
\begin{align}
P_s(s_0)=\sum_{\xi^\bot\in \cL_{\hexagon,r}^\bot} f_Z\left(\xi^\bot+s_0 \right)
=&\sum_{n_1,n_2\in\mathbb{Z}} \frac{e^{-\frac{\left((c/2)(2\sqrt{\pi r}n_1+\sqrt{\pi/ r}n_2)+x\right)^2+\left((c/2)\sqrt{3\pi/r} n_2+z\right)^2}{2\sigma^2}}}{2\pi\sigma^2}\\
=&\ehex(r,x,z)\ .\label{eq:denominatorcondprobhex}
\end{align}
Furthermore, writing 
\begin{align}
\xi_{\overline{P}}^\bot=\frac{c}{2}\tvector{2\sqrt{\pi r}\cdot 1_{\overline{P}\in\{\overline{X},\overline{Y}\}}+\sqrt{\pi/ r}\cdot1_{\overline{P}\in\{\overline{Z},\overline{Y}\}}}{ \sqrt{3\pi/r}\cdot 1_{\overline{P}\in\{\overline{Z},\overline{Y}\}} }
\end{align}
(cf.\ Appendix \ref{app:compcosetprobnndgauss}), we get
\begin{align}
\sum_{\xi \in \cL_{\hexagon,r}} f_Z\left(s_0+\xi+\xi^\bot_{\overline{P}} \right)
=&\sum_{n_1,n_2 \in \mathbb{Z} } 
\frac{e^{-\frac{\left(2c\sqrt{\pi r}\left(n_1+\frac{1_{\overline{P}\in\{\overline{X},\overline{Y}\}}}{2}\right)+c\sqrt{\pi/ r}\left(n_2 +\frac{1_{\overline{P}\in\{\overline{Z},\overline{Y}\}}}{2}\right)+x\right)^2+\left(c\sqrt{3\pi/r} \left(n_2+\frac{ 1_{\overline{P}\in\{\overline{Z},\overline{Y}\}}}{2}\right)+z\right)^2}{2\sigma^2}}
}{2\pi\sigma^2}\\
=&\ehex\left(4r,x+c\sqrt{\pi r}1_{\overline{P}\in\{\overline{X},\overline{Y}\}}+\frac{c}{2}\sqrt{\pi/ r}1_{\overline{P}\in\{\overline{Z},\overline{Y}\}},z+\frac{c}{2}\sqrt{3\pi/r} 1_{\overline{P}\in\{\overline{Z},\overline{Y}\}}\right)\ ,\label{eq:nominatorcondprobhex}
\end{align}
whence the coset probabilities with GKP information \eqref{eq:bayesiandecodingpriornndapp} can be written
\begin{align}\label{eq:condprobhex}
p_{\overline{P}}^{s_0}
=&P_s(s_0)^{-1}\sum_{\xi \in \cL_{\hexagon,r}} f_Z\left(s_0+\xi+\xi^\bot_{\overline{P}} \right)\\
=&\frac{\ehex\left(4r,x+c\sqrt{\pi r}\cdot 1_{\overline{P}\in\{\overline{X},\overline{Y}\}}+\frac{c}{2}\sqrt{\pi/ r}\cdot 1_{\overline{P}\in\{\overline{Z},\overline{Y}\}},z+\frac{c}{2}\sqrt{3\pi/r}\cdot 1_{\overline{P}\in\{\overline{Z},\overline{Y}\}}\right)}{\ehex(r,x,z)}\ .
\end{align}

\newpage
\section{Pseudocode for the Monte-Carlo simulation algorithms~\label{sec:simulationpseudocode}}
Here we give pseudocode for the algorithms used to obtain our numerical results in Section~\ref{sec:numerical}. 
The  routine~\textsc{MCWithoutSideInfo} shown in Fig.~\ref{fig:montecarlowithoutsideinfo} simulates one Monte-Carlo step when decoding without side information\footnote{Our actual implementation differs  slightly from the pseudocode in Fig.~\ref{fig:montecarlowithoutsideinfo} for ease of implementation: We directly use the distribution~$\pi$ to sample the errors~$E_{j,k}$. That is, instead of sampling a displacement error vector~$\nu$ from a normal distribution and then computing the corresponding Pauli error~$E_{j,k}$ as in Steps~\ref{step:samplenu} and~\ref{step:computeEjk}, we first compute the distribution~$\pi$ as in Step~\ref{step:computep} and then  sample each $E_{j,k}$ independently and identically from~$\pi$. These two descriptions are of course equivalent because~$\pi$ is the induced distribution over (single-qubit) errors.\label{foot:actualimplement}}. In this description, the errors~$E_{j,k}$ are sampled exactly from the distribution~$\pi$ and the BSV decoder is given the exact probabilities~$\pi$.
  \begin{figure*}[ht!]
  \begin{algorithmic}[1]
  \Statex
  \Function{MCWithoutSideInfo}{$\sigma^2$,$d$,$r$}\\ \hrulefill\\
  \textbf{Input:} Variance~$\sigma^2$, code distance $d$, asymmetry ratio~$r$\\
  \textbf{Output:} Residual logical Pauli error $\overline{P}\in \{\overline{I},\overline{X},\overline{Y},\overline{Z}\}$     \\ \hrulefill
\For{each qubit $(j,k)$}
   \State{Sample displacement error vector $\nu_{j,k}\in\mathbb{R}^2$ according to~$\mathsf{N}({\bf 0},\sigma^2 I_{2})$.\label{step:samplenu}}
   \State{Compute the (logical) Pauli error $E_{j,k}\in \{I,X,Y,Z\}$ from $\nu_{j,k}$ according to~\eqref{eq:residualerrornlpd} and \eqref{array:nontrivialerror} (for $\GKP(\cL_r)$).\label{step:computeEjk}}
   \EndFor
   \State{Set $E_s:=\otimes_{j,k} E_{j,k}$.}
      \State{Compute the distribution  $\pi=(p_{\overline{I}},p_{\overline{X}},p_{\overline{Y}},p_{\overline{Z}})$ according to~\eqref{eq:Gaussianresidualerr} (for $\GKP(\cL_r)$).\label{step:computep}
   }
      \State{Use the BSV decoder with input $\pi^n$ to compute an $n$-qubit Pauli correction $C$.}
   \State{Compute $\overline{P}=\arg \max_{\overline{P}\in \{\overline{I},\overline{X},\overline{Y},\overline{Z}\}} 1_{E_sC\in \overline{P}\cS}$  (i.e., decide which coset $E_sC$ belongs to).}
   \EndFunction
 \end{algorithmic}
\caption{Subroutine \textsc{MCWithoutSideInfo}  simulates one instance of the error-recovery process when GKP side information is ignored. It returns the residual logical error.  
 \label{fig:montecarlowithoutsideinfo}}
\end{figure*}

 We note that Fig.~\ref{fig:montecarlowithoutsideinfo} does not yet constitute a practical algorithm: Because  expression~\eqref{eq:Gaussianresidualerr} for $\pi$~involves infinite sums, an  approximation has to be made in our implementation, i.e., we work with a suitably chosen  approximation~$\tilde{\pi}$ to $\pi$. We argue in Section~\ref{app:cutoff} that using this approximation~$\tilde{\pi}$  does not impact the validity of our conclusions about the effect of asymmetry.

In Fig.~\ref{fig:montecarlowithsideinfo}, we give pseudocode for the simulation of  error-correction  when using GKP side information. Again, in our implementation, Step~\ref{step:computationapproximate} in this algorithm is replaced by an approximate computation described and analyzed  in Section~\ref{app:cutoff}.

 \begin{figure*}[ht!]
  \begin{algorithmic}[1]
  \Statex
  \Function{MCWithSideInfo}{$\sigma^2$, $d$, $r$}\\ \hrulefill\\
  \textbf{Input:} Variance~$\sigma^2$, code distance~$d$, asymmetry ratio~$r$\\
  \textbf{Output:} Residual logical Pauli error $\overline{P}\in \{\overline{I},\overline{X},\overline{Y},\overline{Z}\}$     \\ \hrulefill
  \For{each qubit $(j,k)$}
   \State{Sample displacement error vector $\nu_{j,k}\in\mathbb{R}^2$ according to~$\mathsf{N}({\bf 0},\sigma^2 I_2)$.}
   \State{Compute the (logical) Pauli error $E_{j,k}\in \{I,X,Y,Z\}$ from $\nu_{j,k}$ according to~\eqref{eq:residualerrornlpd} and \eqref{array:nontrivialerror} (for $\GKP(\cL_r)$ respectively $\GKP(\cL_{\hexagon,r})$).}
   \State{Compute the syndrome $s_0=s(\nu_{j,k})=\tvector{z}{y}$
   according to~\eqref{eq:syndromecomputation}.}
   \State{Compute the conditional distribution  $\pi_{j,k}=(p^{s_0}_{\overline{I}},p^{s_0}_{\overline{X}},p^{s_0}_{\overline{Y}},p^{s_0}_{\overline{Z}})$ according to~\eqref{eq:GaussianresidualerrGKPinfo} (for $\GKP(\cL_r)$)
   respectively~\eqref{eq:condprobhex} (for $\GKP(\cL_{\hexagon,r})$).\label{step:computationapproximate}
   }
   \EndFor
   \State{Set $E_s:=\otimes_{j,k} E_{j,k}$.}
   \State{Use the BSV decoder with input $\prod_{j,k}\pi_{j,k}$ to compute an $n$-qubit Pauli correction $C$.}
   \State{Compute $\overline{P}=\arg \max_{\overline{P}\in \{\overline{I},\overline{X},\overline{Y},\overline{Z}\}} 1_{E_sC\in \overline{P}\cS}$  (i.e., decide which coset $E_sC$ belongs to).}
   \EndFunction
 \end{algorithmic}
\caption{Subroutine \textsc{MCWithSideInfo} Monte-Carlo-simulates one instance of the error-recovery process when GKP side information is used. It returns the residual logical error.
\label{fig:montecarlowithsideinfo}}
\end{figure*}

\section{Cutoff analysis}\label{app:cutoff}
For the computation of the distributions $\pi$ (cf. Step~\ref{step:computep} in  Fig.~\ref{fig:montecarlowithoutsideinfo}) and $\pi_{j,k}$ (cf. Step~\ref{step:computationapproximate} in  Fig.~\ref{fig:montecarlowithsideinfo}),
 infinite sums are approximated by finite ones by introducing a cutoff~$\kappa$. More precisely, the probability distribution~$\pi$  
is computed using the expressions \eqref{eq:1minqx}, \eqref{eq:1minqz} with a  cutoff~$\kappa=10$, meaning that all summands with indices of absolute value $\geq 10$ are neglected. Similarly, for the conditional distribution~$\pi_{j,k}$, a cutoff~$\kappa=15$ is chosen when evaluating expressions \eqref{eq:GaussianresidualerrGKPinfo} 
   respectively~\eqref{eq:condprobhex}.
    Let us denote the corresponding approximate distributions by~$\tilde{\pi}$ and~$\tilde{\pi}_{j,k}$, respectively. 
   Here we discuss the impact of using these approximations instead of the actual distributions~$\pi$ respectively $\pi_{j,k}$ on the accuracy of our simulation.

We first observe that these approximations do not enter in the sampling procedure generating the errors~$E_{j,k}$ when following the description of Fig.~\ref{fig:montecarlowithoutsideinfo} or Fig.~\ref{fig:montecarlowithsideinfo}. That is, the simulation algorithm produces the correct distribution over errors\footnote{For the modified algorithm described in Footnote~\ref{foot:actualimplement},
this is only approximately the case. But the difference is negligible as $\tilde{\pi}$ is a good approximation to~$\pi$ as argued here.}. This means that  only the effect of using $\tilde{\pi}$ instead of the exact distribution~$\pi$ (respectively~$\tilde{\pi}_{j,k}$ instead of $\pi_{j,k}$) as input to the BSV decoder has to be considered.

If the BSV decoder receives an approximation of the actual a priori probabilities~$\pi$ (or $\pi_{j,k}$),
its  action is not that of a maximum likelihood decoder even in the limit of infinite bond dimension~$\chi$. Nevertheless, it still acts as a (hopefully decent)  decoder: the fact that the bond dimension is finite and the input probabilities are approximate as a consequence of a finite cutoff~$\kappa$ can  only possibly affect the performance of the decoder. In other words, the simulation algorithms shown in Figs.~\ref{fig:montecarlowithoutsideinfo} and~\ref{fig:montecarlowithsideinfo}  allow us to compute estimates of logical  error probabilities of {\em some }decoder (with and without side information). 

Running these simulation  algorithms  hence establishes lower bounds on achievable error thresholds  obtained by using non-ideal decoders. Since our general goal is to show that asymmetry is beneficial, this means that we do not need to worry about the choice of cutoff~$\kappa$ as long as we observe an improvement over symmetric codes (i.e., asymmetry ratio~$r=1$). 

In this comparison, it is of course important to have a reliable estimate of the threshold for the symmetric case~$r=1$ since this is our reference point. We make three observations concerning this point:

First, from an operational viewpoint, it is natural to optimize the decoding success probability while restricting attention to efficient decoders. We use the BSV decoder with particularly high bond dimension (see Section~\ref{sec:bondimensiontest}). This should guarantee that 
the decoder provides a good approximation to maximum likelihood decoding while still being efficient.

Second, our threshold estimate for the symmetric case is comparable  to previously obtained threshold values (cf.\ Section \ref{sec:results}) in this well-studied scenario, increasing confidence in their reliability.

Third, we argue here that the approximate probability distributions~$\tilde{\pi}$  are -- for the considered parameter regime and cutoff~$\kappa$ -- very close to the actual probability distributions~$\pi$.   That is, the input to the BSV decoder only differs by a small amount from the ideal input. While we do not make any claims about the continuity properties of the BSV decoder, this is at least some indication that using these approximate distributions is justified.

For this analysis, define  the quantities
\begin{align}
\alpha_+\coloneqq & \max_{r,\sigma}\left\{\sqrt{\frac{2\pi r}{\sigma^2}} \sqrt{\frac{2\pi }{r\sigma^2}}\right\}\ \qquad\textrm{ and }\qquad 
\alpha_-\coloneqq  \min_{r,\sigma}\left\{\sqrt{\frac{2\pi r}{\sigma^2}}, \sqrt{\frac{2\pi }{r\sigma^2}}\right\}\ ,
\end{align}
where the optimizations are over the parameter values $(r,\sigma)$~considered in our simulations. The logical error probabilities without GKP side information \eqref{eq:Gaussianresidualerr} are computed via the probabilities $q_{\overline{X}}$, $q_{\overline{Z}}$, i.e., the expressions \eqref{eq:1minqx}, \eqref{eq:1minqz}. 
As the slope of the error function $\mathrm{erf}(\tau)$ is decreasing with increasing $|\tau|$, the contribution of the summands in \eqref{eq:1minqx}, \eqref{eq:1minqz} decreases with $n\in\mathbb{Z}$. For $n\not=0$, the function 
\begin{align}
\mathrm{inc}(n)\coloneqq \frac{1}{2}\left(\mathrm{erf}\left(\alpha_-n+\frac{\alpha_+}{4}\right)-\mathrm{erf}\left(\alpha_-n-\frac{\alpha_+}{4}\right)\right)
\end{align}
is thus an upper bound on the increment caused by the $n$-th summand in both cases \eqref{eq:1minqx}, \eqref{eq:1minqz}. Note that the error function is an odd function and thus $\mathrm{inc}(n)=\mathrm{inc}(-n)$. We can therefore bound the error introduced by a cutoff $\kappa\geq 1$ by
\begin{align}\label{eq:bounderrorrect}
\max_{\overline{P}\in\{\overline{X},\overline{Z}\}}(1-q_{\overline{P}})-(1-\tilde{q}_{\overline{P}})&\leq 2\sum_{n=\kappa}^{\infty}\mathrm{inc}(n)\\
&=\frac{2}{\sqrt{\pi}}\sum_{n=\kappa}^{\infty}\int_{\alpha_-n-\alpha_+/4}^{\alpha_-n+\alpha_+/4}e^{-\tau^2}d \tau\\
&\leq  \frac{2}{\sqrt{\pi}}\int_{\alpha_-\kappa-\alpha_+/4}^{\infty}e^{-\tau^2}d \tau\\
&\leq  \frac{2}{\sqrt{\pi}}\int_{\alpha_-\kappa-\alpha_+/4}^{\infty}\frac{\tau}{\alpha_-\kappa-\alpha_+/4}e^{-\tau^2}d \tau\\
&=  \frac{1}{(\alpha_-\kappa-\alpha_+/4)\sqrt{\pi}}\ e^{-(\alpha_-\kappa-\alpha_+/4)^2}\ .
\end{align}
Here $\tilde{q}_{\overline{X}}$ and $\tilde{q}_{\overline{Z}}$ are given by the expressions~\eqref{eq:1minqx}, \eqref{eq:1minqz} with the sum over $n\in\mathbb{Z}$ replaced by a sum from $-\kappa+1$ to $\kappa-1$. On the other hand, $1-q_{\overline{X}}$ and $1-q_{\overline{Z}}$ as given in \eqref{eq:1minqx}, \eqref{eq:1minqz} can be bounded from below by their summand $n=0$ respectively. The latter in turn is in both cases  greater than or equal to  $\mathrm{erf}\left(\alpha_-/4\right)$. Thus the ratio of the approximation error in the calculation of \eqref{eq:1minqx}, \eqref{eq:1minqz} and the true values is 
\begin{align}\label{eq:bounderrorrect}
\max_{\overline{P}\in\{\overline{X},\overline{Z}\}}\frac{(1-q_{\overline{P}})-(1-\tilde{q}_{\overline{P}})}{1-q_{\overline{P}}}\leq\left(\sqrt{\pi}(\alpha_-\kappa-\alpha_+/4)\ \mathrm{erf}\left(\alpha_-/4\right)\ e^{(\alpha_-\kappa-\alpha_+/4)^2}\right)^{-1}.
\end{align}
For the given parameter values $r\in [1,4]$ and $\sigma\in [4/10,7/10]$, the parameters $\alpha_+$ and $\alpha_-$ evaluate to $(\alpha_+,\alpha_-)=5\sqrt{2\pi}(1,1/7)$. Inserting this into~\eqref{eq:bounderrorrect} shows that with a cutoff~$\kappa=10$, the approximate values~$1-\tilde{q}_{\overline{X}},1-\tilde{q}_{\overline{Z}}$ are within $10^{-93}\%$ of the correct values~$1-q_{\overline{X}},1-q_{\overline{Z}}$. Since the probabilities of interest are given by degree-$2$
 polynomials in $(q_{\overline{X}},q_{\overline{Z}})$, see Eq.~\eqref{eq:ourpxpypzdependence}, this shows that the introduction of a cutoff leads 
 to a negligible error in the computation of~$\pi$.

\section{Choice of parameters}\label{sec:choiceofparams}
In this appendix, we discuss the choice of parameters
used in our simulation. This includes physical parameters such as the  code size~$d$ (Section~\ref{sec:codesizes}), the noise strength as quantified by the variance~$\sigma$ (Section~\ref{sec:noiselevels}), and the asymmetry ratio~$r$ (Section~\ref{sec:asymmetryratioused}).

We also discuss parameters related to the simulation, including the number of Monte-Carlo iterations (Section~\ref{sec:nomontecarlo}), as well as the necessary bond dimension~$\chi$ in the BSV-decoder (Section~\ref{sec:bondimensiontest}). 

\subsection{Code sizes\label{sec:codesizes}}
We consider code sizes $d\in\{9,13,17,21\}$, which is comparable to what has been used in~\cite{tuckettetal18, tuckettetal19} to study the performance of the surface code under biased noise using the BSV decoder.

\subsection{Noise strengths (variance)\label{sec:noiselevels}}
To identify noise levels of interest, we use previously known results: In~\cite{vuillotetal}, it was shown that (standard)
square lattice toric-GKP codes can tolerate isotropic Gaussian displacement noise with standard deviations up to values between $\sigma\approx 0.54$ and $\sigma\approx 0.55$ when a certain weighted minimum-matching decoding is used without GKP side information, and around $\sigma\approx 0.61$ when GKP side information is employed. Based on this, we simulate noise levels given by standard deviations $\sigma\in [0.4,0.7)$ in increments of~$0.02$ (a higher resolution of~$0.01$ is used in the vicinity of empirical thresholds). 

\subsection{Asymmetry ratios\label{sec:asymmetryratioused}}
As discussed in Section~\ref{sec:biasednoiserectGKP}, a lattice asymmetry ratio $r$ corresponds to an increase of $10\log_{10}(r)$ in squeezing w.r.t.\ the square lattice GKP code (we consider $r\geq 1$). 
 The relevant quantity  to decide whether an amount of  squeezing is physically reasonable is the \emph{total} squeezing resulting from the isotropic noise variance~$\sigma^2$ and the ratio~$r$. The physically reasonable range of~$r$ therefore depends on the value of~$\sigma$ under consideration.
Here we consider asymmetry ratios~$r\in [1,4]$ with increments of $0.5$, which translate to a squeezing $<11$ dB for the considered range of variances $\sigma\in [0.4,0.7]$. Note that this is around the amount of squeezing within reach of near term experimental setups, see e.g.\ the discussion in~\cite{fukuietal18}. Ratios up to $r=4$ turned out to be sufficient for the numerical demonstration of the beneficial effect of asymmetry.

\subsection{Number of Monte Carlo iterations\label{sec:nomontecarlo}}
The number of simulations completed for every tuple $(\sigma,d,r)$ depends on the distance $d$: the distances $9,13,17$, and $21$ are simulated approximately $50000, 30000, 30000$, and $10000$ times respectively; for higher distances the fluctuations decrease and hence the empirical estimates converge faster. For the bond dimension tests (see Section \ref{sec:bondimensiontest}) the number of simulations for all distances is approximately $30000$. 

\subsection{Necessary bond dimension\label{sec:bondimensiontest}}
 For given choices of $\sigma, r$ and $d$, the bond dimension~$\chi$ has to be chosen suitably large in order for the BSV decoder to be sufficiently accurate.

\begin{figure}
 \begin{minipage}{0.48\linewidth}
    \subfloat[Ratio $r=2.0$\label{bdRatio15}]{  \includegraphics[width=.99\textwidth]{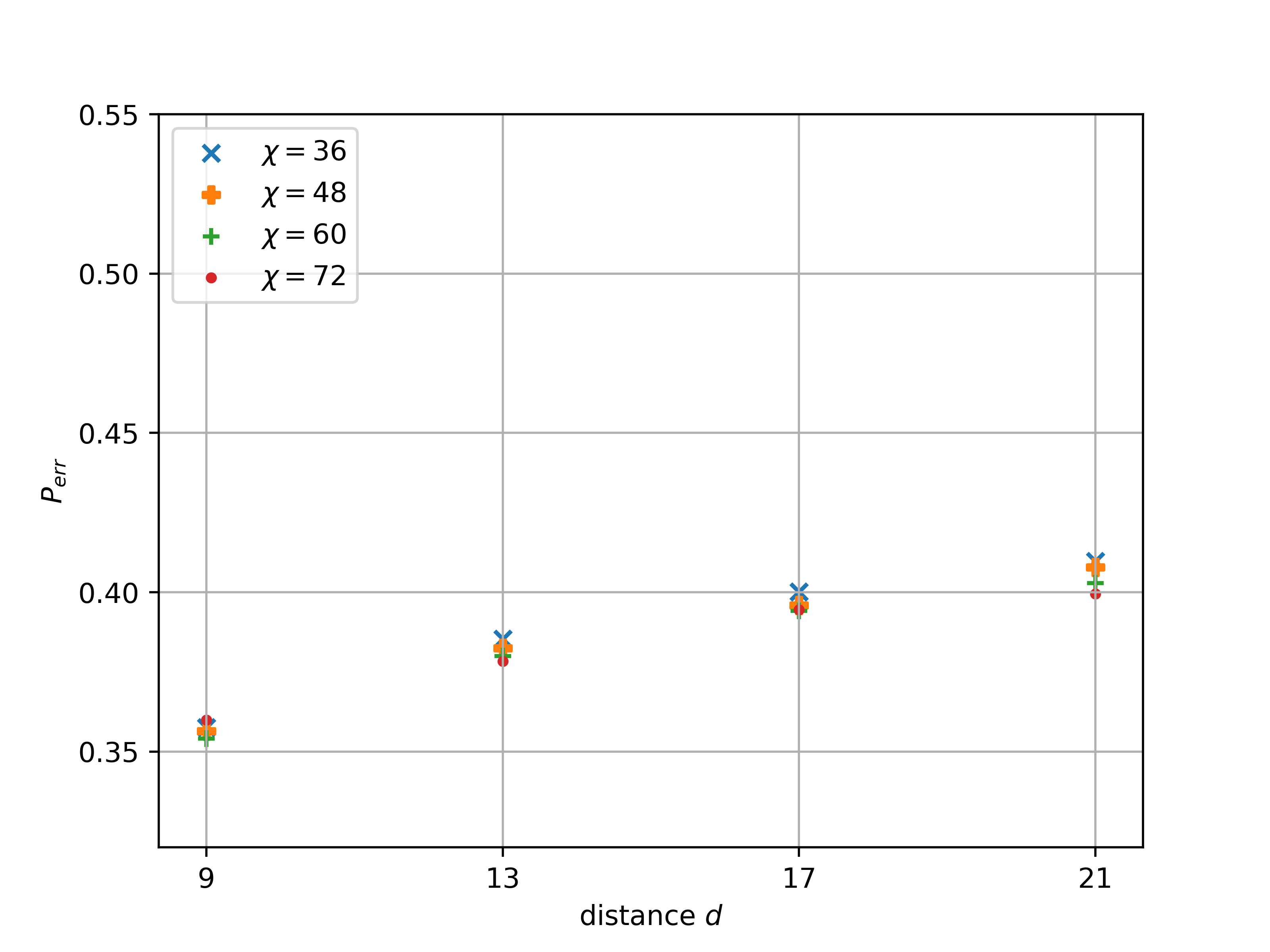}}\\
    \subfloat[Ratio $r=3.0$\label{bdRatio2}]{  \includegraphics[width=.99\textwidth]{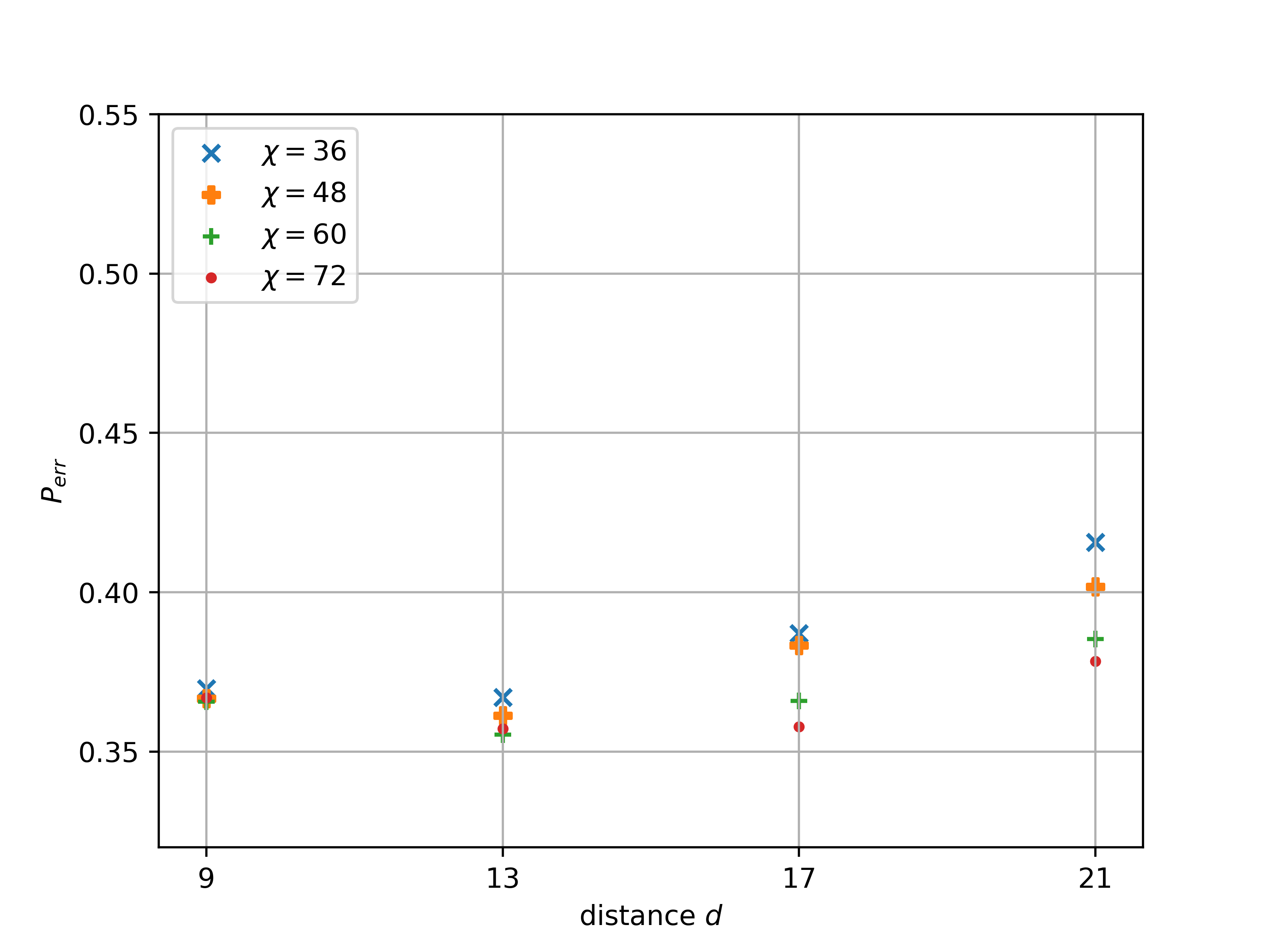}}\\
    \subfloat[Ratio $r=4.0$\label{bdRatio3}]{  \includegraphics[width=.99\textwidth]{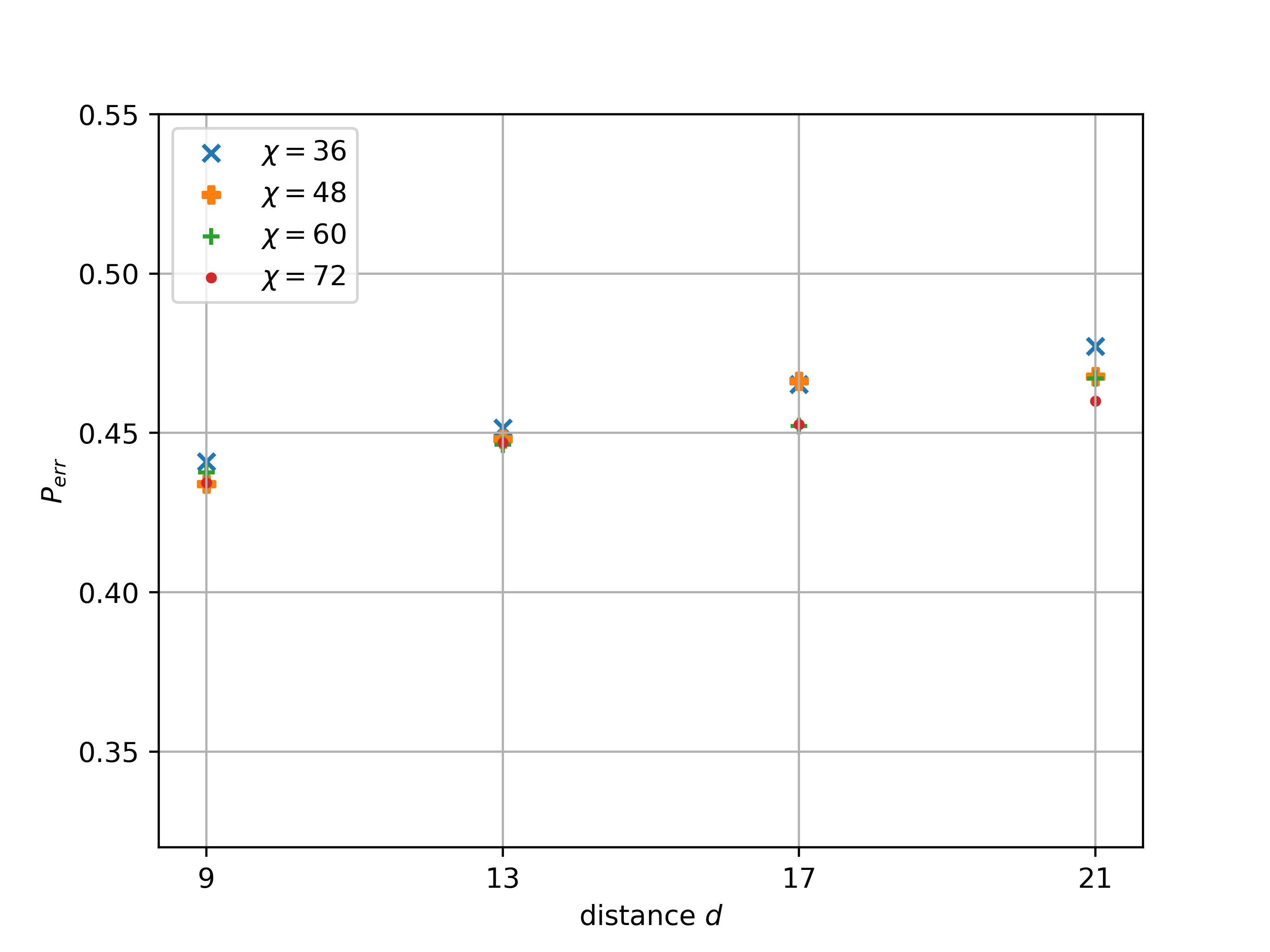}}
\end{minipage} 
\begin{minipage}{0.48\linewidth}
    \subfloat[Distance $d=9$\label{bdDist9}]{  \includegraphics[width=.99\textwidth]{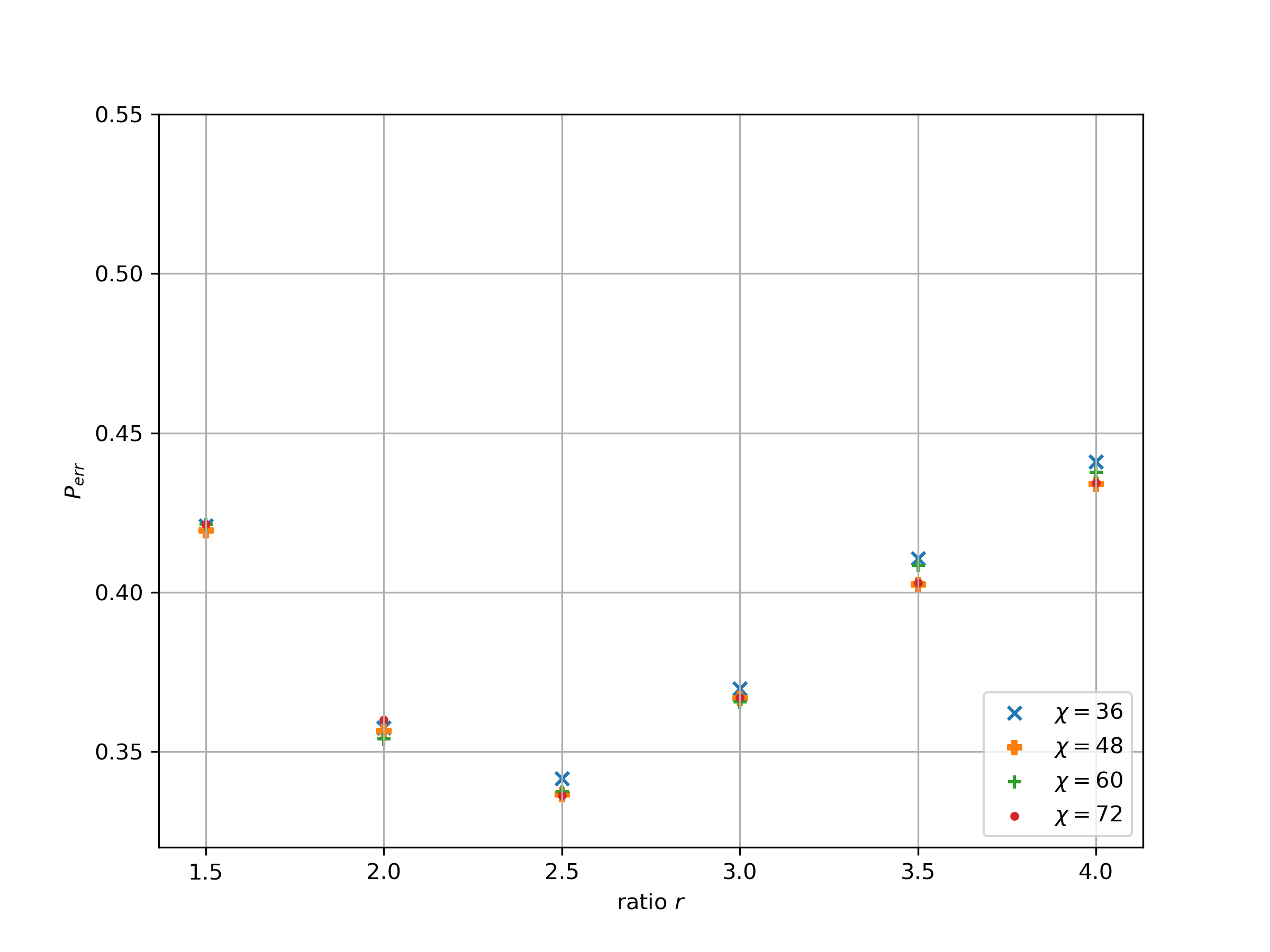}}\\
    \subfloat[Distance $d=13$\label{bdDist13}]{  \includegraphics[width=.99\textwidth]{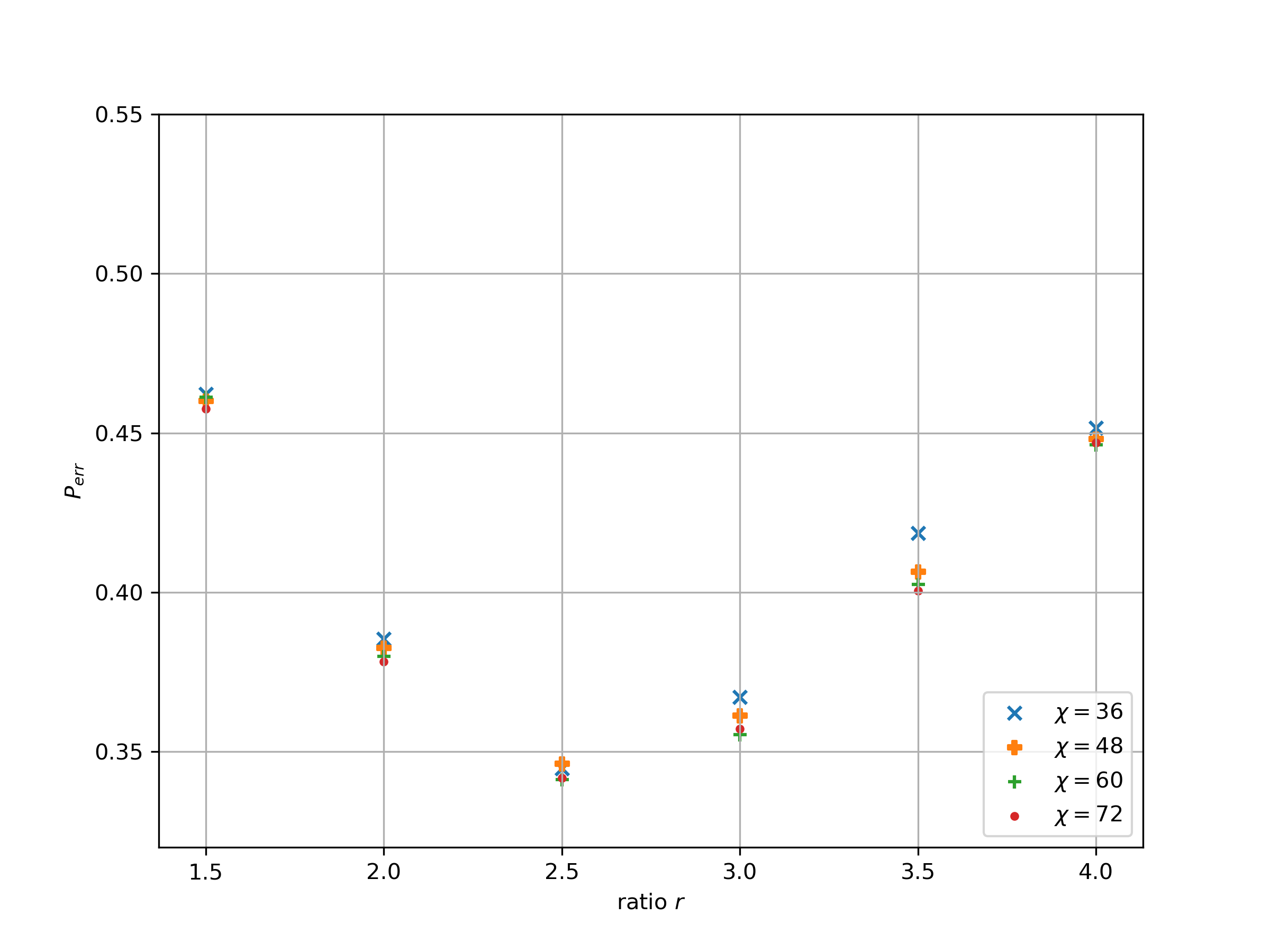}}\\
    \subfloat[Distance $d=17$\label{bdDist17}]{  \includegraphics[width=.99\textwidth]{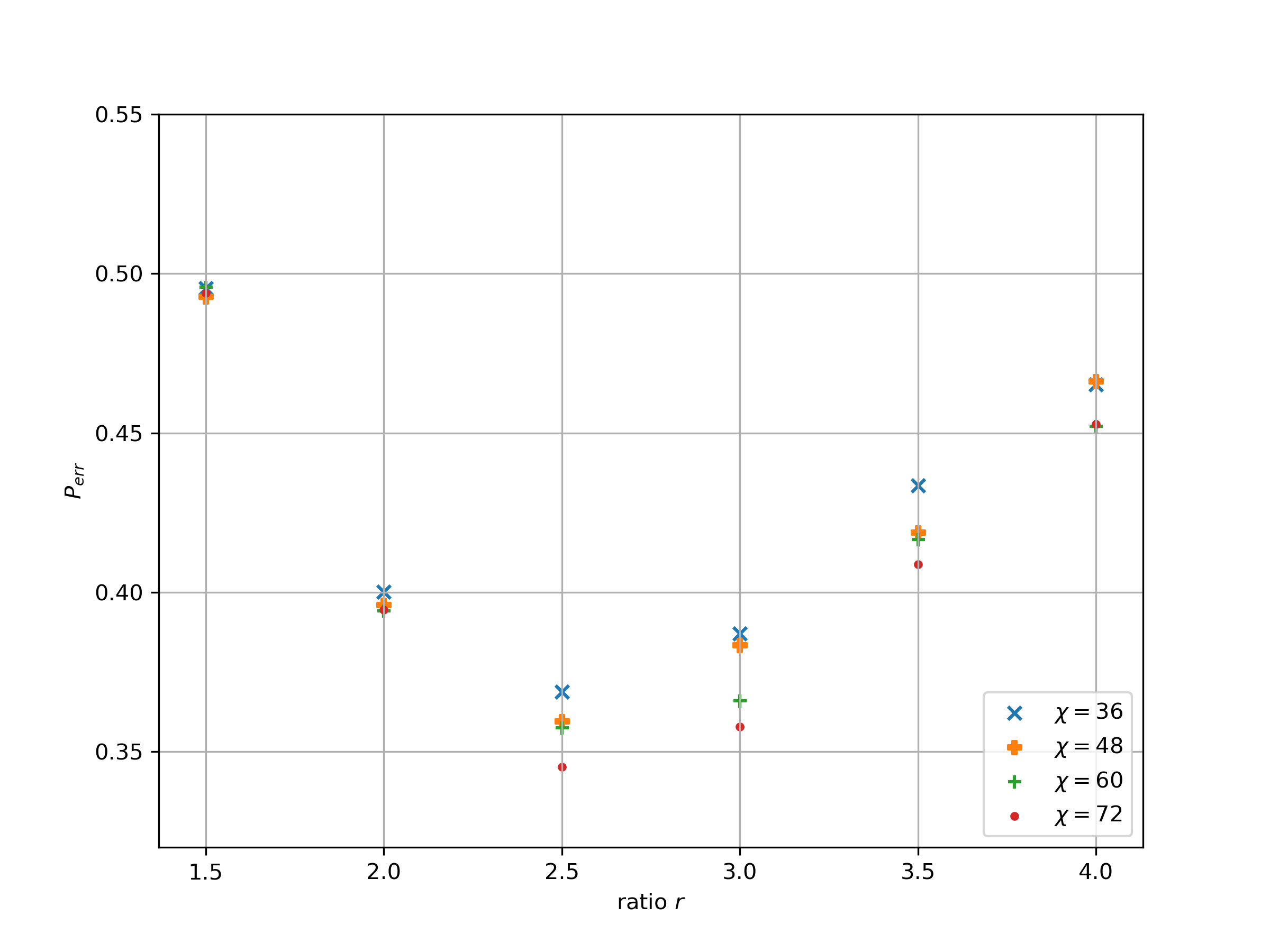}}
\end{minipage}
     
 \caption{Test of the necessary bond dimension for the GKP code (without side information): Error probabilities are estimated for different bond dimensions~$\chi$. Heuristically, a bond dimension~$\chi$ suffices if ~$P_{err}$ does not decrease significantly when the bond dimension is increased further. We observe that 
 the required bond dimension~$\chi$ appears to be non-monotonic as a function of the asymmetry ratio~$r$, being maximal in some intermediate regime of~$r$ (left hand side).
  The required bond dimension increases with increasing distance~$d$ (right hand side).}\label{fig:GKPBDtest}
\end{figure}

To determine appropriate bond dimensions for our model, we conducted test simulations for distances $d\in\{9,13,17,21\}$ and ratios $r\in [1.5,4]$ (in increments of $0.5$), for a \emph{fixed} standard deviation~$\sigma=0.58$ (i.e., in the vicinity of the expected threshold estimate), and for bond dimensions $\chi\in \{36,48,60,72\}$. Note that $\chi=36$ and $\chi=48$ are the bond dimensions used in \cite{tuckettetal19} and \cite{tuckettetal18}, respectively. Examples are depicted in Fig.~\ref{fig:GKPBDtest}. We find a similar behaviour as observed in~\cite{tuckettetal18}: intermediate ratios (biases) require higher bond dimensions. Furthermore, the necessary bond dimension increases with the distance $d$. Our tests show that for distances $d=9,13,17$ the bond dimensions $\chi=48,60,72$ are sufficient respectively, for all considered ratios. For distance $d=21$ we choose~$\chi=100$, since  the tests indicate that $\chi=72$ is insufficient. Also, the values for ratio $r=1$, which serve as benchmarks, are simulated with bond dimension $\chi=100$ to prevent any bias towards our interpretation. 
As explained in the main text, for the asymmetric $(r>1)$ cases choosing
lower bond dimensions~$\chi$ suffices. This is because any choice of~$\chi$ provides estimates for the logical error probability of some decoder. This suffices to demonstrate improvements over the symmetric $(r=1$) case.

\newpage

%\bibliographystyle{plain}
%\bibliography{q}

\end{document}